\input harvmac.tex

%\draftmode
%\input toc.tex
%\writetoc
%\listtoc
\input epsf.tex
\def\figin{\epsfcheck\figin}\def\figins{\epsfcheck\figins}
\def\epsfcheck{\ifx\epsfbox\UnDeFiNeD
\message{(NO epsf.tex, FIGURES WILL BE IGNORED)}
\gdef\figin##1{\vskip2in}\gdef\figins##1{\hskip.5in}% blank space instead
\else\message{(FIGURES WILL BE INCLUDED)}%
\gdef\figin##1{##1}\gdef\figins##1{##1}\fi}
\def\DefWarn#1{}
\def\figinsert{\goodbreak\midinsert}
\def\ifig#1#2#3{\DefWarn#1\xdef#1{fig.~\the\figno}
\writedef{#1\leftbracket fig.\noexpand~\the\figno}%
\figinsert\figin{\centerline{#3}}\medskip\centerline{\vbox{\baselineskip12pt
\advance\hsize by -1truein\noindent\footnotefont{\bf
Fig.~\the\figno:} #2}}
\bigskip\endinsert\global\advance\figno by1}
%%% TO PUT FIGURES INSERT:
%\ifig\fivelegs{ Write caption
%  } {\epsfxsize2.5in\epsfbox{fivelegs.eps}}

\input epsf
% +--------------------------------------------------------------------+
% |                                                                    |
% |                           TABLES.TEX                               |
% |                                                                    |
% |                     Ray F. Cowan  15-Feb-85                        |
% |                                                                    |
% |                       Princeton University                         |
% |                                                                    |
% |          Present Address:  Laboratory for Nuclear Science          |
% |                            M.I.T.                                  |
% |                            Cambridge, MA 02139                     |
% |                                                                    |
% |                   E-mail:  rfc@slacvm.slac.stanford.edu            |
% |                                                                    |
% |                                                                    |
% |                     Last Revision: 17-Apr-86                       |
% |                                                                    |
% |   Macros I find handy for making tables.  See TABLEDOC TEX for     |
% |   a longer description.  The token-counting macros are straight    |
% |   from the TeXbook's "Dirty Tricks" appendix.                      |
% |                                                                    |
% +--------------------------------------------------------------------+
%
\newbox\hdbox%
\newcount\hdrows%
\newcount\multispancount%
\newcount\ncase%
\newcount\ncols% This is the number of primary text columns in the table.
\newcount\nrows%
\newcount\nspan%
\newcount\ntemp%
\newdimen\hdsize%
\newdimen\newhdsize%
\newdimen\parasize%
\newdimen\spreadwidth%
\newdimen\thicksize%
\newdimen\thinsize%
\newdimen\tablewidth%
\newif\ifcentertables%
\newif\ifendsize%
\newif\iffirstrow%
\newif\iftableinfo%
\newtoks\dbt%
\newtoks\hdtks%
\newtoks\savetks%
\newtoks\tableLETtokens%
\newtoks\tabletokens%
\newtoks\widthspec%
%
%  Book-keeping stuff--see how often these macros are called.
%
%  MOD RFC 900221.
%  Removed usage logging:  it's too complicated under VM/XA.
%\immediate\write15{%
%CP SMSG GJMSINK TEXTABLE --> TABLE MACROS V. 851121 JOB = \jobname%
%}%
%
%  Turn on table diagnostics.
%
\tableinfotrue%
\catcode`\@=11%  Allows use of "@" in macro names, like PLAIN.TEX does.
%  Debugging aid.  Writes #1 on the
%                                    user's terminal and in the log file.
%
%  Define the \tstrut height, depth in terms of the x_height parameter.
%
\def\tstrut{\vrule height3.1ex depth1.2ex width0pt}%
\def\and{\char`\&}%  Allows us to get an `&' in the text.  This is the
%                    same as using the PLAIN TeX macro \&.
\def\tablerule{\noalign{\hrule height\thinsize depth0pt}}%
\thicksize=1.5pt%  Default thickness for fat rules.  The user should feel
%                  free to change this to his preference.
\thinsize=0.6pt%   Default thickness for thin rules.
\def\thickrule{\noalign{\hrule height\thicksize depth0pt}}%
\def\ctr#1{\hfil\ #1\hfil}%
%
%
%
%  Here are things for controlling the width of the finished table.
%
\tablewidth=-\maxdimen%
\spreadwidth=-\maxdimen%
\def\tabskipglue{0pt plus 1fil minus 1fil}%
%
%  Stuff for centering or not.
%
\centertablestrue%
%
%
%
%  \vctr vertically centers its argument in the row.
%
\parasize=4in%
\gdef\ARGS{########}%  Produces the correct number of #'s in the preamble
%                      by the time eveything is expanded and \halign sees
%                      it.
\gdef\headerARGS{####}%  Same as \ARGS, but used in \header macros.
\def\@mpersand{&}%  Allows us to get alignment tab characters later
%                   when we have made the character "&" an active macro.
{\catcode`\|=13%  Make |'s locally active.
\gdef\letbarzero{\let|0}%  Globally define a macro that allows us to
%                          keep active |'s from being expanded in edef's.
\gdef\letbartab{\def|{&&}}%
\gdef\letvbbar{\let\vb|}%
%  This \def will cause active |'s read by
%                            \ruledtable to be converted into double
%                            alignment tabs.
}%  End of locally active |'s.
{\catcode`\&=4%  Make these alignment tabs.
\def\ampskip{&\omit\hfil&}%  This local macro skips a vertical rule.
\catcode`\&=13%  Now make &'s into active macros.
\let&0%  This allows us to expand \ampskip in the next \xdef without
%        attempting to expand the & and getting an "undefined control
%        sequence" error.
\xdef\letampskip{\def&{\ampskip}}%
\gdef\letnovbamp{\let\novb&\let\tab&}
%  This will cause active &'s read by
%                                   \ruledtable to be converted into
%                                   double tabs and an \omit'ted \vrule.
}%  End of locally active &'s.
\def\begintable{%  Here we make |'s and &'s active characters so we can
%                  interpret them as macros.  Note that this action is
%                  true only until we encounter the matching \endgroup
%                  token later at the end of the \ruledtable macro.
   \begingroup%
   \catcode`\|=13\letbartab\letvbbar%
   \catcode`\&=13\letampskip\letnovbamp%
   \def\multispan##1{%  We must redefine \multispan to count the number
%                       of primary columns, not physical columns.
      \omit \mscount##1%
      \multiply\mscount\tw@\advance\mscount\m@ne%
      \loop\ifnum\mscount>\@ne \sp@n\repeat%
   }%  End of \multispan macro.
   \def\|{%
      &\omit\widevline&%
   }%
   \ruledtable%  Now we call \ruledtable to do the real work.
}%  End of \begintable macro.
\long\def\ruledtable#1\endtable{%
%
%  This macro reads in the user's data entries
%  and converts them into a ruled table.
%
%  Important note:  Many macros and parameters are re-defined here, and
%  these must be kept local to the table macros to avoid conflict with
%  their use outside of tables.  This is done by the \begingroup token
%  macro \begintable and the \endgroup token at the end of
%  this macro.
%
   \offinterlineskip%  Needed to make rules touch each other.
   \tabskip 0pt%  Needed for same reason as \offinterlineskip.
   \def\widevline{\vrule width\thicksize}%  Make outer \vrule's wider.
   \def\endrow{\@mpersand\omit\hfil\crnorm\@mpersand}%
   \def\crthick{\@mpersand\crnorm\thickrule\@mpersand}%
   \def\crthickneg##1{\@mpersand\crnorm\thickrule
          \noalign{{\skip0=##1\vskip-\skip0}}\@mpersand}%
   \def\crnorule{\@mpersand\crnorm\@mpersand}%
   \def\crnoruleneg##1{\@mpersand\crnorm
          \noalign{{\skip0=##1\vskip-\skip0}}\@mpersand}%
   \let\nr=\crnorule%  A shorter abbreviation.
   \def\endtable{\@mpersand\crnorm\thickrule}%
   \let\crnorm=\cr%  Allows us to use \cr for our own purposes.
%
%  Cause user-typed \cr's to follow a row with a \tablerule.
%
   \edef\cr{\@mpersand\crnorm\tablerule\@mpersand}%
   \def\crneg##1{\@mpersand\crnorm\tablerule
          \noalign{{\skip0=##1\vskip-\skip0}}\@mpersand}%
   \let\ctneg=\crthickneg
   \let\nrneg=\crnoruleneg
   \the\tableLETtokens%  Get the user's extra \let's, if any.
%
%  Put the data entries into a token register so we can scan through them
%  and see what the user is asking us to do.
%
   \tabletokens={&#1}%  We add an extra alignment tab to the beginning
%                       of the first row to allow for the first \vrule.
%
%  Now count how many rows are in the table and return the result in
%  count register \nrows; do the same for columns, and return that
%  in register \ncols.
%
   \countROWS\tabletokens\into\nrows%
   \countCOLS\tabletokens\into\ncols%
%
%  Now do a little arithmetic to convert the number of primary columns
%  into the number of physical columns that the alignment preamble must
%  prepare for;  similarly for rows.
%
   \advance\ncols by -1%
   \divide\ncols by 2%
   \advance\nrows by 1%
%
%  Tell the user how many rows and columns we found in his data, if he
%  wants to know.
%
   \iftableinfo %
      \immediate\write16{[Nrows=\the\nrows, Ncols=\the\ncols]}%
   \fi%
%
%  Now we actually go ahead and produce the table.
%
   \ifcentertables
      \ifhmode \par\fi%  Make sure we are in vertical mode.
      \line{%  The final table comes out as an \hbox of width the \hsize.
      \hss%  The final table will be centered left-to-right.
   \else %
      \hbox{%
   \fi
      \vbox{%
         \makePREAMBLE{\the\ncols}%  Generate the preamble.
         \edef\next{\preamble}%  This line and the next line force the
         \let\preamble=\next%    expansion of all \ARGS tokens into the
%                                appropriate number of #'s.
         \makeTABLE{\preamble}{\tabletokens}%  Go do the \halign here.
      }%  End of \vbox.
      \ifcentertables \hss}\else }\fi%  Finish the centering effect.
%                                       It is important that no spaces
%                                       follow the two `}' here.
%  }%  End of \line.
   \endgroup%  Return all local macros and parameters to their outside
%              values.
   \tablewidth=-\maxdimen%  Reset \tablewidth to normal.
   \spreadwidth=-\maxdimen% Same for \spreadwidth.
}%  End of macro \ruledtable.
\def\makeTABLE#1#2{%  Does an \halign for the \ruledtable macro.
   {%  Start of local parameter values.
   \let\ifmath0%     These macros would cause trouble if they were to be
   \let\header0%     expanded in the following \xdef; we \let them be
   \let\multispan0%  equal to a digit, because digits can't be expanded.
%
%  Set up the width specification here.
%
   \ncase=0%
   \ifdim\tablewidth>-\maxdimen \ncase=1\fi%
   \ifdim\spreadwidth>-\maxdimen \ncase=2\fi%
   \relax%  This \relax is absolutely necessary, without it the following
%           \ifcase will always take \ncase=0.
%
   \ifcase\ncase %
      \widthspec={}%
   \or %
      \widthspec=\expandafter{\expandafter t\expandafter o%
                 \the\tablewidth}%
   \else %
      \widthspec=\expandafter{\expandafter s\expandafter p\expandafter r%
                 \expandafter e\expandafter a\expandafter d%
                 \the\spreadwidth}%
   \fi %
%\out{Widthspec=[\the\widthspec]}%
%\out{Preamble=[\preamble]}%
   \xdef\next{%  We must force the preamble to be expanded BEFORE the
      \halign\the\widthspec{%
%        \halign is done;  this \edef\next{...}\next construction
%                does the trick.
      #1%  This is the preamble text.
      \noalign{\hrule height\thicksize depth0pt}%  Makes the top \hrule.
      \the#2\endtable%  This is the main body.
%
%     \noalign{\hrule height0.7pt depth0pt}%  Makes the last \hrule.
      }%  End of \halign.
   }%  End of \next.
   }%  End of local values.
   \next%  This \next must be outside of the local values, because now
%          we want those troublesome macros in the \let's above to have
%          their normal actions.
}%  End of macro \makeTABLE.
\def\makePREAMBLE#1{%  This macro generates the necessary preamble for a
%                      ruled table with #1 primary columns.
%                      (Primary columns means the number of columns NOT
%                       counting those used for vertical rules.)
   \ncols=#1%  Get the number of columns desired.
   \begingroup%  Start local parameter definitions.
   \let\ARGS=0%  This is the key to the whole thing; it prevents \ARGS
%                from being expanded in the following \edef's.
   \edef\xtp{\widevline\ARGS\tabskip\tabskipglue%
   &\ctr{\ARGS}\tstrut}%  A 1-column preamble.  Gets the sizing right.
   \advance\ncols by -1%  One column has been generated; decrement the
%                         counter.
   \loop%  Append as many further columns as needed to the preamble.
      \ifnum\ncols>0 %
      \advance\ncols by -1%
      \edef\xtp{\xtp&\vrule width\thinsize\ARGS&\ctr{\ARGS}}%
   \repeat
   \xdef\preamble{\xtp&\widevline\ARGS\tabskip0pt%
   \crnorm}%  Adds the last \vrule.
   \endgroup%  End of local parameters.
}%  End of macro \makePREAMBLE.
\def\countROWS#1\into#2{%  This counts the number of rows in #1 by
%                          looking for control sequences that end a row,
%                          e.g., \cr, \crthick, etc., and puts the result
%                          into count register #2.
   \let\countREGISTER=#2%
   \countREGISTER=0%
%  \out{In countROWS:  tokens are [\the#1]}%
   \expandafter\ROWcount\the#1\endcount%
}%
\def\ROWcount{%
   \afterassignment\subROWcount\let\next= %
}%
\def\subROWcount{%
%  \out{In subROWcount:  next is [\meaning\next]}%  Debugging aid.
   \ifx\next\endcount %
      \let\next=\relax%
   \else%
      \ncase=0%
      \ifx\next\cr %
         \global\advance\countREGISTER by 1%
         \ncase=0%
      \fi%
      \ifx\next\endrow %
         \global\advance\countREGISTER by 1%
         \ncase=0%
      \fi%
      \ifx\next\crthick %
         \global\advance\countREGISTER by 1%
         \ncase=0%
      \fi%
      \ifx\next\crnorule %
         \global\advance\countREGISTER by 1%
         \ncase=0%
      \fi%
      \ifx\next\crthickneg %
         \global\advance\countREGISTER by 1%
         \ncase=0%
      \fi%
      \ifx\next\crnoruleneg %
         \global\advance\countREGISTER by 1%
         \ncase=0%
      \fi%
      \ifx\next\crneg %
         \global\advance\countREGISTER by 1%
         \ncase=0%
      \fi%
      \ifx\next\header %
%     \out{In subROWcount:  next=header, ncase set=1}%
         \ncase=1%
      \fi%
%     \out{In subROWcount:  ncase is [\the\ncase]}%
      \relax%
      \ifcase\ncase %
         \let\next\ROWcount%
%        \out{subROWcount---> ncase=\the\ncase}%
      \or %
         \let\next\argROWskip%
%        \out{subROWcount---> ncase=\the\ncase}%
      \else %
      \fi%
   \fi%
%  \out{subROWcount---> NEXT=\meaning\next}%
   \next%
}%  End of macro \subROWcount.
\def\counthdROWS#1\into#2{%
\dvr{10}%
   \let\countREGISTER=#2%
   \countREGISTER=0%
\dvr{11}%
%  \out{In counthdROWS:  tokens are [\the#1]}%
\dvr{13}%
   \expandafter\hdROWcount\the#1\endcount%
\dvr{12}%
}%
\def\hdROWcount{%
   \afterassignment\subhdROWcount\let\next= %
}%
\def\subhdROWcount{%
%\out{In subhdROWcount:  next is [\meaning\next]}%
   \ifx\next\endcount %
      \let\next=\relax%
   \else%
      \ncase=0%
      \ifx\next\cr %
         \global\advance\countREGISTER by 1%
         \ncase=0%
      \fi%
      \ifx\next\endrow %
         \global\advance\countREGISTER by 1%
         \ncase=0%
      \fi%
      \ifx\next\crthick %
         \global\advance\countREGISTER by 1%
         \ncase=0%
      \fi%
      \ifx\next\crnorule %
         \global\advance\countREGISTER by 1%
         \ncase=0%
      \fi%
      \ifx\next\header %
%\out{In subhdROWcount:  next=header, ncase set=1}%
         \ncase=1%
      \fi%
%\out{In subhdROWcount:  ncase is [\the\ncase]}%
\relax%
      \ifcase\ncase %
         \let\next\hdROWcount%
%\out{subhdROWcount---> ncase=\the\ncase}%
      \or%
         \let\next\arghdROWskip%
%\out{subhdROWcount---> ncase=\the\ncase}%
      \else %
      \fi%
   \fi%
%\out{subhdROWcount---> NEXT=\meaning\next}%
   \next%
}%
{\catcode`\|=13\letbartab
\gdef\countCOLS#1\into#2{%
%  \out{In countCOLS:  tokens are [\the#1]}
   \let\countREGISTER=#2%
   \global\countREGISTER=0%
   \global\multispancount=0%
   \global\firstrowtrue
   \expandafter\COLcount\the#1\endcount%
   \global\advance\countREGISTER by 3%
   \global\advance\countREGISTER by -\multispancount
%  \out{countCOLS-->[\the\countREGISTER]}
}%
\gdef\COLcount{%
   \afterassignment\subCOLcount\let\next= %
}%
{\catcode`\&=13%
\gdef\subCOLcount{%
%\out{In subCOLcount: next is [\meaning\next]}
   \ifx\next\endcount %
      \let\next=\relax%
   \else%
      \ncase=0%
      \iffirstrow
         \ifx\next& %
            \global\advance\countREGISTER by 2%
            \ncase=0%
         \fi%
         \ifx\next\span %
            \global\advance\countREGISTER by 1%
            \ncase=0%
         \fi%
         \ifx\next| %
            \global\advance\countREGISTER by 2%
            \ncase=0%
         \fi
         \ifx\next\|
            \global\advance\countREGISTER by 2%
            \ncase=0%
         \fi
         \ifx\next\multispan
            \ncase=1%
            \global\advance\multispancount by 1%
         \fi
         \ifx\next\header
            \ncase=2%
         \fi
         \ifx\next\cr       \global\firstrowfalse \fi
         \ifx\next\endrow   \global\firstrowfalse \fi
         \ifx\next\crthick  \global\firstrowfalse \fi
         \ifx\next\crnorule \global\firstrowfalse \fi
         \ifx\next\crnoruleneg \global\firstrowfalse \fi
         \ifx\next\crthickneg  \global\firstrowfalse \fi
         \ifx\next\crneg       \global\firstrowfalse \fi
      \fi%  End of \iffirstrow.
\relax%\out{subCOL-->  ncase=[\the\ncase]}
% \out{subCOL-->  next=\meaning\next}
      \ifcase\ncase %
         \let\next\COLcount%
      \or %
         \let\next\spancount%
      \or %
         \let\next\argCOLskip%
      \else %
      \fi %
   \fi%
%  \out{subCOL-->  countREGISTER=[\the\countREGISTER]}
   \next%
}%
\gdef\argROWskip#1{%
%  Deletes the next balanced, undelimited argument from a
%                 token list.
% \out{---> Entering argROWskip <---}
% \out{In argROWskip:  deleted arg is [#1]}%
   \let\next\ROWcount \next%
}%  End of macro \argskip.
\gdef\arghdROWskip#1{%
%  Deletes the next balanced, undelimited argument from a
%                 token list.
% \out{---> Entering arghdROWskip <---}
% \out{In arghdROWskip:  deleted arg is [#1]}%
   \let\next\ROWcount \next%
}%  End of macro \arghdROWskip.
\gdef\argCOLskip#1{%
%  Deletes the next balanced, undelimited argument from a
%                 token list.
% \out{---> Entering argCOLskip <---}
% \out{In argCOLskip:  deleted arg is [#1]}%
   \let\next\COLcount \next%
}%  End of macro \argskip.
}%  End of active &'s.
}%  End of active |'s.
\def\spancount#1{%\out{spancount--->\meaning#1}
   \nspan=#1\multiply\nspan by 2\advance\nspan by -1%
   \global\advance \countREGISTER by \nspan
%  \out{number spancount--->\the\nspan; \the\countREGISTER}
   \let\next\COLcount \next}%
\def\dvr#1{\relax}%
% \omit\hfil%
% \parindent=0pt\hsize=1.1in\valign{%
% \vfil#\vfil&\vfil#\vfil\cr\hfil\hbox{\ Added to\ }\hfil&%
% \hfil\hbox{\ empty events\ }\hfil\cr}\hfil%
\def\header#1{%
\dvr{1}{\let\cr=\@mpersand%
\hdtks={#1}%
%\out{In header:  hdtks=[\the\hdtks]}%
\counthdROWS\hdtks\into\hdrows%
\advance\hdrows by 1%
\ifnum\hdrows=0 \hdrows=1 \fi%
%\out{In header:  Nhdrows=[\the\hdrows]}%
\dvr{5}\makehdPREAMBLE{\the\hdrows}%
%\out{In header:  headerpreamble=[\headerpreamble]}%
\dvr{6}\getHDdimen{#1}%
%\out{In header:  hdsize=[\the\hdsize]}%
%\striplastCR{#1}%
{\parindent=0pt\hsize=\hdsize{\let\ifmath0%
\xdef\next{\valign{\headerpreamble #1\crnorm}}}\dvr{7}\next\dvr{8}%
}%
}\dvr{2}}%  End of macro \header.
\def\makehdPREAMBLE#1{%This macro generates the necessary preamble for a
\dvr{3}%
%                      ruled table with \ncols primary columns.
%                      (Primary columns means the number of columns NOT
%                       counting those used for vertical rules.
\hdrows=#1%  Get the number of columns desired.
{%  Start local parameter definitions.
\let\headerARGS=0%
%  This is the key to the whole thing; it prevents \ARGS
\let\cr=\crnorm%
%                from being expanded in the followin \edef's.
\edef\xtp{\vfil\hfil\hbox{\headerARGS}\hfil\vfil}%
\advance\hdrows by -1%  One row has been generated; decrement the
%                         counter.
\loop%  Append as many further rows as needed to the preamble.
\ifnum\hdrows>0%
\advance\hdrows by -1%
\edef\xtp{\xtp&\vfil\hfil\hbox{\headerARGS}\hfil\vfil}%
\repeat%
\xdef\headerpreamble{\xtp\crcr}%
}%  End of local parameters.
\dvr{4}}%  End of \makehdPREAMBLE.
\def\getHDdimen#1{%
%\out{In getHDdimen:  Arg 1=[#1]}%
\hdsize=0pt%
\getsize#1\cr\end\cr%
}%  End of macro getHDdimen.
\def\getsize#1\cr{%
%\out{In getsize:  Arg 1=[#1]}%
%  Here we have to check arg#1 and see if the first token in #1 is an
%    \end; if so, we stop, else we check the width of arg#1.
%  We recall that each arg#1 will be terminated with a \cr token.
\endsizefalse\savetks={#1}%
%\out{In getsize:  the savetks = [\the\savetks]}%
\expandafter\lookend\the\savetks\cr%
%\out{In getsize:  ifendsize = [\meaning\ifendsize]}%
\relax \ifendsize \let\next\relax \else%
\setbox\hdbox=\hbox{#1}\newhdsize=1.0\wd\hdbox%
\ifdim\newhdsize>\hdsize \hdsize=\newhdsize \fi%
%\out{In getsize:  hdsize=[\the\hdsize]}%
%\out{In getsize:  newhdsize=[\the\newhdsize]}%
\let\next\getsize \fi%
\next%
}%
\def\lookend{\afterassignment\sublookend\let\looknext= }%
\def\sublookend{\relax%
%\out{In sublookend:  looknext = [\looknext]}%
\ifx\looknext\cr %
%\out{In sublooknext:  looknext=cr}%
\let\looknext\relax \else %
%\out{In sublooknext:  looknext/=cr}%
   \relax
   \ifx\looknext\end \global\endsizetrue \fi%
   \let\looknext=\lookend%
    \fi \looknext%
}%
%
%  Allow the user to make his own names for crthick, etc.
%
\def\tablelet#1{%
   \tableLETtokens=\expandafter{\the\tableLETtokens #1}%
}%
\catcode`\@=12%  Change @'s back to their normal category code.
%

%%%%%%%%%%%%%%%%%%%%%%%%   END OF TABLES MACRO %%%%%%%%%%%%%%%%%%%%%%%

%%%%%%%%% These convert mathematica output
\def\frac#1#2{ { #1 \over #2 } }
\def\text{}

\lref\KorchemskySpin{ A.~V.~Belitsky, A.~S.~Gorsky and G.~P.~Korchemsky,
  %``Logarithmic scaling in gauge / string correspondence,''
  Nucl.\ Phys.\  B {\bf 748}, 24 (2006)
  [arXiv:hep-th/0601112].
  %%CITATION = NUPHA,B748,24;%%
}

%\AldayDV
\lref\AGM{
  L.~F.~Alday, D.~Gaiotto and J.~Maldacena,
%  ``Thermodynamic Bubble Ansatz,''
  arXiv:0911.4708 [hep-th].
  %%CITATION = ARXIV:0911.4708;%%
}
%\AldayYN
\lref\octagon{
  L.~F.~Alday and J.~Maldacena,
 % ``Null polygonal Wilson loops and minimal surfaces in Anti-de-Sitter space,''
  JHEP {\bf 0911}, 082 (2009)
  [arXiv:0904.0663 [hep-th]].
  %%CITATION = JHEPA,0911,082;%%
}

\lref\AMSV{
  L.~F.~Alday, J.~Maldacena, A.~Sever and P.~Vieira,
%  ``Y-system for Scattering Amplitudes,''
  arXiv:1002.2459 [hep-th].
}

%\GubserTV
\lref\GKP{
  S.~S.~Gubser, I.~R.~Klebanov and A.~M.~Polyakov,
  %``A semi-classical limit of the gauge/string correspondence,''
  Nucl.\ Phys.\  B {\bf 636}, 99 (2002)
  [arXiv:hep-th/0204051].
  %%CITATION = NUPHA,B636,99;%%
}

\lref\AMtwo{ L.~F.~Alday and J.~M.~Maldacena,
  %``Comments on operators with large spin,''
  JHEP {\bf 0711}, 019 (2007)
  [arXiv:0708.0672 [hep-th]].
  %%CITATION = JHEPA,0711,019;%%
}

\lref\Basso{B. Basso, to appear.}

%\AlexandrovPP
\lref\AlexandrovPP{
  S.~Alexandrov and P.~Roche,
  %``TBA for non-perturbative moduli spaces,''
  arXiv:1003.3964 [hep-th].
  %%CITATION = ARXIV:1003.3964;%%
}
  %\NekrasovRC
\lref\NekrasovRC{
  N.~A.~Nekrasov and S.~L.~Shatashvili,
  %``Quantization of Integrable Systems and Four Dimensional Gauge Theories,''
  arXiv:0908.4052 [hep-th].
  %%CITATION = ARXIV:0908.4052;%%
}

   %\KruczenskiFB
\lref\Kruczenski{
  M.~Kruczenski,
  %``A note on twist two operators in N = 4 SYM and Wilson loops in Minkowski
  %signature,''
  JHEP {\bf 0212}, 024 (2002)
  [arXiv:hep-th/0210115].
  %%CITATION = JHEPA,0212,024;%%
}
%\KorchemskyXV
\lref\KorchemskyMar{
  G.~P.~Korchemsky and G.~Marchesini,
  %``Structure function for large x and renormalization of Wilson loop,''
  Nucl.\ Phys.\  B {\bf 406}, 225 (1993)
  [arXiv:hep-ph/9210281].
  %%CITATION = NUPHA,B406,225;%%
}

%\GaiottoHG
\lref\GMNtwo{
  D.~Gaiotto, G.~W.~Moore and A.~Neitzke,
  %``Wall-crossing, Hitchin Systems, and the WKB Approximation,''
  arXiv:0907.3987 [hep-th].
  %%CITATION = ARXIV:0907.3987;%%
}
 %\BerensteinJQ
\lref\BMN{
  D.~E.~Berenstein, J.~M.~Maldacena and H.~S.~Nastase,
  %``Strings in flat space and pp waves from N = 4 super Yang Mills,''
  JHEP {\bf 0204}, 013 (2002)
  [arXiv:hep-th/0202021].
  %%CITATION = JHEPA,0204,013;%%
}

%\GaiottoCD
\lref\GMNone{
  D.~Gaiotto, G.~W.~Moore and A.~Neitzke,
  %``Four-dimensional wall-crossing via three-dimensional field theory,''
  arXiv:0807.4723 [hep-th].
  %%CITATION = ARXIV:0807.4723;%%
}

\lref\SokWard{ J.~M.~Drummond, J.~Henn, G.~P.~Korchemsky and E.~Sokatchev,
  %``Conformal Ward identities for Wilson loops and a test of the duality with
  %gluon amplitudes,''
  Nucl.\ Phys.\  B {\bf 826}, 337 (2010)
  [arXiv:0712.1223 [hep-th]].
  %%CITATION = NUPHA,B826,337;%%
}
%\BernIZ
\lref\BDS{
  Z.~Bern, L.~J.~Dixon and V.~A.~Smirnov,
  %``Iteration of planar amplitudes in maximally supersymmetric Yang-Mills
  %theory at three loops and beyond,''
  Phys.\ Rev.\  D {\bf 72}, 085001 (2005)
  [arXiv:hep-th/0505205].
  %%CITATION = PHRVA,D72,085001;%%
}

%\BraunRP
\lref\ConformalQCD{
  V.~M.~Braun, G.~P.~Korchemsky and D.~Mueller,
  %``The uses of conformal symmetry in QCD,''
  Prog.\ Part.\ Nucl.\ Phys.\  {\bf 51}, 311 (2003)
  [arXiv:hep-ph/0306057].
  %%CITATION = PPNPD,51,311;%%
}

%\BrandhuberYX
\lref\brandhuber{
  A.~Brandhuber, P.~Heslop and G.~Travaglini,
  %``MHV Amplitudes in N=4 Super Yang-Mills and Wilson Loops,''
  Nucl.\ Phys.\  B {\bf 794}, 231 (2008)
  [arXiv:0707.1153 [hep-th]].
  %%CITATION = NUPHA,B794,231;%%
}
\lref\BES{N.~Beisert, B.~Eden and M.~Staudacher,
  %``Transcendentality and crossing,''
  J.\ Stat.\ Mech.\  {\bf 0701}, P021 (2007)
  [arXiv:hep-th/0610251].
  %%CITATION = JSTAT,0701,P021;%%
}

\lref\Zarembo{
 K.~Zarembo,
  %``Worldsheet spectrum in AdS(4)/CFT(3) correspondence,''
  JHEP {\bf 0904}, 135 (2009)
  [arXiv:0903.1747 [hep-th]].
  %%CITATION = JHEPA,0904,135;%%
}

%\BelavinVU
\lref\BPZ{
  A.~A.~Belavin, A.~M.~Polyakov and A.~B.~Zamolodchikov,
  %``Infinite conformal symmetry in two-dimensional quantum field theory,''
  Nucl.\ Phys.\  B {\bf 241}, 333 (1984).
  %%CITATION = NUPHA,B241,333;%%
}

\lref\FL{
  N.~Beisert and M.~Staudacher,
%  ``The N=4 SYM Integrable Super Spin Chain,''
  Nucl.\ Phys.\  B {\bf 670} (2003) 439
  [arXiv:hep-th/0307042].
  $\bullet$ %\cite{Beisert:2004ry}
  N.~Beisert,
 % ``The dilatation operator of N = 4 super Yang-Mills theory and
 % integrability,''
  Phys.\ Rept.\  {\bf 405}, 1 (2005)
  [arXiv:hep-th/0407277].
  %%CITATION = PRPLC,405,1;%%
  }

  \lref\ZhangTR{
  J.~H.~Zhang,
 % ``On the two-loop hexagon Wilson loop remainder function in N=4 SYM,''
  arXiv:1004.1606 [hep-th].
  %%CITATION = ARXIV:1004.1606;%%
}

%\AldayHR
\lref\AldayHR{
  L.~F.~Alday and J.~M.~Maldacena,
 % ``Gluon scattering amplitudes at strong coupling,''
  JHEP {\bf 0706}, 064 (2007)
  [arXiv:0705.0303 [hep-th]].
  %%CITATION = JHEPA,0706,064;%%
}

%\DrummondAUA
\lref\DrummondAUA{
  J.~M.~Drummond, G.~P.~Korchemsky and E.~Sokatchev,
  %``Conformal properties of four-gluon planar amplitudes and Wilson loops,''
  Nucl.\ Phys.\  B {\bf 795}, 385 (2008)
  [arXiv:ward [hep-th]].
  %%CITATION = NUPHA,B795,385;%%
}

%\BrandhuberYX
\lref\BrandhuberYX{
  A.~Brandhuber, P.~Heslop and G.~Travaglini,
  %``MHV Amplitudes in N=4 Super Yang-Mills and Wilson Loops,''
  Nucl.\ Phys.\  B {\bf 794}, 231 (2008)
  [arXiv:0707.1153 [hep-th]].
  %%CITATION = NUPHA,B794,231;%%
}

%\DrummondAQ
\lref\DrummondAQ{
  J.~M.~Drummond, J.~Henn, G.~P.~Korchemsky and E.~Sokatchev,
  %``Hexagon Wilson loop = six-gluon MHV amplitude,''
  Nucl.\ Phys.\  B {\bf 815}, 142 (2009)
  [arXiv:0803.1466 [hep-th]].
  %%CITATION = NUPHA,B815,142;%%
}

%\AldayHE
\lref\AldayHE{
  L.~F.~Alday and J.~Maldacena,
  %``Comments on gluon scattering amplitudes via AdS/CFT,''
  JHEP {\bf 0711}, 068 (2007)
  [arXiv:0710.1060 [hep-th]].
  %%CITATION = JHEPA,0711,068;%%
}

%\DrummondBM
\lref\DrummondBM{
  J.~M.~Drummond, J.~Henn, G.~P.~Korchemsky and E.~Sokatchev,
  %``The hexagon Wilson loop and the BDS ansatz for the six-gluon amplitude,''
  Phys.\ Lett.\  B {\bf 662}, 456 (2008)
  [arXiv:0712.4138 [hep-th]].
  %%CITATION = PHLTA,B662,456;%%
}

%\BernAP
\lref\BernAP{
  Z.~Bern, L.~J.~Dixon, D.~A.~Kosower, R.~Roiban, M.~Spradlin, C.~Vergu and A.~Volovich,
  %``The Two-Loop Six-Gluon MHV Amplitude in Maximally Supersymmetric Yang-Mills
  %Theory,''
  Phys.\ Rev.\  D {\bf 78}, 045007 (2008)
  [arXiv:0803.1465 [hep-th]].
  %%CITATION = PHRVA,D78,045007;%%
}

%\AnastasiouKNA
\lref\AnastasiouKNA{
  C.~Anastasiou, A.~Brandhuber, P.~Heslop, V.~V.~Khoze, B.~Spence and G.~Travaglini,
  %``Two-Loop Polygon Wilson Loops in N=4 SYM,''
  JHEP {\bf 0905}, 115 (2009)
  [arXiv:0902.2245 [hep-th]].
  %%CITATION = JHEPA,0905,115;%%
}

%\DelDucaZG
\lref\DelDucaZG{
  V.~Del Duca, C.~Duhr and V.~A.~Smirnov,
  %``The Two-Loop Hexagon Wilson Loop in N = 4 SYM,''
  JHEP {\bf 1005}, 084 (2010)
  [arXiv:1003.1702 [hep-th]].
  %%CITATION = JHEPA,1005,084;%%
}

%\HodgesHK
\lref\HodgesHK{
  A.~Hodges,
  %``Eliminating spurious poles from gauge-theoretic amplitudes,''
  arXiv:0905.1473 [hep-th].
  %%CITATION = ARXIV:0905.1473;%%
}

%\FrolovAV
\lref\FrolovAV{
  S.~Frolov and A.~A.~Tseytlin,
  %``Semiclassical quantization of rotating superstring in AdS(5) x S(5),''
  JHEP {\bf 0206}, 007 (2002)
  [arXiv:hep-th/0204226].
  %%CITATION = JHEPA,0206,007;%%
}

%\draftmode

%%%%%%%%%%%%%%%%%%%%%%%%%%%%%%%%%%%%%%%%%%%%%%%%%%%%%%%%%%%%%%%%%%%%
\Title{\vbox{\baselineskip6pt \hbox{} \hbox{
} }} {\vbox{\centerline{
  }
\centerline{
An Operator Product Expansion }
\vskip .5cm
\centerline{ for Polygonal null Wilson Loops}
}}
\bigskip
\centerline{Luis F. Alday$^{1}$, Davide Gaiotto$^{1}$ , Juan Maldacena$^{1}$, Amit Sever$^{2}$, Pedro Vieira$^{2}$}
\bigskip
\centerline{ \it  $^1$School of Natural Sciences, Institute for
Advanced Study,
% } \centerline{\it
 Princeton, NJ 08540, USA}
\centerline{ \it  $^2$ Perimeter Institute for Theoretical Physics, Waterloo,
Ontario N2J 2W9, Canada}

\bigskip
\bigskip
We consider polygonal Wilson loops with null edges in conformal gauge theories.
We derive an OPE-like expansion when several successive lines of the polygon are becoming aligned.
The limit corresponds to a collinear, or multicollinear, limit and we explain the systematics
of all the subleading corrections, going beyond the leading terms that were previously considered.
These subleading corrections are governed by excitations of high spin operators, or excitations
of a flux tube that goes between two Wilson lines. The discussion is valid for any conformal gauge
theory, for any coupling and  in any dimension.

For ${\cal N}=4$ super
Yang Mills we
 check this expansion at strong coupling and at two loops at weak coupling .
 We also make predictions for the remainder function at higher loops.

 In the process, we also derived a new version for the TBA integral equations
 that determine the strong
 coupling answer and present the area as the associated Yang-Yang functional.

\vskip .3in \noindent
%%%%%%%%%%%%%%%%%%%%%%%%%%%%%%%%%%%%%%%%%%%%%%%%%%%%%%%%%%%%%%%%%%%%%%%%%%%%%%%%%%%%%%%%%%%%
%%%%%%%%%%%%%%%%%%%%%%%%%%%%%%%%%%%%%%%%%%%%%%%%%%%%%%%%%%%%%%%%%%%%%%%%%%%%%%%%%%%%%%%%%%%%

 \Date{ }

%%%%%%%%%%%%%%%%%%%%%%%%%%%%%%%%%%%%%%%%%%%%%%%%%%%%%%%%%%%%%%%%%%%%%%%%%
\vskip 10cm

\listtoc\writetoc

\newsec{Introduction}

The Operator Product Expansion (OPE) is a powerful tool for studying correlation functions in
conformal field theories. The expansion is controlled by the spectrum of local operators of the theory.
In particular, their dimensions control the powers of the expansion parameter in the OPE, which is simply
the separation between operators.
In some cases, such as two dimensional minimal models, it is possible to completely fix the correlation
functions by demanding this property in all possible channels \BPZ.

In this paper we  derive a similar OPE-like expansion  for
polygonal Wilson loops with light-like edges. Our main
motivation   is the relation between Wilson loops and
amplitudes in ${\cal N}=4$ super Yang Mills
\refs{\AldayHR,\DrummondAUA,\BrandhuberYX,\DrummondAQ,\BernAP}.
Furthermore, similar looking Wilson loops appear in a
 variety of high energy processes
in gauge theories. The OPE expansion is valid for any conformal
field theory and  in any dimension where we can define null
polygonal Wilson loops. It is valid whenever the Wilson lines
produce a conserved
 flux that cannot
be screened.
  It is  useful to consider first
 large $N$ gauge theories in the planar approximation. We will later make some remarks beyond the planar limit.
One challenging aspect of the problem is that  Wilson loops with light-like edges
are eminently Lorentzian observables without
an obvious Euclidean counterpart.

  \ifig\Cut{ We consider a general Wilson loop with null edges. We select two of its edges and we
  place and imaginary cut along a line connecting these two selected edges. We then expand the answer in
  terms of states propagating across this cut.
 } {\epsfxsize1in\epsfbox{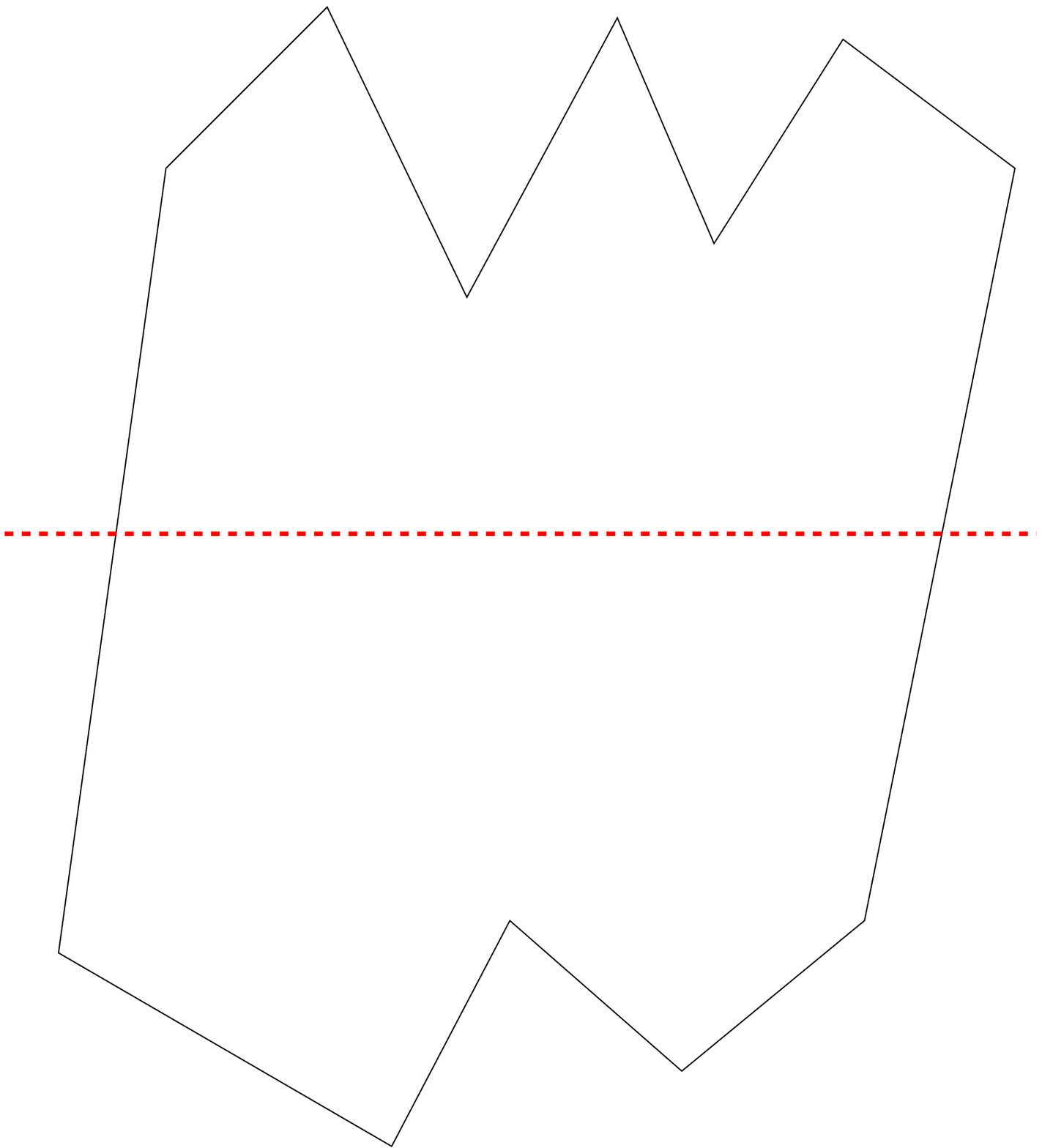}}

 Our OPE expansion is performed as follows, see \Cut.
First we select  two non-consecutive lines of the polygonal Wilson loop.
 We then cut the Wilson loop into a top part and a bottom part and   we
 expand it in terms of the states propagating through the cut. The states that propagate through the
 cut correspond to excitations of a flux tube that ends on two light-like lines.
 Fortunately these states have been considered before.  In fact, they
 are excitations around the infinite spin limit of high spin operators. In theories with gravity/string duals
 they are excitations around the spinning string (GKP) considered
 in \GKP .
  In ${\cal N}=4$ super Yang Mills one can compute these dimensions for all values of the
 coupling \refs{\BES,\Basso}.
 Thus, we find that the Wilson loop has an OPE-like expansion in terms of operators, or states, which
 are excitations of the infinite spin limit of the GKP string. These states can also be viewed as
 created by local operator insertions along a null Wilson line.

 %We emphasize that this is a very general result, and it holds for any conformal gauge theory in any
 %dimension where null polygonal Wilson loops can be defined.

  \ifig\CollinearExpansion{  The collinear expansion including subleading terms. The first term is the
  usual statement that in the collinear limit we recover the Wilson loop with one less line. The second
  term corresponds to the insertion of one operator along the contour, the second to two operators, etc.
 } {\epsfxsize3.5in\epsfbox{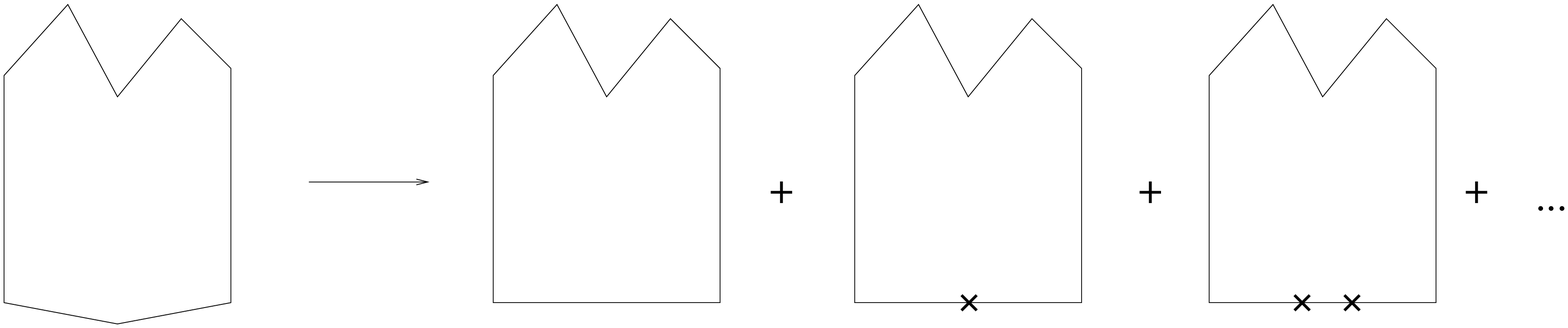}}

 The simplest example of this OPE is a collinear limit, where two consecutive lines become
 parallel, see   \CollinearExpansion . This collinear limit was considered previously and its leading divergent and
 constant terms were understood \BDS .
  These leading terms are determined by the  special conformal symmetry
 anomalous Ward identities \SokWard \foot{Though this statement is believed to be true, to our knowledge it hasn't been proven. We will see that it follows from our discussion at
 the end of section 3.}. We are now saying that we can continue the expansion into the
 subleading terms. The existence of this expansion places   constraints on the possible form of the
 Wilson loop correlator. In fact, it constrains  the so called remainder
 function \refs{\AldayHE,\DrummondBM,\AnastasiouKNA,\BernAP}
 which contains the conformal invariant
 information on the Wilson loop correlator.
   We will demonstrate this explicitly below for the case of the six sided Wilson loop in
   ${\cal N}=4$ super Yang Mills.
    In fact we will
   verify that the strong coupling answer and the two loop weak coupling answer have the form required
   by the OPE. We will also make a prediction, using \Basso, for the expansion for all values of the
   coupling.  At strong coupling we can go further and check that the expansion has the required form for
   general polygons.

 It is quite likely, that the existence of this expansion in all possible channels, together with
 integrability would completely constrain the correlator. This would simply be the extension of
 the Bootstrap procedure for correlation functions  \BPZ\ to Wilson loop correlators.

  Our paper is organized as follows.
  In section 2 we discuss the symmetries of the problem. We will use them heavily for
  deriving the existence of the expansion. The key idea is to identify a ``Hamiltonian'' which organizes
  the expansion.
  In section 3 we explain the form of the OPE expansion for Wilson loops.
  In section 4 we discuss the form of the expansion for the case of the hexagonal Wilson loop at strong and weak
  coupling. Finally in section 5 we briefly discuss higher
  order predictions.

The paper contains several appendices. In appendix A we describe our setup in terms of momentum twistors.   In appendix B we present further arguments for the identification of the states we encounter in
  the expansion and excitations around the flux tube connecting two null Wilson lines. In appendix C we discuss the dispersion relation of the eigenvalues of the ``Hamiltonian". In appendix D we present some details on the one loop computation of some ``form factors".

Finally, in appendices E and F we give a new form for the integral equations described in \refs{\AGM,\AMSV}.
   In the new form, the equations involve only physical cross ratios. Furthermore, the expression
   for the area is given by the Yang-Yang functional of the associated TBA-like integral equations.
   This form is particularly useful for analyzing our limit. It is very likely that it will also
   be useful for further studies on integrability and Wilson loops/amplitudes. Thus these appendices
   can be read on their own.

\newsec{ Symmetries }

 The expansion we are studying is   governed by the symmetries preserved by two null lines
 and also by the symmetries preserved by the square Wilson loop, i.e. a Wilson loop with four null sides.

 To justify the need to understand the symmetries, let us recall first how symmetries constrain and
 determine the ordinary OPE for correlation functions. In that case the main symmetry in question is
 dilatations. If we have a correlator $\langle O(1) O(2)...O(n) \rangle$. Then we can set $O(1)$
 at the origin and   consider the limit where $O(2)$ approaches it. This operation corresponds to
 a dilatation applied to operators $O(1)$ and $O(2)$, but not to the rest of the operators. This becomes
 more clear if we map this to the (Euclidean) cylinder. Then $O(1)$ is at $\tau_1 = - \infty$ and
 $O(2)$ is at some value of $\tau_2$ and the rest are at fixed values of $\tau_i$. Here $\tau$ is the
 (Euclidean) time coordinate on the cylinder. We also demand that $\tau_2 < {\rm min}\{ \tau_i \}$.
 The OPE expansion follows by cutting the cylinder between $\tau_2$ and $ {\rm min}\{ \tau_i \}$ and
 inserting a complete basis of eigenstates of the dilatation operator. The fact that the OPE is convergent
 follows from the asymptotic growth of states\foot{For a $d$ dimensional CFT we have
 $\log N(\Delta) \sim ({\rm constant}) \Delta^{d-1\over d }  \ll \Delta$ as $\Delta \to \infty$.}.

 \subsec{ The symmetries preserved by null lines and the   square Wilson loop  }

 Let us start with two generic non-intersecting null lines and consider the symmetries that leave these
 two null lines invariant. Working in $R^{1,3}$,
 by conformal transformations we send a  null line to null infinity and the other to a null line extended
 along $x^-$ and passing through the origin. We can do this so that both null lines are in an $R^{1,1}$ subspace,
 with one of the lines at null infinity.
 The symmetries that leave this configuration invariant include dilatations $D$, boosts in $x^\pm$, $M^{+-}$,
  and a rotation in the two transverse directions $M^{12}$.
   In addition we have translations $P_-$ along $x^-$ and special conformal transformations $K^-$.
 The symmetries $P_-$, $K^-$ and $D + M^{+-} $ form an $SL(2,R)$ subgroup, which commutes with the two remaining abelian
 symmetries: $D-M^{+-}$ and $M^{12}$.
 Note that these two remaining symmetries leave
 individual points on the lines invariant, while the $SL(2,R)$ symmetry maps points on a given line to other
 points on the same line, but does not leave them invariant in a pointwise fashion.

 For this counting of symmetries it   is important
 that we have the line at null infinity. If we only had the line
 that goes through the origin we would have more symmetry generators such as $K^i$, $K^+$,  and boosts $M^{+i}
 \sim x^+ \partial_i - x^i \partial_{x^-}$, where $i$ is a transverse index. However, these transformations
 do not leave the line at infinity invariant. This can be checked by using an inversion to map the
 line at infinity to the origin. For example,
 this turns $K^i$ into $P^i$ which indeed does not leave the line at
 the origin fixed.    These symmetries
 preserve one of the null lines but move the other one. So they are not true symmetries of the two line
 system. Nevertheless we will see that they have some interesting consequences.

 \ifig\TwoLines{   We start with two null lines, denoted by the solid lines. By picking a point $A$ on
 the first line, we determine a unique point $C$ on the second line which is light-like separated from $A$.
 Similarly for $B$. In this way we construct a square.   In the figure we projected onto two spatial
directions and suppressed the time direction.
 } {\epsfxsize1.6in\epsfbox{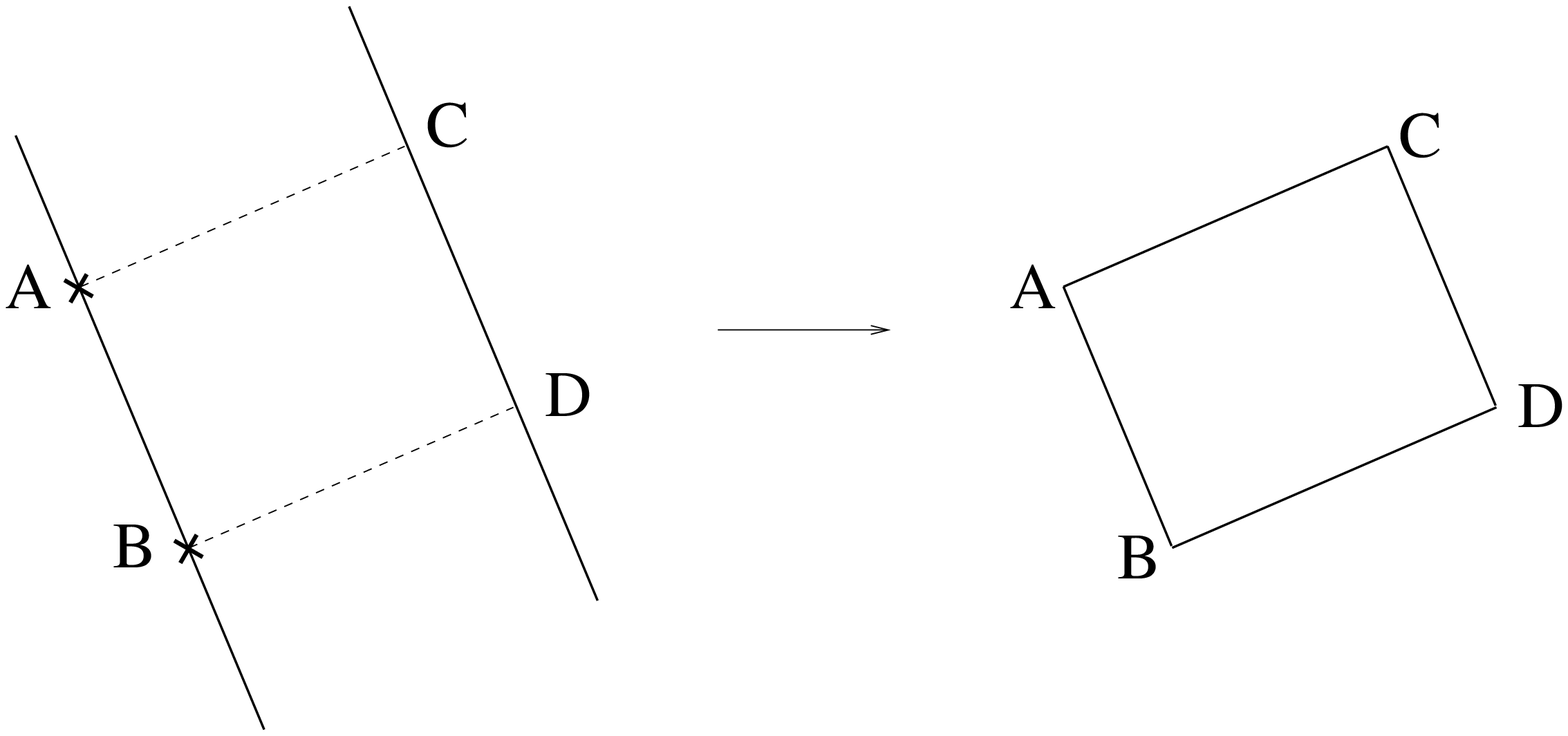}}

 \ifig\Square{   (a) The square Wilson loop projected onto two spatial
directions. We suppressed the time direction.  The ``Hamiltonian''
symmetry $\partial_\tau$ moves points along the
 arrow direction. It leaves points on the thin black lines fixed, but
moves points along the thick red lines.
 The ``momentum'' symmetry $\partial_\sigma$ is also indicated. (b) We
inserted an operator on the bottom line. We
 can integrate the operator along the bottom edge in such a way that
it has a definite momentum with respect
 to the $\partial_\sigma$ symmetry.
 If it creates a single particle state, then it will also have a
definite energy. (c) We have  mapped the
 square to an $R^{1,1}$ subspace by a conformal transformation that
sent a cusp to spatial infinity.
Two of the null lines are at null infinity. One of the cusps is at the
 origin. Operators inserted along the top and bottom lines in (b)
correspond to operator insertions which
 are spacelike separated, and indicated by a cross.  The
``Hamiltonian'' and ``momentum'' correspond to the generators
$D \pm M^{+-}$.
 } {\epsfxsize3.5in\epsfbox{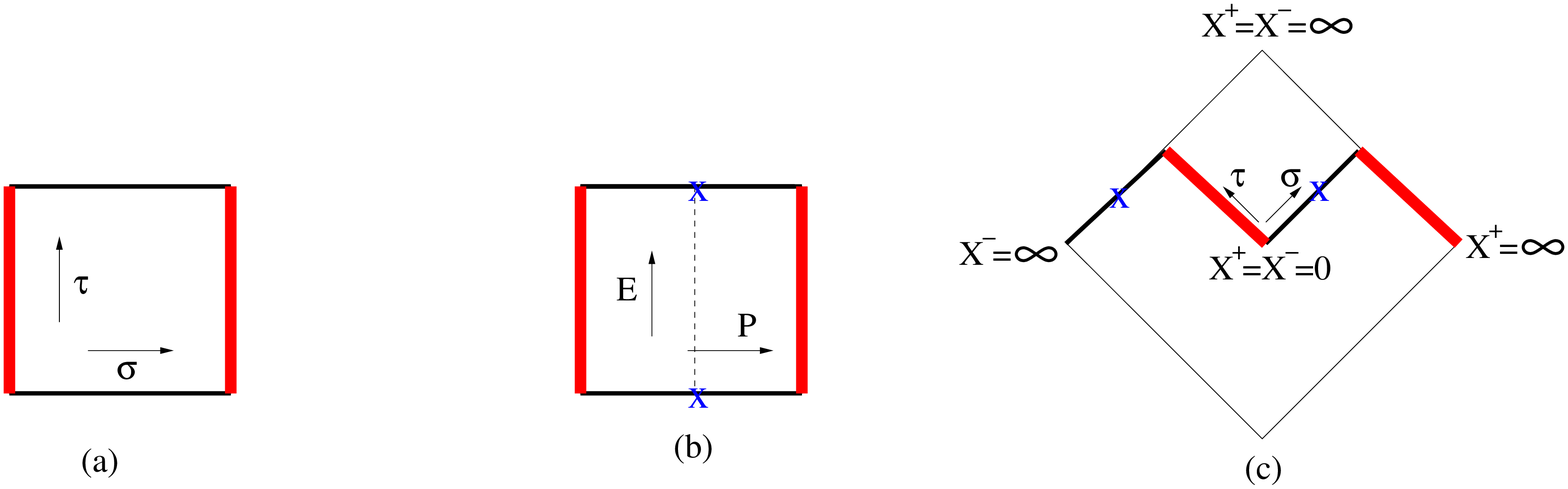}}

 Once we have understood the symmetries of two lines, we now want to understand the symmetries of the
 square. To define the square we need to pick two points along one of the null lines where the vertices of
 the square will be sitting. Once we pick a point along a null line to be a vertex of the square, the
 corresponding point along the other null line is automatically fixed, since it is the point on the other
 null line that is null separated with respect to the point chosen on the first line, see \TwoLines .
  Thus, picking a square, now corresponds to picking two points along a null line. This breaks the $SL(2,R)$
  symmetry group  to a single generator.  We are going to view this remaining generator of $SL(2,R)$ as a
  ``Hamiltonian''. This Hamiltonian moves points on both segments of the original null lines towards the
  same side of the square. In fact, we will see that, because the new sides of the square are spacelike
  separated, this ``Hamiltonian'' is a Euclidean Hamiltonian such that the evolution operator
  looks like $e^{- \tau E}$, as opposed to $e^{ - i \tau E}$, where $E$ are the real energy eigenvalues.
  The second non-compact symmetry moves points along the bottom and top lines of the square,
   see  \Square (a).
  We can view this as a ``momentum'' generator. In fact, the ``Hamiltonian'' has  a continuous spectrum
  due to the existence of this second non-compact symmetry.

  In summary, the square has three commuting symmetries. Two noncompact symmetries and one  transverse rotation symmetry.
  In (2,2) signature the transverse symmetry becomes a boost and is also non-compact.

  The symmetries of the square can also be understood directly as follows.
 By a conformal transformation we can map three of the points to the boundary of Minkwoski space.
 One of the points is mapped to spatial infinity, and the other two are
 mapped to some points on null infinity.  The
  fourth point can be set at the origin.
 In that case we just have a null cusp with lines along $x^+x^-=0$,
 $x^\pm > 0$. See \Square (c).  Let us understand the symmetries that leave this invariant. We clearly have three
 commuting symmetries: dilatations,   boosts in the $x^\pm$ plane and the rotation in the
 transverse plane. In fact, these are all the symmetries of this Wilson loop. It is fairly clear that
 there are no further Poincare symmetries of $R^{1,3}$ that leave it invariant. There are no
 special conformal symmetries that leave it invariant because any special conformal
 symmetry moves the point at spatial infinity, which is one of the vertices of the polygon.

\vskip3em
\vbox{
\begintable
{\rm Single Line}   | $ D -M^{+-}$  |$ M^{12}$ | $D+ M^{+-}$ | $K^-$ |$P_- $|$ M^{+i}$ |$K^+ $| $K^{i}$
  \cr
  {\rm Two Lines}   |$ D -M^{+-} $ | $M^{12}$ |$ D+ M^{+-}$ | $K^-$ |$P_- $| ~|~ | ~
  \cr
  {\rm Square }   | $D -M^{+-}$  | $M^{12}$ |$ D+ M^{+-} $| ~ |~ |~ |~ |~
 \cr
  {\rm Name }   |  $ P $, ~~$\partial_\sigma $ | $\tilde S ,  ~~ \partial_\phi $ | $E$, ~~$\partial_\tau$ | ~ |~ |~ |~ |~
  \endtable}
%\inparg
\centerline{{ Table~1:\/}   Symmetries preserved by one and two lines and the square. }
\centerline{
In the last line
we introduced some notation for each of the three symmetries. }
% \centerline{Namely:  Energy, Spin, Momentum. }
\medskip
%\outparg

  It is also useful to understand how the ``Hamiltonian'' acts on the ``bottom'' side of the square, see
  \Square .
  It leaves it invariant on a point by point basis and it rescales the transverse directions. More precisely,
  if we say that this bottom side is along $x^+$, then it rescales $x^- \to \lambda^2 x^-$ and $y_\perp \to
  \lambda y_\perp $, see \Square (c). This action is sometimes called ``twist'' operator.
  More concretely, we will find that in our expansion procedure we will need to consider operators
  inserted along this ``bottom'' null line, see \Square (b) (c).
  Similar operators arise also in the study of high energy
  processes in QCD, for a review see \ConformalQCD . Such an operator is characterized by a ``momentum''
  along the direction of the line.  This is the same ``momentum'' generator that we mentioned above.
  In the QCD literature  on operators involving null Wilson lines this
  generator is called the spin of the state (e.g. in \ConformalQCD).
  It turns out that if we insert a local operator along the
  null Wilson line, this operator will not be an eigenstate of the twist operator (or our ``Hamiltonian'').
  However, single particle states with definite ``momentum'' are indeed eigenstates.

  The states that are contributing to the OPE are the states created by inserting local operators along
  the bottom side of the Wilson loop and propagating all the way to the top, see \Square (b).
  These states have a continuous
  spectrum, due to the existence of a second non-compact symmetry. This is in contrast to the ordinary
  operator product expansion where the spectrum of dimensions is discrete. However, due to the momentum
 symmetry we can certainly isolate the contributions from ``single particle'' states propagating along
 the square.

 \newsec{ The collinear  operator product expansion }

 In this section we state more precisely how to perform the OPE expansion.
 In the case of ordinary correlation functions the expansion can be organized in terms of the
 dimensions and the spins of the intermediate operators. As we mentioned above, the operators
 that   appear in our expansion are characterized by their dimension (or energy) and also by
 another continuous label which we will call the momentum. There is a third discrete label which
 is the spin in the transverse dimension.  In order to isolate the various quantum numbers it is
 convenient to introduce the following three parameter family of polygons.

 \subsec{A three parameter family of polygons}

\ifig\Family{ (a)  Given a general polygon we select two
segments. We prolong them and choose two points on one of these
segments $A$ and $D$. These determine points $B$ and $C$ on the
other line. We use them to form a reference square ABCD. We now
cut the polygon into a top part and a bottom part. We apply a
combination of symmetries of the square to the bottom part. In
the limit $\tau \to \infty$ the bottom part is flattened out
into the straight edge of the reference square. In that limit
we get the polygon in (b) which we will call the ``top''
polygon. Alternatively, we could have applied the symmetry to
the top half of the polygon in (a). In the limit we would get
the polygon in (c) which we call the ``bottom'' polygon.
 } {\epsfxsize2.5in\epsfbox{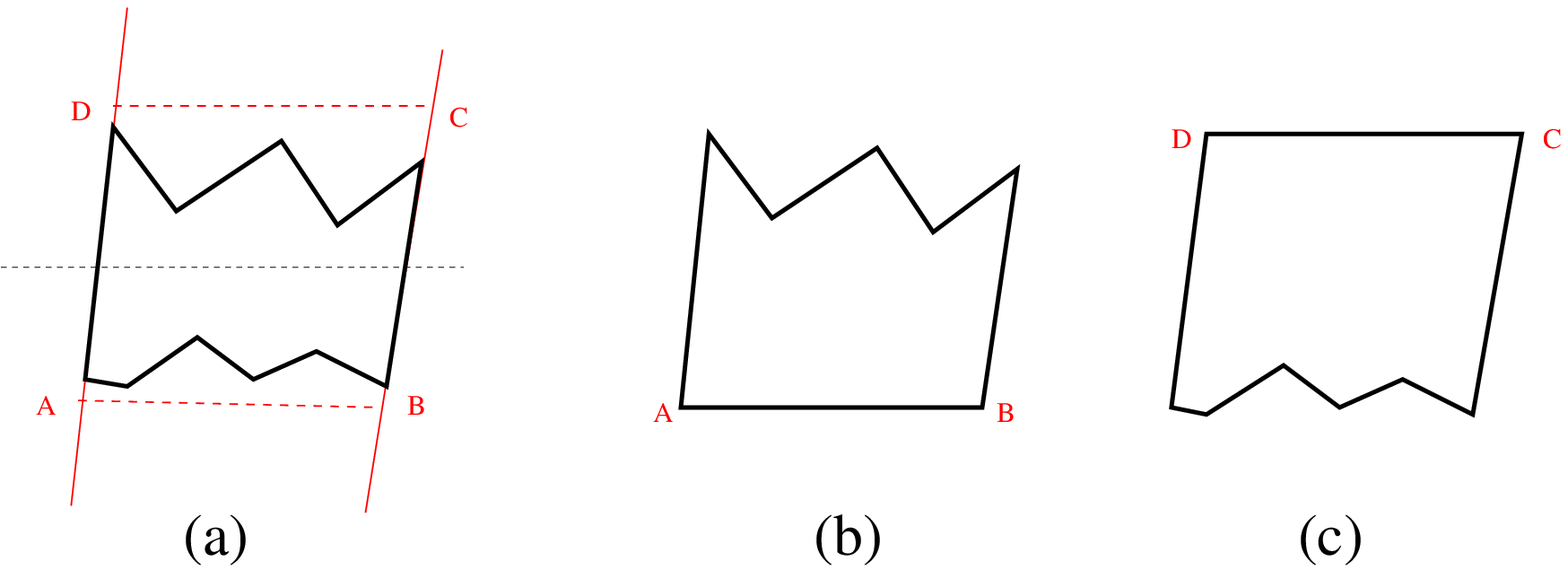}}

 We would now like to introduce  an interesting family of
 polygons.
 The family is specified as follows. Let us assume that all the points on the polygon
 are spacelike separated, except for the obvious null separations that define the polygon.
 We first pick two non-consecutive null sides of the polygon, and extend them to two null lines.
 We then pick two points along the null line which lie outside the segment that belongs to the polygon
 on that null line. These two points, together with the two null lines define a certain reference square.
 The symmetries of the square involve the three symmetries we discussed above, see \Family .
 Let us say the two lines we picked are lines $i$ and $j$. This splits the polygon into
 a ``top'' part and a ``bottom'' part.  We now consider a symmetry generator $M$ which leaves the reference
 square invariant. We leave the ``top'' invariant and act with $M$ on the bottom part of the original polygon.
 The cross-ratios for this family of polygons can be described more
 explicitly in terms of momentum twistors, see appendix A.

 The family has three parameters $M = e^{ \tau  \partial_\tau  + \sigma \partial_\sigma +
 \phi \partial_\phi}= e^{ -\tau E + i \sigma P + i \phi S } $.
  The
 three parameters $\tau ,\sigma,\phi$ are conjugate to the three symmetries of the problem.
 The $\tau $ parameter is the one that is coupling to the symmetry that was included in the $SL(2,R)$ symmetry
 of the two null lines. When we send $\tau  \to \infty$ we are mapping the whole set of points between lines
 $j$ and $i$ to one of the sides of the square. We can view this as a multicollinear limit, see \Family .
The collinear limit corresponds to the special case where we have only two segments between segments $j$ and
 $i$  and we take the $\tau \to \infty$ limit, see \CollinearExpansion .

 The idea is that we can expand the whole Wilson loop in terms of eigenstates of these three operators.
 These eigenstates corresponds to excitations of the flux tube (or string) that goes between the two selected
  null lines.
   Thus, we are cutting the Wilson loop. The bottom and top parts
 of the loop correspond to two particular superpositions of states. Introducing the
 family of polygons is helpful for isolating the contributions from different eigenstates.
 To perform this expansion we
 are simply using the conformal symmetry of the problem. Thus, the expansion makes sense for null
 polygons in any dimension
 where we can define Wilson loops and for any value of the coupling.

 Again, given a polygon we fix a reference square and introduce a family of polygons parametrized by $\tau,\ \sigma$ and $\phi$. These coordinates are associated to the unbroken symmetries of the square.  The original polygon may be associated to the point $\tau=\sigma=\phi=0$ but that is of course not important,  we then expand at large $\tau$ and resum to get $\tau=0$ if we want. The choice of reference square is the analog of the choice of point for an operator product expansion and the points $\tau,\ \sigma$ and $\phi$ are the analog of the coordinates with respect to that  chosen point.

 So far we are describing the kinematics of the expansion we want to define.
 In order to flesh this out a bit more we will need to understand some more aspects of the
 dynamics so that we have a clearer picture of the states that appear in the expansion.

\subsec{States propagating in the expansion }

 What is interesting about this family of polygons is that we can view the symmetry conjugate to $\tau$
 as a Hamiltonian. In that case we expect that the result has an expression of the form
 \eqn\limitexp{
  \langle W \rangle = \int d n e^{ - \tau E_n } C_n ~,~~~~~~~~C_n = C^{\rm top} _n C^{\rm bottom}_n
  }
   where $n$ is some set of labels for the different
 ``energy'' eigenstates.  We have also emphasized the fact that the coefficients $C_n$ factorize into two
 contributions which are the overlap of the intermediate state with the top or bottom polygon.
 In order to make a precise statement we will have to take into account that
 UV divergencies of the Wilson loop break these symmetries. Fortunately, this breaking is well understood and
 we will be able to take care of the those effects in a simple fashion. So, let us treat it as an exact
 symmetry for   the time being.

  Now, let us explain  better what kind of states we expect in \limitexp .
  In the case of the ordinary OPE of local operators we can surround the two operators we are considering
  by a three sphere. We can then view the states appearing in the OPE as energy eigenstates of the theory
  on the sphere. To be more precise, we could consider the conformal field theory in $R\times S^3$ and find
  the energy eigenstates.

  In our case, the procedure is very similar, we could consider a constant $\tau$ surface. We see that the
  surface is pierced by two null Wilson lines, see \Cut .   The Wilson lines that pierce
  the surface create a color electric flux tube. The
   states can be understood as excitations of this flux tube. In other words, the flux with no excitations
   gives a flux vacuum. We need to understand the excitations of the field theory around this vacuum.

    \ifig\cylinder{ (a) Field theory on $R \times S^3$ in the presence of two null Wilson lines. Dotted lines
    are in the back of the cylinder. Of course the circle of the cylinder is really an $S^3$.
    (b) Double analytic continuation sends the initial and finial states of (a) to two null lines, the
    two blue null lines. Different states will contain different insertions along these two blue null lines.
 } {\epsfxsize2.5in\epsfbox{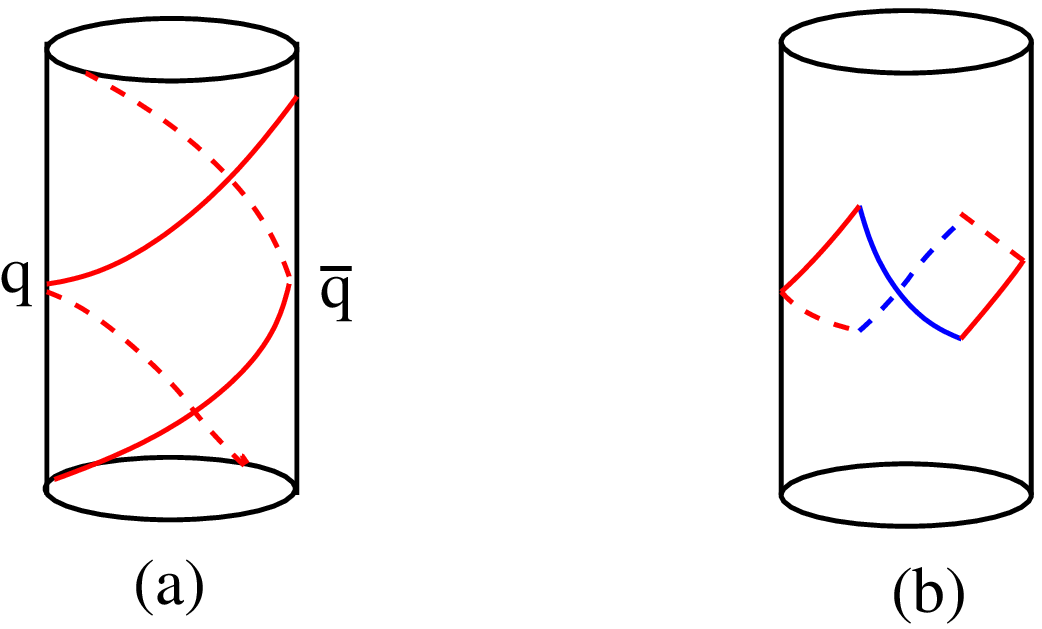}}

   More explicitly the states propagating can be understood
  as follows.
  We start with the Yang Mills theory on $R \times S^3$. We now add two null Wilson lines moving along a great
  circle of the $S^3$, see \cylinder (a). If $\varphi$ is the coordinate
  of that circle, then the lines are along $t=\varphi$
  and $t= \varphi + \pi$ where $t$ is the time coordinate. The generator $\Delta - S =
   i \partial_t + i \partial_\varphi $ is
  a symmetry of the configuration and we will call it the ``Energy''. It is also called the ``twist''.
   One line corresponds to a quark and the other to
  an anti-quark. These lines create a color electric field. The configuration is invariant under
  the ``momentum'' symmetry, see \AMtwo , generated by $P=-i\partial_{\sigma}$.
  This is not an obvious geometric symmetry of $R \times S^3$, it is a combination
  involving conformal Killing vectors. The flux is extended along the $\sigma$ direction and has a
  constant energy density along $\sigma$. Thus its energy diverges.
  The energy density is simply the cusp anomalous dimension \AMtwo .  The flux emanates from
  a point in the extra    $S^2 \subset S^3$ and the transverse $SO(2)$ symmetry is the rotation that leaves this point invariant.
  In appendix B we show how to choose
  a conformal frame where both the $\tau$ and $\sigma$ translations are manifest.
  This is the ground state of the configurations we consider. We can then add excitations which propagate
  on top of this configuration. In the planar theory these are excitations of the color electric flux and
  the indices of the particles we create are contracted with the indices of the background flux.
  This is a Lorentzian picture for the states appearing in the OPE. See appendix B for further discussion.

  We can perform an analytic continuation of the configuration with two Wilson lines on $R\times S^3$ to
  a configuration with four Wilson lines describing the square Wilson loop, see \cylinder  (b).
   Upon this transformation the
  Energy generator becomes a generator which acts in a Euclidean fashion. It becomes the $\partial_\tau$
  generator we mentioned before for the square. The two lines of the Lorentzian picture become two opposite lines
  of the square. The excitations are produced by adding operators on
  the other two lines of the square, see \cylinder (b).

  \ifig\Expansion{ Expansion of a Wilson loop in terms of states propagating on the square. The first
  term corresponds to the expectation values of the top and bottom Wilson loops. We can define the
  expectation value of the square as being one. The second term corresponds to the exchange of a single
  particle. The top and bottom Wilson loops give rise to the factors $C^{\rm top} $ and $C^{\rm bottom} $
   in the OPE expansion.
  The third term contains two particle states, and possible bound states, etc.
 } {\epsfxsize3.5in\epsfbox{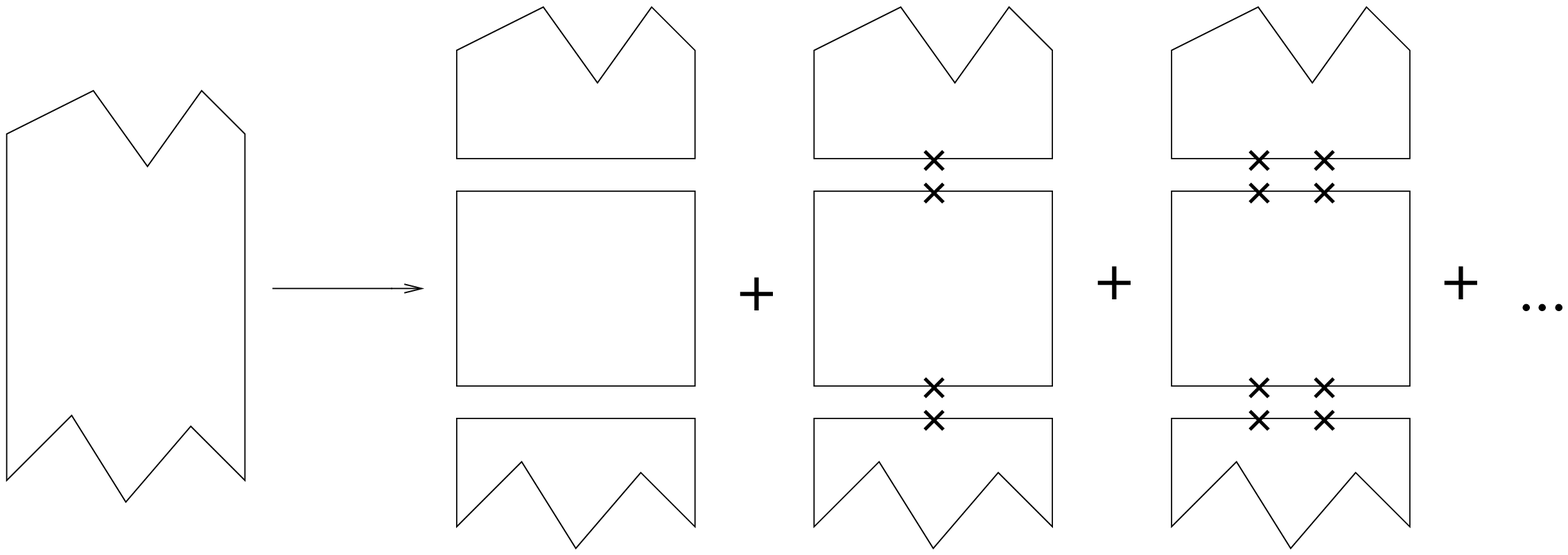}}

   These states
  form representations under the two other commuting symmetries.
  One of them, $\partial_\sigma$, is a non-compact
  symmetry. Thus, we expect a continuous spectrum of excitations. We can thus separate states according
  to their ``momentum'' in the $\sigma$ direction. In addition, we have some angular momentum number $m$ under
  the transverse rotation. Thus  a more refined statement has the form
  \eqn\limtmore{
   \langle W \rangle = \int d n e^{ - \tau E_n + i p_n \sigma + i m_n \phi } C_n
   }

  Now that we have defined this family we can simply take the large $\tau$ limit. This would select  the lowest lying states from
  \limtmore . But the expansion makes sense for any $\tau$ and we conjecture that
  it  converges for all values of $\tau$ such that
   all points in the polygon remain spacelike separated  for all $\sigma$ and $\phi$ for that
   fixed $\tau$.

 % Since the expansion in \limtmore\ involves the ``energies'' of some states, one can ask the question
 % of whether we can remove the quotation marks and make these real energies of a system. In fact, a similar
 % situation arises when we do an ordinary OPE. We have anomalous dimensions of operators. These are mapped
 % to energies of the field theory on $R\times S^3$. Here there is a similar mapping. The
 % energies we are taking about can be obtained by first mapping $R^{1,3}$ to $AdS_3 \times S^1$. Then
 % the energy $\partial_\tau$ is an energy in $AdS_3$. The second symmetry $\chi$ is again another
 % direction inside the $AdS_3$ subspace. In addition we have a constant color electric field $F_{\tau \chi}$
 % along the $\tau,\chi$ directions. We are considering excitations of this color flux configuration.
 % The ground state is simply given by the ground state energy of this color electric flux. SEE ???.

  % The lowest lying state is given by the configuration for the square
  % Wilson line. This has a certain ground state energy, which diverges due to the infinite length
  % of the $\chi$ direction. This divergence is well understood and it is characterized by the cusp
  % anomalous dimension $\Gamma_{cusp}$. We will simply subtract this well understood term.

  %  This paper
  % is about further subleading corrections. In the particular case of the collinear limit
  % the leading terms were understood in CITE . Here we are pointing out that we can make a non-trivial
  % statement about the further subleading corrections.

   The first subleading correction comes from a single excitation that is moving on the background of
   the square Wilson loop. This corresponds to the lowest lying excitation of the color electric flux
   we discussed above. In other words,
   taking the leading order limit corresponds to replacing the whole bottom of the Wilson
   loop by the single line of the square,  \Family (b). Taking into account the fact that the line is not really straight
     corresponds to insertions of $F_{+i}$ along that line, where $+$ is the direction along the line.
   This is the case, because deformations of a Wilson loop can be understood in terms of insertions of
   such   operators along the original loop. Thus, the simplest deformation corresponds to a single
   insertion of the field strength $F_{+i}$ along the loop.
   By an insertion along the loop, what we really mean is that we have an expression of the form
   \eqn\exprf{
 %  W F_{+i }(x) =
  Tr[   P e^{\oint_x^x A } F_{+i }(x) ]
   }
   where $F$ is inside the trace and inserted along the contour at point $x$. $P  e^{\oint_x^x A }$ denotes
   the integral of the connection along the loop starting at $x$, going around the loop and ending again at
   $x$.

   Note that $F$ is inserted along a null line. We can now consider the generator
   $\Delta -S = D - M^{+-}$ which leaves points on this line invariant. This is the so called ``twist''
   generator. In fact, operators which correspond to Wilson lines along a null direction with operators
   inserted on them were analyzed quite extensively in gauge theories, including QCD, because they govern
   many interesting high energy processes, going back to the classic analysis of deep inelastic scattering. See
   \ConformalQCD\ for a review.

   Thus, one of the interesting points is that the operators that appear are fairly well understood and
   have been studied in the past. Note that the operator \exprf\ breaks the $\sigma$ translation symmetry, since
   $\partial_\sigma$ moves points along the null line where $F_{+i}$ is inserted. Thus, we can consider
   superpositions with definite momentum $p$ along the $\sigma$ direction.  In addition, it has
   charge $\pm 1$ under the transverse $SO(2)$ symmetry. There is a unique state with momentum $p$
   and thus this state is automatically an eigenstate of the Hamiltonian $\partial_\tau$, with energy
   $\epsilon(p)$. We can think of $\epsilon(p)$ as the single particle dispersion relation.
    Our discussion so far, has been valid for any conformal planar gauge theory. In the particular
    case of ${\cal N}=4$ super Yang mills the function $\epsilon(p, \lambda)$ has been computed
    exactly in \Basso . The square has a symmetry under the exchange of $\tau$ and $\sigma$. This
    amounts to a ``Wick'' rotation which exchanges the spatial and time direction. This implies certain
    constraints on the dispersion relation. More can be found in appendix C.

    For general non-planar conformal theories this discussion continues to be valid as long as the flux
    is conserved. For example, in non planar ${\cal N}=4$ super Yang Mills the discussion continues
    to be valid if we consider Wilson lines in the fundamental. The flux vacuum is well defined, and so
     are its
    excitations.  If we consider Wilson lines in the adjoint, then
    the flux can be screened and the discussion will need some modifications.

    In summary, the leading order expression in the large $\tau$ limit is given by\foot{Taking the
    logarithm reorganizes the expansion in the usual fashion.}
    \eqn\leadexp{
   \log \langle  W \rangle = \log \langle W\rangle^{\rm top} + \log \langle W\rangle^{\rm bottom} +
    \sum_{\pm} \int dp C_{\pm }(p,\lambda)
    e^{ \pm i \phi } e^{i p \sigma } e^{ - \epsilon(p, \lambda) \tau }  + \cdots
    }
    where $C_{\pm}(p,\lambda)$ are some unknown functions, analogous to OPE coefficients. And $\log \langle
    W \rangle^{\rm top,~bottom}$ are the expectation values of the top and bottom Wilson loops, see
    \Family\ (b)(c).
    The function $\epsilon(p,\lambda)$ is the exact dimension (or twist)  for a single insertion of
    $F_{+i}$ along a null line. In fact,
    $C_\pm = C^{\rm top}_\pm \times C^{\rm bottom}_\pm$ where $C^{\rm top}_\pm $ and
    $  C^{\rm bottom}_\pm $ correspond to
    the expectation values of the Wilson loop
    contours in \Expansion\ with an $F_{+i}$ insertion on the top or bottom line\foot{More precisely, they
    are the ratio of the expectation value with an insertion divided by the expectation value without the
    insertion.}.
     In other words, as usual, $C^{\rm top}_\pm$ is the overlap between the intermediate state
    that is propagating and the state created by top  part of the Wilson loop while $C^{\rm bottom}_\pm$ is the
    overlap with  the bottom Wilson loop.

    \subsec{Taking care of violations of conformal symmetry: the remainder function}

    So far we have ignored the UV divergencies of the Wilson loop and we have treated the
    symmetries as if they were unbroken. Since the breaking of the symmetries is a well understood
    phenomenon \SokWard , it is clear that we should be able to take this breaking into account.
    In other words, there is an anomalous ward identity which tell us how the Wilson loop changes when
    we apply a special conformal transformation. Thus, we could go through the above argument adding the
    corresponding terms due to the anomalous ward identity.

    We will follow a slightly different route which we found more convenient. This is based on the
    observation that the anomalous ward identity is also obeyed by the same Wilson loop correlator but
    in a free $U(1)$ theory.
    More precisely, we simply need to take the free $U(1)$ result and replace the cusp anomalous dimension
    of the $U(1)$ theory, $\Gamma_{1,cusp}$,
     by the full cusp anomalous dimension, $\Gamma_{cusp}$
      of the interacting theory in question.
    So we can write
    \eqn\answe{
    \langle  W\rangle_{\widetilde U(1)}    =  \left[ \langle W\rangle_{U(1)} \right]^{\Gamma_{cusp} \over \Gamma_{1,cusp} }
    = e^{   \Gamma_{cusp} w_{U(1)} }
     }
     where $w_{U(1)}$ is the result for a $U(1)$ theory.
      This function is also related to the one loop maximally helicity violating amplitude once the tree level contribution is stripped out \refs{\brandhuber,\BDS}.  It is given by a single gluon (or rather photon) exchange
     between all pairs of lines. It is a completely explicit
     function of the distances between
     various cusps and we will not need its explicit expression\foot{
      Its explicit form can be found in formula (4.58) of \BDS . In this context this function is known as ``the
      BDS expression''. }. The tilde in $\widetilde U(1)$ just reminds us
      that we have put the cusp anomalous dimension of the interacting theory.

     Thus the ratio
     \eqn\remfun{
     R = \log \left[ { \langle W \rangle \over \langle W\rangle_{\widetilde U(1)}} \right]
     }
     is a conformal invariant function. This is also called the remainder function. By definition, its
     first non-zero contribution is at two loops \refs{\DrummondBM,\BernAP,\AnastasiouKNA}.
     This definition clearly takes care of the divergence problem and leads to an explicitly conformal
     invariant answer. Though, this ratio takes care of the double logarithmic divergencies, there are
     single logarithmic divergencies that should be taken care of. These are regulator dependent. Thus, we
     adjust the regulator in the $\tilde U(1)$ theory so that the single logarithmic divergencies match
     those of the full theory. Similarly there is a finite   constant   which is proportional
     to the number of cusps and is also regulator dependent \foot{We thank E. Sokatchev for pointing out the
     omission of these single logarithmic terms from the first version of this paper. }.
     These single log divergent terms or the constant terms are not important for this paper.

     This ratio gives a nice conformal invariant expression,
      however, we should understand how
     it modifies the expectations from the point of view of the Hamiltonian interpretation of the
     family of polygons and the expansion for large $\tau$.

     For this purpose, it is useful to note that $\langle W\rangle_{\widetilde U(1)}$ does have its
     own Hamiltonian interpretation
     since it is a computation in a $U(1)$ theory. Thus, we have an expansion of the form in
     \limtmore\ where
     the anomalous dimensions vanish, and the energies appearing in \limtmore\ are simply the twist of
     the corresponding operators in the free $U(1)$ theory. In the free $U(1)$ theory the twists are just
      integers.
      Thus the expansion in \remfun\ contains the ratio of the two
     expansions, or difference once we take the log.

     In other words, the final version of the expansion takes the form
     \eqn\expansf{
      R = R^{\rm top} + R^{\rm bottom}  + \int d n  C_n e^{ - E_n \tau + i p_n \sigma + i m_n \phi} -
      \int d n  C^0_n e^{ - E^0_n \tau + i p^0_n \sigma + i m^0_n \phi}
       }
       where the subindex 0 indicates that we are considering the $U(1)$ theory. $ R^{\rm top}$ and
       $   R^{\rm bottom} $
       are the remainder functions for the two polygons in figure \Family (b),(c).
        We are ignoring here a
       possible constant piece which is independent of the kinematics.

       Now, in practice, we can distinguish the terms that come from the free theory because
       the energies are those of a free theory.

       For example, if we consider the terms arising from a single insertion of the field $F_{+i}$ then
       the expansion takes the simple form
       \eqn\expansf{
      R =R^{\rm top} + R^{\rm bottom}   + \sum_{\pm }
     e^{\pm i \phi } \left[ \int d p  C_{\pm}(p, \lambda)  e^{ - \epsilon(p,\lambda) \tau + i p  \sigma } -
       e^{-\tau } \int dp  C^0_{\pm}(p) e^{  i p \sigma  } \right]
       }
       The last term is the contribution from the $U(1)$ theory and it has a simple $\tau$ dependence. The
       first term includes the full dispersion relation   and this  leads to a more complicated $\tau$
       dependence.
       In the next section we will check that for the case of an hexagon the strong coupling answer \AGM ,
        as well as the two loop
weak coupling  answer \refs{\DrummondBM,\DelDucaZG,\ZhangTR}
has the form predicted by \expansf . We will also discuss the
general structure of $n$-sided polygons at strong coupling.
       We see that \expansf\ implies a particular structure for the Wilson loop correlator.

       The dispersion relation $\epsilon(p, \lambda)$ in \expansf\ has been computed for all
       values of the coupling in \Basso .  Using that result, then \expansf\ gives a prediction for the
       Wilson loop (or the MHV amplitude if they are equal)
        that should hold for all values of the coupling. Notice that we do not know what $C$ is for all values
        of $\lambda$. Nevertheless we have a concrete prediction in terms of the $\tau$ and $\sigma$ dependence
        for the answer.

       In appendix C we make further remarks on the spectrum of excitations around the flux tube.
       It would be useful for the reader to consult this appendix if he/she is not familiar with these
       excitations.

  \subsec{Taking care of violations of conformal symmetry without using the $U(1)$ or one loop answer}

It would be nice to be able to strip off completely the $U(1)$ theory contribution
from our expansion. This is actually possible, by regulating the anomalies in a different
way.  It is useful first to understand how this can be done in a $U(1)$ theory.
We start with the following ratio of Wilson loop expectation values in a $U(1)$ theory.
\eqn\rbds{
r_{U(1)} =\log\left(
\frac{\langle W\rangle_{U(1)}\langle W_{square} \rangle_{U(1)}}{\langle W^{\rm top}
\rangle_{U(1)}\langle W^{\rm bottom}\rangle_{U(1)}}\right)
}
 is finite and conformal invariant, if $W^{\rm top}$ is the Wilson loop with the
 bottom part replaced by the single bottom line of the square
 and $W^{\rm bottom}$ is
 the Wilson loop with the top part replaced by the single top line of the square, see \Family (b),(c).

\ifig\Familytopbot{ (a) The original contour with the reference square given by its vertices $ABCD$.
(b) The OPE expansion in the $U(1)$ theory can be written in terms of $ r_{U(1)}$ which is computed
by a single photon exchange between the top and bottom Wilson loops in this figure.
 In (c) and (d) we see the same, but with a choice of points that removes the divergencies. Points D and B
 of the reference square coincide with points on the original polygon.
 } {\epsfxsize2.5in\epsfbox{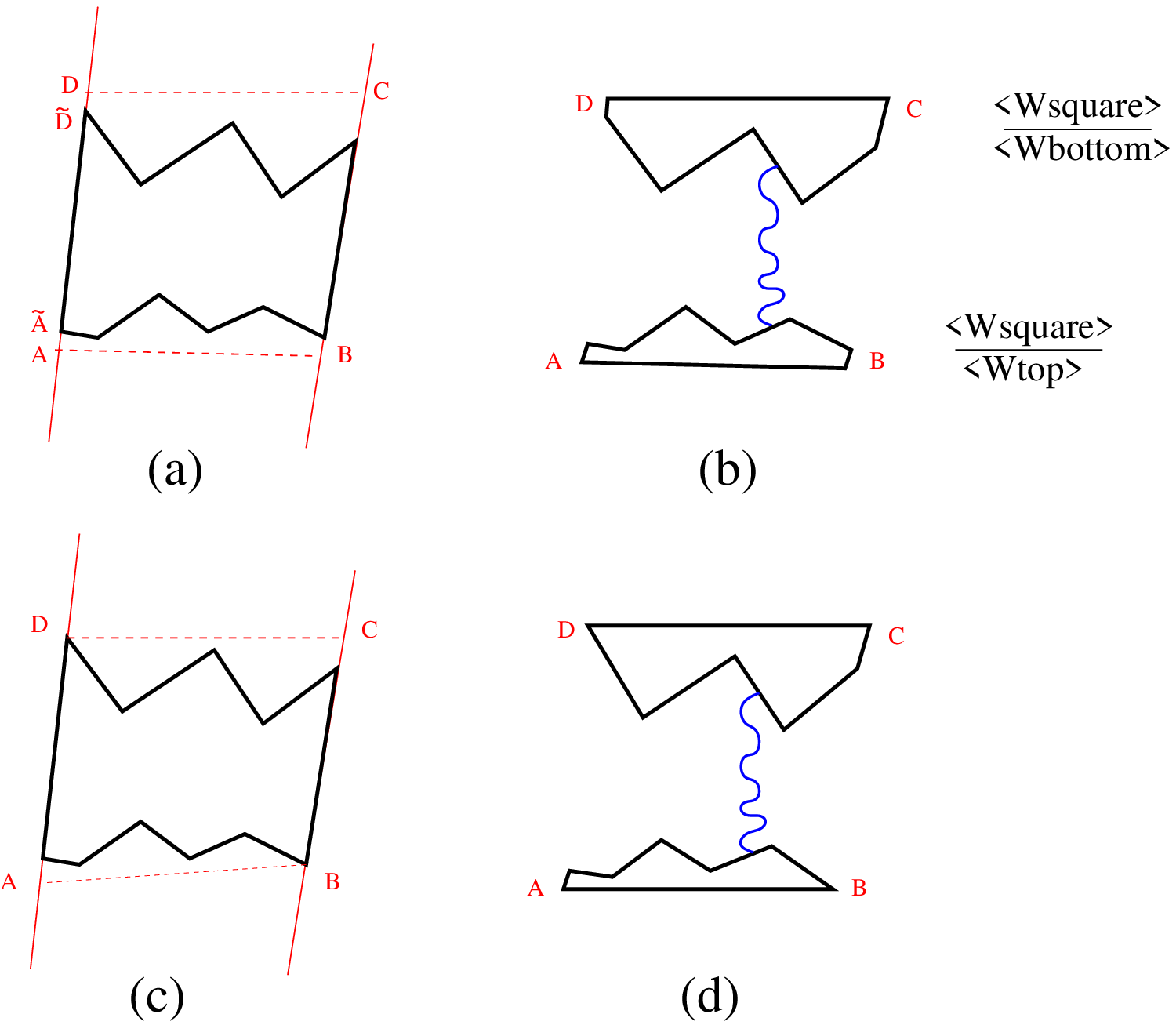}}

The $U(1)$ Wilson loop is the double integral of the free propagator
\eqn\bdsdef{\log \langle W\rangle_{U(1)} = \half \oint_W dx^\mu \oint_{W'} dx'^\nu G_{\mu \nu}(x,x') }
where we regularized the integral by shifting $W$ slightly in the second contour integral.
Then it is easy to see that
\eqn\rbdspath{
r_{U(1)} =  \oint_{W_{square}-W_{top}} dx^\mu \oint_{W_{square}-W_{bottom}}
 dx'^\nu G_{\mu \nu}(x,x') }
Here the contours $W_{square}-W_{top}$ and $W_{square}-W_{bottom}$
 are well separated and one is tempted to say that this expression is finite.
  However, there is a remaining divergence from propagators that connect
 the short portion of the null line in the bottom diagram with the segments that touch the extension of
 same
 null line in the top , see \Familytopbot . This leads to divergent terms that
  depend only on the location of the points
 along the null lines though AD , for example\foot{There is a single log divergence proportional to the
 log of the cross ratio of the four points that are sitting along the null line passing through AD.
 In other words, we have a term
 $\log \mu \log { x_{A D}^2 x_{\tilde A \tilde D }^2 \over x^2_{\tilde A D} x^2_{A \tilde D} } $, where
 $A D$ are points on the reference square and $\tilde A, \tilde D$ are points on the original polygon along
 the same line, see \Familytopbot (a).  There
 a similar one from BC. These cross ratios depend only on $\tau$ and not on $\sigma$. They vanish if
  $D = \tilde D$ as in \Familytopbot (d). }.
 Such divergencies are independent of $\sigma$. They
 can be interpreted as coming from a condensate of Goldstone bosons of the broken $SL(2,R)$ symmetry
 that acts on $\tau$. These divergencies can also be removed by choosing points B and D  (or A and C)
 of the reference polygon to coincide with the vertices of the original polygon, see \Familytopbot (c) (d).
 Thus, we can either remove these divergencies by an appropriate choice of the reference polygon, or we
 can just simply ignore them because they are independent of $\sigma$.

The expansion of $r_{U(1)}$
has the correct form to be identified with
     \eqn\expansfrbds{
     r_{U(1)} =  \int d n  C^0_n e^{ - E^0_n \tau + i p^0_n \sigma + i m^0_n \phi}
       }
Indeed we can expand the propagator in eigenfunctions of the symmetries of the square, including the action of $M$ on the bottom Wilson loop:
$$G_{\mu \nu}(x,x')d x^\mu d x'^\nu = \int dn\ dx^\mu\psi_\mu(x,n)\ dx'^\nu\psi_\nu(x',n)\ e^{ - E^0_n \tau + i p^0_n \sigma + i m^0_n \phi}$$
and compute the sources
 \eqn\rbdssource{
c^{top}(n)  = \oint_{W_{square}-W_{top}} dx^\mu\psi_{\mu}(x,n)}
and $c^0_{bottom}(n)$ so that $c^0_n = c_{top}(n) c_{bottom}(n)$.
In the next section we check this result for the hexagon Wilson loop.
Note that this gives an explicit formula for the coefficients
$C^0 = { \Gamma_{cusp} \over \Gamma_{1,cusp} } c^0_n $
appearing in the expansion \expansf .

With this $U(1)$ result as an inspiration we are lead to define an alternative conformal invariant and
finite expression\foot{ Again we choose $D =\tilde D$ and $B = \tilde B$ to eliminate a single log divergence.
Alternatively, we could consider a more general reference square and ignore the $\sigma$ independent divergence.}
\eqn\rbds{
%r = r_{\widetilde U(1)} R  = \log\left[\frac{\langle W\rangle\ \langle W_{square} \rangle_{\widetilde U(1)}
% }{\langle W_{top}\rangle_{\widetilde U(1)} \langle W_{bottom}\rangle_{\widetilde U(1)} }\right]
r =   \log\left[\frac{\langle W\rangle\ \langle W_{square}
\rangle
 }{\langle W_{top}\rangle  \langle W_{bottom}\rangle  }\right]
}
where now all expectation values are in the full interacting theory.
This leads to a simple expansion   involving only the physical excitations
     \eqn\expansfr{
    % r= R^{\rm top} + R^{\rm bottom} + \int d n\  C_n e^{ - E_n \tau + i p_n \sigma + i m_n \phi}\ .
      r=  \int d n\  C_n e^{ - E_n \tau + i p_n \sigma + i m_n \phi}\ .
       }
The combination in \rbds\ is subtracting the contribution from the unexcited flux tube,
or flux vacuum contribution.
The remainder function defined in \remfun\ is simply given by $R = R_1 + R_2 + r - r_{\tilde U(1)}$.
Notice that the fact that \rbds\ has an expansion in terms of subleading terms only also explains why
the leading terms in the ordinary collinear limit are fixed by conformal symmetry. In that case $W_{\rm bottom}$
is a pentagon whose expectation value is fixed by the anomalous ward identities \SokWard . This pentagon leads to the
terms in the splitting function computed previously \BDS .

       \newsec{ Detailed checks for the hexagon }

       A simple case that we can consider is the hexagonal Wilson loop. In this case there are three
       cross ratios $u_1,u_2,u_3$. Since our family involves three parameters, then it is clear
       that we can parametrize the three cross ratios in terms of the three parameters $\tau, \sigma ,\phi$.
       In this case, the limit $\tau \to \infty$ is a collinear limit which leaves behind a pentagon, which
       does not have any remaining cross ratios.

       There are different ways to parametrize the $u_i$. An important choice is the choice of
       two opposite lines of the hexagon. There are three different ways of doing this which   represent
       the
       three different channels in term of which we can expand the correlator.
       Once we chose a pair of opposite lines we can now choose the reference square in slightly
       different ways which amount to the action of the generators of $SL(2,R)$ that leaves the two lines
       fixed. Different choices give slightly different parametrizations which only affect results at higher
       orders. These different choices change the answer in a predictable way, as is the case for the ordinary
       OPE. We can make a choice so that the  cross ratios are, see appendix F.3,
\eqn\writecr{\eqalign{
u_2 = &  { 1 \over \cosh^2\tau  }
%= e^{ - 2 \tilde u } + \cdots
\cr
u_1 = &  { e^\sigma \sinh \tau \tanh \tau  \over 2 ( -\cos \phi + \cosh\tau \cosh\sigma ) }
%= { 1 \over 2 } { e^v \over
%\cosh v } - e^v { \cosh f \over \cosh^2 v} \epsilon + o(\epsilon^2 )
\cr
u_3 = &  { e^{-\sigma} \sinh \tau \tanh \tau  \over 2 ( -\cos \phi  + \cosh\tau \cosh\sigma ) }
%= { 1 \over 2 } { e^{-v} \over
%\cosh v } -e^{-v} { \cosh f \over \cosh^2 v} \epsilon + o (\epsilon^2 )
}}
The expression for $u_1$ and $u_3$ appears rather complicated,
 but one may note that $u_1/u_3 = e^{ 2 \sigma }$ and  that the parameter $\mu$ appearing in the
integral equations in \AGM\ is $\mu = -e^{ i \phi }$. In (2,2) signature we can set $\mu = e^f$ and
$\cos \phi \to - \cosh f $.

 Once we have this parametrization we can expand the remainder function for large $\tau$.

 \subsec{ Expansion of the hexagon at strong coupling }

 In this section we quote the result of the expansion of the hexagon at strong coupling.
 We leave the details to appendix F,
  where we show that the remainder function at strong coupling does indeed admit
 an expansion of the form \expansf\ involving a reasonable spectrum of operators,
  thanks to the cancellation of several unwanted terms.

 We expand the answer to order $e^{- 2 \tau }$.
 We find
 \eqn\resex{ \eqalign{
 R &=R_{1} + R_{\sqrt{2} } + R_2 + \cdots
\cr
R_1 &=   - \cos \phi \,  e^{-\tau }  (   \cosh \sigma  \log[ 2 \cosh\sigma]
- \sigma \sinh \sigma )
\cr
R_{\sqrt{2} } &=    4 \cos \phi \int { d \theta \over 2 \pi } { 1 \over ( \cosh 2 \theta )^2 }
e^{ - \tau\sqrt{2} \cosh \theta + i \sigma {\sqrt{2}}  \sinh \theta  }
\cr
R_{2} &=  { e^{- 2 \tau } } \left[ { \log ( 2 \cosh\sigma ) - \sigma \over 2 } +
 \cos 2 \phi \, g(\sigma) \right] +
2 \int { d \theta \over 2 \pi } {  e^{ -  2 \tau \cosh \theta + 2 i \sigma \sinh \theta }\over [\sinh (2 \theta + i 0 ) ]^2 }
 }}
 Let us explain the interpretation of the various terms.
 The terms multiplied by $\cos \phi$ correspond to the propagation of a particle created
 by the insertion of an excitation $F_{+i}$ which carries unit charge under the transverse $SO(2)$. The $R_1$ term has energy (or twist) one, independent of the momentum. This
 is then interpreted as coming from the $U(1)$ subtraction term.
  The term $R_{\sqrt{2}}$ is the contribution from the
 propagation of the corresponding particle at strong coupling. At strong coupling this particle has a
 relativistic dispersion relation with mass $\sqrt{2}$; its momentum is $p=\sqrt2\sinh\theta$.
 The term $R_2$ contains terms going like $e^{- 2 \tau}$ which come from the contribution of the $U(1)$
 theory. One of them is independent of $\phi$. This term could come from the insertion of two twist one
 fields or one twist two field. The strong coupling contribution is again coming from the exchange of a
 single particle with a relativistic dispersion relation, last term in $R_2$.  This particle has mass two. We do not know
 whether this particle remains present at all values of the coupling. It might decay into two fermions
 as it was found in a similar context in \Zarembo .

 There is an interesting interplay between  the terms with integer powers of $e^{-\tau}$ and
the integral terms.    As we analytically continue $\tau$ and $\sigma$,   there are poles that cross
 the integration contours  and we can get extra contributions which contain integer powers of $e^{-\tau}$.
   In fact, they combine with the terms already present  to ensure that the expansion
 has the right properties. For example, the term in $R_2$ going like $e^{-2 \tau}$ is not
 symmetric under $\sigma \to - \sigma$. This lack of symmetry is cured by the lack of symmetry of the
 integral which is introduced by the $i0$ prescription.
 There are similar effects that occur when we go to large $\sigma$ and then to $\sigma > \tau$.
  The   result is invariant under the ``Wick'' rotation which exchanges $\tau$ and
 $\sigma$.

   \subsec{Expansion of the hexagon at weak coupling }

  When we expand \expansf\ at weak coupling we generate terms going like $\lambda^n \tau^{k} e^{-\tau } $,
   $k\leq n-1$,
  where we took into account that $C$ starts at order $\lambda$.
  These are the usual logs that we get in perturbation theory and arise from the expansion
  of the exponent
  $\epsilon(p,\lambda) = 1 + \lambda \gamma_1(p) + \cdots $, where $\gamma_1(p)$ is the one loop anomalous dimension
  for the excitation.
   More concretely, consider the lowest order term, which is proportional to $\cos \phi$.
  \eqn\weakexp{
R = \cos \phi\int dp\  e^{i p \sigma } \left[  ( \lambda C^{(1)} + \lambda^2 C^{(2)} +\cdots) e^{ - \tau
 - (\lambda \gamma_1 + \cdots) \tau }
- ( \lambda  + \lambda^2 { \Gamma_{2} \over \Gamma_1}  ) e^{-\tau}c^0  \right]
}
where we have expanded everything in powers of $\lambda$ and $C^{(i)}$, $c^0$ are functions of $p$.
$\Gamma_i$ are the coefficients in the expansion
of the cusp anomalous dimension.
Now, the fact that the remainder function is zero at order $\lambda$ implies that
$C^{(1)} =c^0(p)$, which is the result for the $U(1)$ theory, or the result at one loop.

We can now compute the terms of order $\lambda^2$. Expanding \weakexp\ we find
\eqn\weaksquare{
R = \cos \phi e^{-\tau } \lambda^2  \int d p e^{i p \sigma } ( C^{(2)}  - \gamma_1 \tau c^0 + { \Gamma_2 \over \Gamma_1}
 c^0 )
} Thus, we can take the two loop expression and look at the
term $\tau e^{-\tau} h(\sigma)$. If we fourier transform this
term we get $-\gamma_1(p) c^0 (p)$. We can independently
compute $c^0(p)$ from \rbdssource . We can  also compute it
with a trick: we can go to strong coupling, where the real
result and the U(1) result do not mix. The $U(1)$ result is
contained in the term $R_1$. Its fourier transform is
  \eqn\fourtrmain{
c^0(p) \propto  \int d \sigma e^{ip \sigma } (  \cosh\sigma \log[ 2 \cosh \sigma] -\sigma \sinh\sigma )=
 { 1 \over (1 + p^2) }
 {\pi  \over \cosh { p \pi \over 2 } }
}
It is easy to confirm this result from the direct expansion of $r_{U(1)}$ in \rbdspath\
 at the leading order, see appendix D.

We also need to know $\gamma_1(p)$. This is given by the one loop anomalous dimension and for an
$F_{+i}$ propagating on a sea of derivatives. It was computed in \KorchemskySpin\  and is given by
\eqn\gammaexp{
\gamma_1(p) =  \psi\left({ 3 \over 2 } + i \frac{p}{2} \right)  + \psi\left({ 3 \over 2 } - i \frac{p}{2} \right) - 2 \psi(1) \,.
}
From \KorchemskySpin\ (equation (3.37) with $s=3/2$)
 we might naively expect $-2\psi(3)$ instead of $-2\psi(1)$.
 However,
  we need to add the anomalous dimension per excitation
  $\frac{\delta\Delta}{L \lambda}=3$ of the state $\tr\left( F_{+i}^{}\right)^L$ \FL\
  to the result of  \KorchemskySpin.

\ifig\numerics{Using the expressions from \ZhangTR\ we computed the functions $h(\sigma)$
appearing in the expansion $\tau e^{-\tau } h(\sigma) $
of the two loop result for the hexagon Wilson loop. The dots in this figure are the values obtained by fitting the numerical data in the large $\tau$ regime. The red curve corresponds to the analytic prediction discussed in the main text. Recall that there is no fit of any parameter in this comparison; hence this check is a very strong check of our predictions.
 } {\epsfxsize3.5in\epsfbox{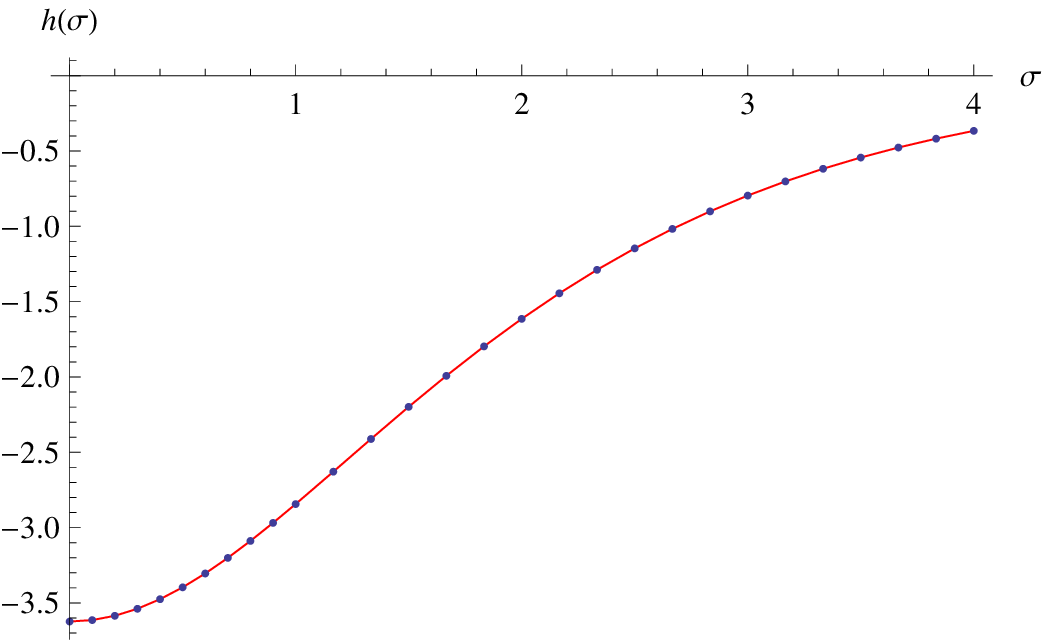}}

We have checked numerically that this prediction is indeed true
using the expressions from \ZhangTR\ for the two loop remainder
function for the hexagon. In other words, the fourier transform
of $-\gamma_1(p) c^0(p)$ gives a function $h(\sigma)$ \foot{This
function can actually be computed analytically: $$h(\sigma) \propto \cosh \sigma \left( 2 \log(1+e^{2\sigma}) \log(1+e^{-2\sigma})-4\log(2 \cosh \sigma)\right) +4 \sigma \sinh \sigma$$} which is precisely
the one appearing in the term $\tau e^{-\tau } h(\sigma) $ in the
expansion of the two loop result for large $\tau$, see figure
\numerics .

It should be clear that a similar prediction can be formulated for a general polygon, simply by replacing
$C^{(1)}(p)$ with the leading part of $\log r_{U(1)}$ for that polygon. It would be interesting to check
this prediction with an explicit two-loop calculation: the $U(1)$ coefficient $c^0(p)$
should capture the full,
intricate dependence on the Wilson loop cross-ratios of the term
$\tau e^{-\tau } h(\sigma) $ in the expansion of the two loop result for large
$\tau$.

Clearly, at higher loops we can make further predictions. For example, at $l$ loops we expect that
the remainder function has a term of the form
\eqn\remllo{
R \sim \cos \phi \,e^{-\tau} { (-1)^{l-1} \tau^{l-1} \over (l-1)!}
\int dp e^{i p \sigma} c^0(p) [ \gamma_1(p)]^l
}
This is the highest power of $\tau$ that appears at this loop order. We also have terms $\tau^k e^{-\tau}$
with $k< l-1$. These also also fixed by the knowledge of the dispersion relation at higher loops, which
is given in \Basso .

\newsec{Higher order predictions}
In principle one can push the expansion of the remainder function to higher
order in $e^{- \tau}$, and consider terms which correspond to the propagation of
several particles. The states which appear in the $U(1)$ answer $C^0_n$
are naively only ``one particle states'', corresponding to insertions of
a single four dimensional field on
 the sea of derivatives. At the 1-loop, in the spin chain language,
such naive ``one particle states'' will mix with the continuum of
true multi-particle excitations, and possibly to actual single particle bound states.
Even without computing the exact 1-loop overlaps, we still get some constraints on
terms of the form $\tau e^{-k \tau } H(\sigma) $ in the expansion of the two loop result of a general polygon for large
$\tau$, at least as long as $k$ is sufficiently small compared to the number of sides of the polygon:
the full functional dependence of $H(\sigma)$ on the cross-ratios is captured by the  twist $k$
coefficients $C^0_k(p) = C^{0 \, {\rm top} }_kC^{0 \, {\rm bottom} }_k$,
multiplied by some unknown function
of $p$ which does not depend on the specific shape of the top and bottom Wilson loops.

\newsec{Conclusions}

In this paper we have derived an OPE expansion for polygonal Wilson loops with light-like edges.
The OPE expansion is performed by picking two non-consecutive null lines in the polygonal Wilson loop.
This divides the Wilson loop into a ``top'' part and a ``bottom'' part with states propagating between
the two. The state that propagates contains  a flux tube going between the two selected null lines.
The states consist of excitations of this flux tube. These states can also be understood as excitations
around high spin operators. The spectrum of states is continuous and consist of many particles propagating
along the flux tube. In  ${\cal N}=4$ Super Yang Mills these
 particles have a calculable dispersion relation  \Basso . Then the OPE expansion leads to predictions
 for the Wilson loop expectation values. Namely, it implies constraints on the subleading terms in the
 collinear limit of the remainder function, which is the function containing the conformal invariant
 information of the Wilson loop expectation value. We have checked these predictions both at strong coupling and at two loops at weak coupling.

 In order to perform this expansion at strong coupling,
  we have derived an alternative presentation of the integral
 equations   \refs{\octagon,\AGM,\AMSV}. In the new representation
 the parameters in the equations are the physical spacetime  cross ratios. In addition, the formula
 for the area is given by the Yang-Yang functional associated to the modified TBA equations.
 This new form of the equations are as useful as the old ones \refs{\octagon,\AGM,\AMSV}
  for numerical computations.

 For single particle exchanges we have been able to characterize completely the state of the particle in
 terms of its momentum. Using the exact formula for the single particle dispersion relation, given in \Basso ,
 one can find an exact prediction for the first subleading term in the collinear expansion. Expanding this
 formula in powers of $\lambda$ one would obtain specific predictions to all orders. Here we have checked
 this prediction with the only available direct computation in the literature, the two loop hexagon
 \refs{\DrummondBM,\BernAP,\AnastasiouKNA,\DelDucaZG,\ZhangTR}.

  When we have multiparticle exchanges the situation is less clear and we will probably
 need the full power of integrability, together with the infinite number of charges to characterize the state.
  It is likely that characterizing the states in this fashion and then demanding a consistent expansion in
  all channels one could determine the full function.

\newsec{Acknowledgments}

We thank Benjamin Basso for discussions and for providing us with a draft of \Basso .
We also thank N. Arkani-Hamed, F. Cachazo, P.Dorey, M. Spradlin,
D.Volin and  K.Zarembo for discussions.

This work was supported in part by   U.S.~Department of Energy
grant \#DE-FG02-90ER40542. Research at the Perimeter Institute is supported in part by the Government of Canada through NSERC and by the Province of Ontario
through MRI. D.G. is supported in part by the Roger Dashen membership in the
Institute for Advanced Study. D.G. is supported in part by the NSF
grant PHY-0503584. A.S. and P.V. thanks the Institute for Advanced Studies for warm hospitality.
J.M. thanks the Perimeter Institute for hospitality.

\appendix{A}{Describing the family of polygons  in terms of momentum twistors}

 One can describe the kinematics of the family of polygons that we considered in section 3.1  by
 twistors. These are sometimes called ``momentum twistors'' and they appear naturally both in weak coupling
  \HodgesHK\ and strong coupling \refs{\octagon,\AGM,\AMSV} calculations.

 Each vertex of the polygon can be represented by a null vector $Z$ in $R^{4,2}$, defined up to rescaling.
 The vector $Z$ can be thought of carrying two antisymmetric spinor indices, and can be rewritten as the bilinear antisymmetric combination of two twistors.
 Schematically, $Z = \lambda \wedge \lambda'$. Both $\lambda$ and $\lambda'$ carry a spinor index of $R^{4,2}$ and are defined up to rescaling.
 A polygon with $N$ null sides can be given as a sequence of $N$ twistors $\lambda_i$, such that the intersection
 of the sides $i$ and $i+1$ is the point $Z_{i+1/2} = \lambda_i \wedge \lambda_{i+1}$. Indeed $Z_{i-1/2} \cdot Z_{i+1/2}=0$ is the condition that the $i$-th side should be null.
 It is also useful to introduce dual momentum twistors $\mu_i$, cospinors which can be defined as $\mu_i = \lambda_{i-1} \wedge \lambda_i \wedge \lambda_{i+1}$,
 and also safisfy  $Z_{i+1/2} = \mu_i \wedge \mu_{i+1}$.

 A twistor $\lambda$ together with an orthogonal dual twistor $\mu$, $(\mu, \lambda)=0$, defines a null line (and viceversa): all points of the form $Z = \lambda \wedge v$ with $(\mu,v)=0$, $v$
 defined up to rescaling and shifts by $\lambda$. Our family of polygons with fixed sides $i,j$ must have fixed $\lambda_{i,j}$ and $\mu_{i,j}$.
 We are now in condition to give an alternative description of the group of transformations fixing two non-intersecting null lines,
 defined by the pairs $\lambda_i$, $\mu_i$ and  $\lambda_j$, $\mu_j$. In order for the lines to be at generic position with respect to each other, $(\mu_i, \lambda_j)$ and
 $(\mu_j, \lambda_i)$ should be both non-zero. A conformal transformation $M$ will leave the lines invariant if $\lambda_{i,j}$ are right eigenvectors of $M$, i.e.  $M \lambda_i = e^{\sigma + \phi} \lambda_i$ and $M \lambda_j = e^{-\sigma + \phi} \lambda_j$, and $\mu_{i,j}$ are left eigenvectors. As long as
 $(\mu_i, \lambda_j)$ and  $(\mu_j, \lambda_i)$ are non-zero, it must be that $\mu_i M = \mu_i e^{-\sigma + \phi}$ and $\mu_j M = \mu_j e^{\sigma + \phi}$.

 If we decompose the twistor space into multiples of $\lambda_i$, multiples of $\lambda_j$ and the orthogonal to
 $\mu_{i,j}$, $M$ will be block diagonal, with elements $e^{\pm \sigma + \phi}$ and $e^{- \phi} R$ for some $SL(2)$ transformation $R$.
 To pick a reference square is the same as picking two more twistors $\lambda_{\pm}$ in the orthogonal to $\mu_{ij}$. The square is left invariant by $M$ iff $\lambda_\pm$ are the two remaining
 eigenvectors of $M$, with eigenvalues $e^{\pm \tau - \phi}$.

 Once we have picked a specific reference square, the family of polygons is defined by the twistors $\lambda_i \cdots \lambda_j$ and $M \lambda_{j+1} \cdots M \lambda_{i-1}$.

  \appendix{B}{The hamiltonian picture and its analytic continuation}

 In this appendix we consider the large spin limit of local operators and we see how they
give rise to the flux vacuum and its excitations.

   We write the coordinates of $R^{1,3}$ as $Z_M = Z_{-1} ,Z_0,Z_1, \cdots Z_4$, with the
   condition $Z^2 =0$, where the indices are contracted with the Minkowski metric of $R^{2,4}$.
   In addition we impose the identification $Z \sim \lambda Z$.
   Usual poincare coordinates are
   \eqn\oord{
   x_\mu =   { Z_\mu \over (Z_{-1} + Z_4 )} ~,~~~~~~{\rm  for} ~~~ \mu =0,1,2,3
   }
   The usual $R^{1,3}$ metric is simply the induced metric on the lightcone $Z^2=0$ with the gauge
   condition $Z_{-1} +Z_4 =1$. Different ``gauge fixing'' conditions lead to Weyl transformations for
   the metric.
   Just for reference, the usual coordinates of $R\times S^3$ are defined by
   \eqn\rtimesr{
     \tan t = { Z_0 \over Z_{-1} } ~,~~~~~~~  n^i = { Z^i   \over \sqrt{ \sum_{i=1}^{4} Z_i^2 } } ~,~~~~
     i=1,\cdots,4
     }
     where $ n^i$ is a unit vector in $R^4$ and describes a point on $S^3$.
   Similarly, if we choose a gauge fixing function $Z_2^2 + Z_3^2 =1$ we get a metric which is that of
   $AdS_3 \times S^1$. \foot{ This $AdS_3$ has no relation to the $AdS$ space that appear in $AdS/CFT$. This $AdS$ space is purely within the boundary theory. This discussion applies to any CFT, whether is has a known gravity dual or not.
    }  This is again conformally related to $R^{1,3}$.
   % In other words, we can
   %write the $R^4$ metric $  - dx^+ dx^- + dr^2 + r^2 d \phi^2 = r^2 \left[ { - d x^+ dx^- + dr^2 \over r^2 }
   %+ d \phi^2 \right]$.

   Within this $AdS_3$ factor we can choose coordinates which are similar to Euler angles for $S^3$ (since
   $AdS_3$ is an analytic continuation of $S^3$).
   We can write
   \eqn\writeal{
   Y^M= { Z^M \over \sqrt{ Z_2^2 + Z_3^2 } } ~, M \not = 2,3~; ~~~~~~~~ \pmatrix{ Y_{-1} + Y_4 & Y_0 + Y_1
   \cr
   -Y_0 +Y_1 & Y_{-1} - Y_4 } = e^{ i \tau_l \sigma_2 } e^{ \beta \sigma_3 } e^{ \sigma \sigma_1 }
   }
   where $\sigma_i$ are the usual Pauli matrices. We have parametrized the space in terms of coordinates
   $\tau_l, ~\beta, \sigma$ and $\phi={\rm arctan}(Z_2/Z_3)$. The metric in these coordinates is
   \eqn\explsym{\eqalign{
    ds^2_{AdS_3 \times S^1} = & -d\tau_l^2 + d\sigma^2 + d\beta^2 + 2
   \sinh 2 \beta d \tau_l d\sigma  + d\phi^2 =
   \cr =& ( d\sigma + \sinh 2  \beta d \tau_l)^2 -
   \cosh^2  2 \beta  d\tau_l^2 + d\beta^2   + d\phi^2
   }}
   These coordinates make manifest the symmetries of the two Wilson line configuration. We see that the
   $\sigma$ direction is fibered over an $AdS_2$ space which realizes the $SL(2,R)$ symmetry. This metric
   differs from the flat  metric of $R^{1,3}$ by an overall conformal factor, which is not important if we
   are dealing with a conformal field theory.
    We can consider two Wilson lines at $Z_2 = Z_3 =0$ and at
   $Z_0/Z_{-1} = Z_1/Z_{4}$. The $SL(2,R)$ that preserves two of the lines acts by multiplication on
   the left in \writeal .
   These Wilson lines lie at the boundary of the $AdS_3$ space and produce a flux
   of color electric field along the $\tau_l , \sigma $ directions, $F_{\tau_l \sigma } =$constant.
   This is a flux tube whose energy is localized in the transverse non-compact $\beta $ direction.
   This is discussed in more detail in \AMtwo .
   Excitations of this flux vacuum  constitute the states that arise in the OPE expansion.
   In any theory where the flux
   cannot ``break'', these states are well defined. For example, in ${\cal N}=4$ super Yang Mills this
   flux is well defined at finite $N$
   for any value of the coupling if the external lines are in the fundamental, so that
   the flux cannot be screened. In theories with fundamentals, these states might be not defined away from
   the planar limit.

   Let us now discuss the analytic continuation $\tau_l \to i \tau$. We also need to set
   $\beta \to i \tilde \beta$. Then the coordinates \writeal\ continue to describe an $AdS_3$ space
   written as an $SU(1,1)$ group element where we have analytically continued $Y_{4} \to i  Y_{0}$ and
   $Y_{0} \to i Y_{4} $. This now maps the Wilson lines as follows.  After the analytic continuation
   the parametrization \writeal\ becomes a parametrization as an $SU(1,1)$ matrix. We can perform a
    simple relabeling of the matrices $\sigma_2 \to \sigma_3$ and $\sigma_3 \to - \sigma_2$ which
    corresponds to the transformation which takes the $SU(1,1)$
    group element back to the usual presentation
    in terms of $SL(2,R)$ matrices.
     After this transformation we end up with a parametrization of
    the form
    \eqn\newpar{
     \pmatrix{ \tilde Y_{-1} +\tilde Y_4 &\tilde Y_0 + \tilde Y_1
   \cr
   -\tilde Y_0 +\tilde Y_1 & \tilde Y_{-1} - \tilde Y_4 } =
    e^{\tau  \sigma_3 } e^{ i  \tilde \beta \sigma_2 } e^{ \sigma \sigma_1 }
   }
   We now see that the transformation $\tau \to \tau + $constant corresponds to a combination of
   an ordinary dilatation and a boost if we go back to the usual Poincare coordinates via \oord .
   This is such that it leaves $x^+$ fixed and it changes $x^-$. We then get a square as in \Square (c).
   The two original Wilson lines are the ones acted on by these shifts of $\tau $ and are the ones that
   are extended along $x^-$. The other two sides of the square have arisen after the analytic continuation
   and they are the lines on which are are inserting the operators creating the flux vacuum together with
   its excitations. If we only have the flux vacuum, these extra lines do not have any further operators
   inserted. See also \cylinder (b).

   \subsec{Analytic continuation at strong coupling}

\ifig\GKPlor{Plot of the  the spinning string solution for small spin and for large spin.
 } {\epsfxsize2.5in\epsfbox{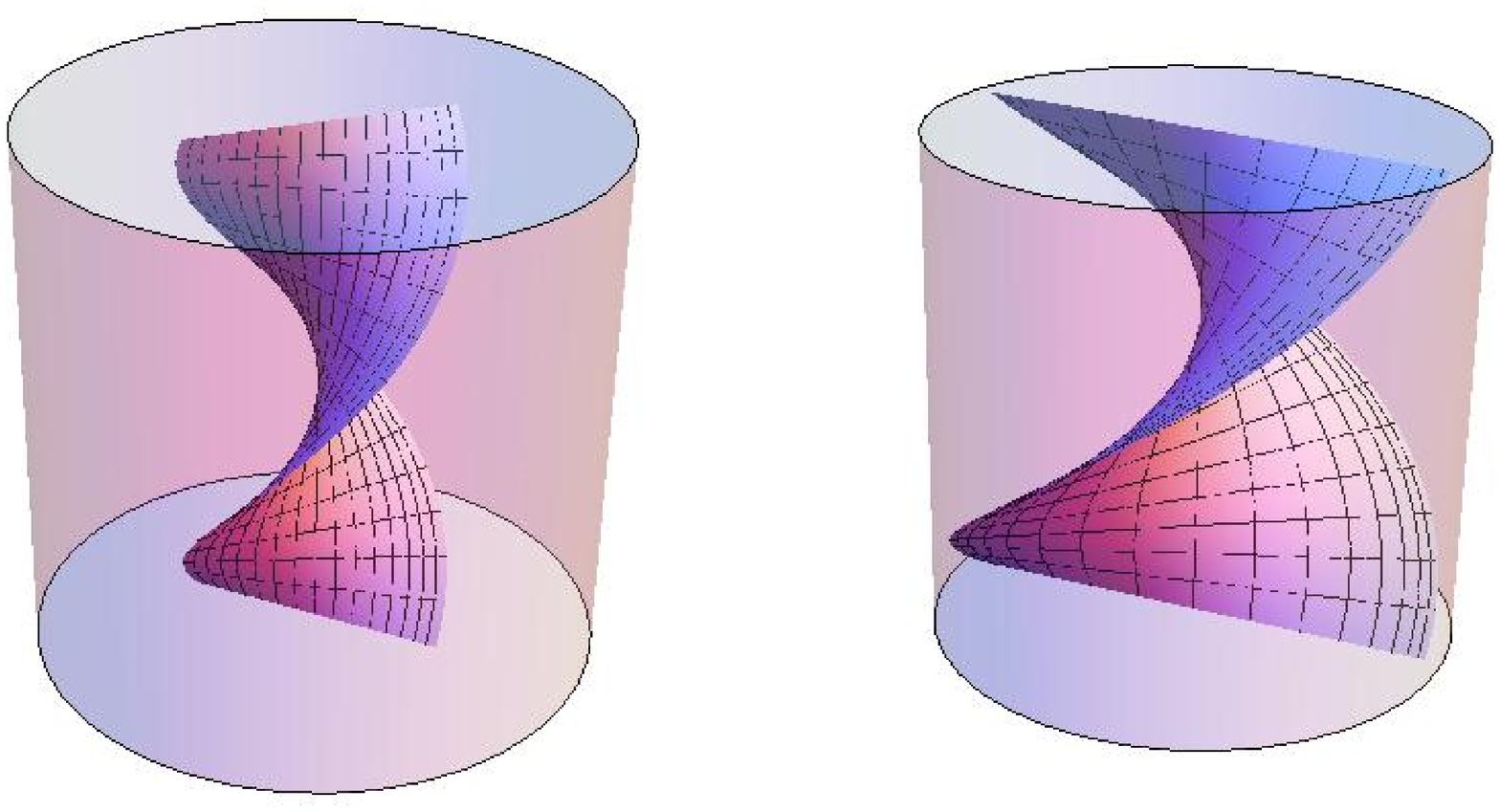}}

\ifig\GKPeucfinite{Plot of the Euclidean continuation of the spinning string solution
for small spin. Side view and top view.} {\epsfxsize2.0in\epsfbox{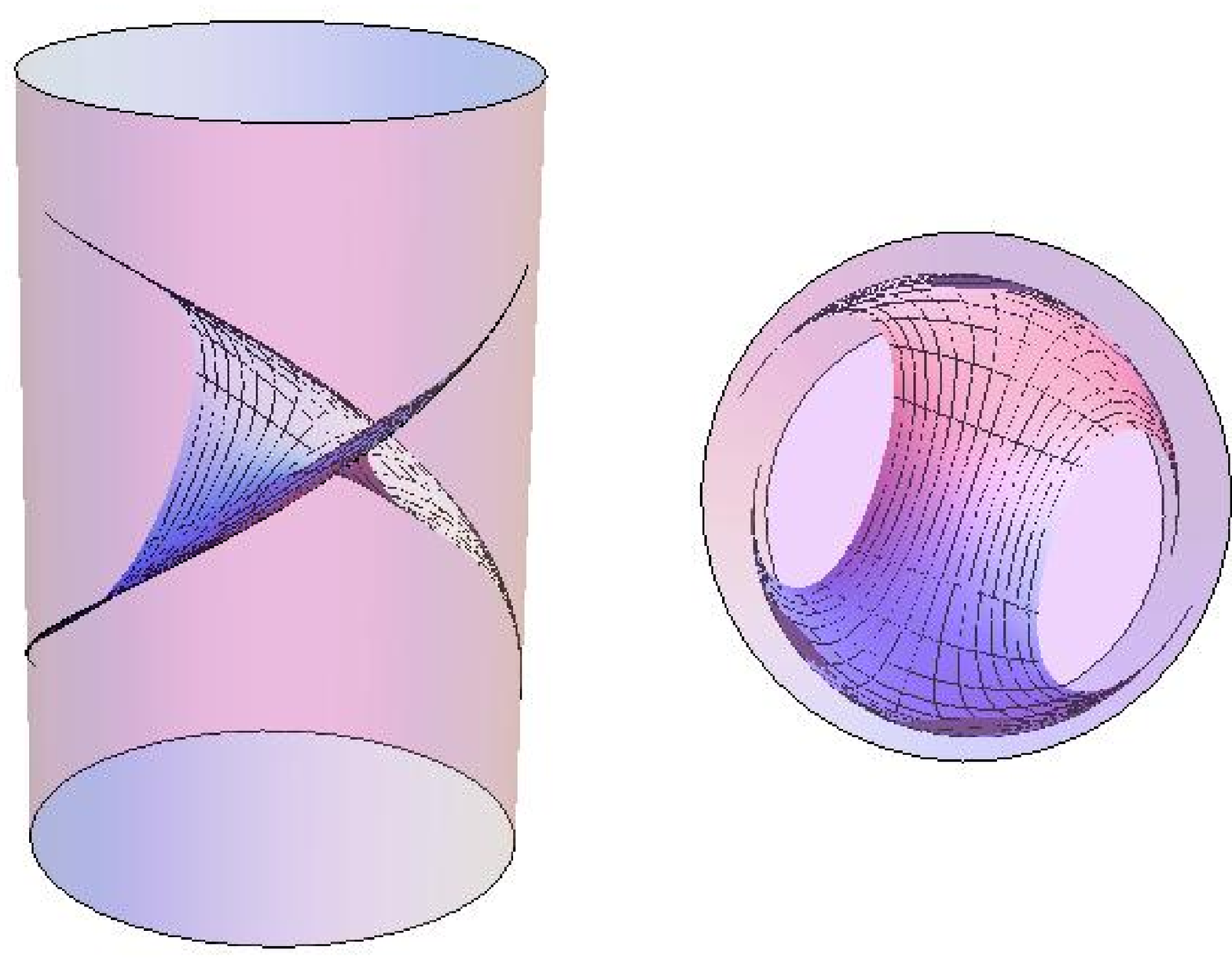}  }

\ifig\GKPeuclarges{Plot of the Euclidean continuation of the spinning string solution
 for very large spin. The edges of the string in the bulk approach the boundary, leading to the two
 other boundaries of the Wilson loop. } {\epsfxsize1.0in\epsfbox{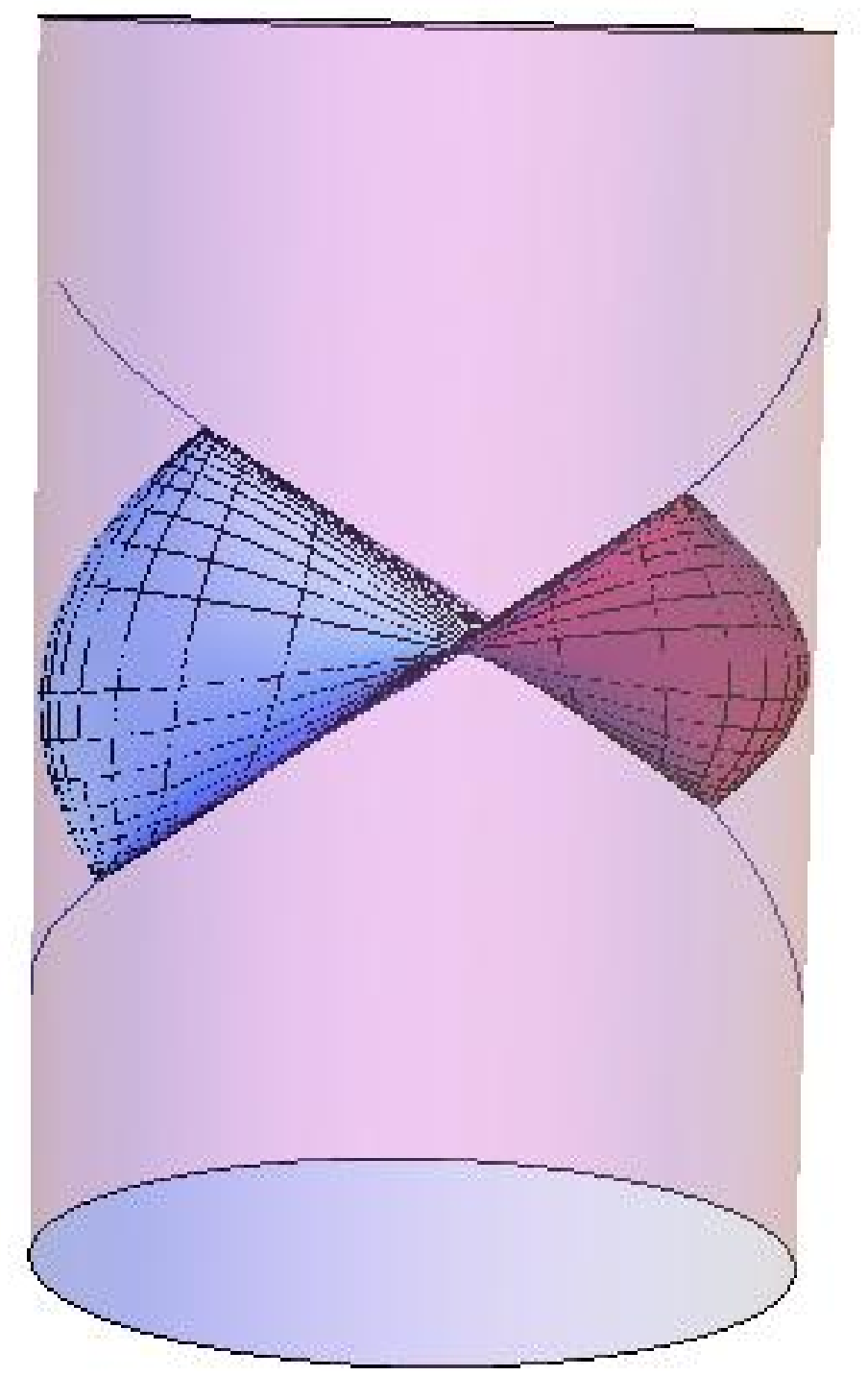} }

   At strong coupling we can consider the string solution that is dual to the high spin
   operators considered in \GKP . In this appendix we will show how the high spin limit, plus an
   analytic continuation, produces the
   square Wilson loop. A closely related discussion can be found in \Kruczenski . The idea that the large
   spin limit can be described in terms of Wilson loops can found in  \KorchemskyMar .
   It is convenient to focus just on an $R^{1,1}$ subspace of the boundary theory and an $AdS_3$ subspace
   of the bulk, with metric
  $
   ds^2 = - \cosh^2 \rho dt^2 + \sinh^2 \rho d\varphi^2 + d\rho^2
 $
    This should not be confused with the $AdS_3$ subspace discussed above which was purely on
   the boundary.  For finite spin we have the solution \GKP\
   \eqn\solut{
    \varphi = w t  ~,~~~~~~~~~   \rho \in [ 0, \rho_{max}] ~,  ~~~~  \tanh \rho_{max} = 1/w
   }
   where $w$ is very large for small spin and $ w \to 1$ when the spin goes to infinity.
   This is the Lorentzian time solution. We can now analytically continue $t\to i \tau$ and
   $\varphi \to i \chi$.
   We then write
   \eqn\tiel{
   \tilde Y_{-1} \pm  \tilde Y_{4}  = e^{\pm \tau} \cosh \rho ~,~~~~~~~~ \tilde Y_{0} \pm \tilde Y_{1}
   = e^{\pm w \tau } \sinh \rho  ~,~~~~~~~~ \rho \in [ 0, \rho_{max}]
   }
   This is plotted in \GKPeucfinite .
   We see that as we approach the boundary and $\tau \to \infty$ the solution approaches two null lines.
   The worldsheet tips, sitting at $\rho = \rho_{max}$,
   move through the bulk and approach the boundary along these null lines.
   As the spin goes to infinity $w\to 1$ and $\rho_{max } \to \infty$.  Then the tips of
   the string get closer to the boundary. In the limit, the tip is joining a point that is a quarter way
   from the tip of the null line in \GKPeucfinite\ to a light-like separated point on the other line.

   At weak coupling we expect a similar picture. In fact, a closely related picture was discussed in
   \KorchemskyMar . The insertion of an operator $Tr[ \Phi \partial_+^S \Phi] $ produces a displacement
   of the field insertions along the $x^+$ direction together with an adjoint Wilson line.\foot{Alternatively, a null Wilson line between two $\Phi$ insertions can be expanded in derivatives. Then, correlators of these Wilson operators are dominated by large spin.}
   As the spin gets larger we expect that the effective displacement of the two fields along the null direction
   becomes larger.
   Now suppose that we have two such operators inserted at antipodal points in the cylinder. These points are spacelike separated. However,
   as we increase the spin, each of the points is splitting into two and they become displaced along the
   null direction. As this displacement grows one of the $\Phi$ insertions of one operator can become light-like
   separated with the $\Phi$ insertion of the other. This happens as the spin goes to infinity.
 Operators that diagonalize the dilation do not have fixed separation along the $x^+$ direction, but are
 a suitably weighted superposition. At strong coupling this manifests itself as the fact that we always
 get full null lines as we approach $ \tau \to \pm \infty$ in \tiel , see \GKPeucfinite .

   This picture is not restricted to the large spin limit of twist two operators, but it would also hold
   for higher twist operators which contain other field insertions among the derivatives, as long as we consider
   the large spin limit and look at the lowest excitations. See \ConformalQCD\ for further discussion and
   further references.

 Finally, notice that by inserting $n$ high spin twist two operators at spacelike separated points and
 by taking their large spin limit we expect to reproduce a null polygonal Wilson loop with $ 2n $ sides.
 In particular, the three point function of high spin operators of the type $Tr[ \Phi \partial^S \Phi]$ should
 produce the hexagonal Wilson loop, when each of the spins goes to infinity. The cross ratios of the
 Wilson loop should come from the orientation of the spins and their ratios as they go to infinity.
 It should be interesting to see if this produces a simpler way to compute Wilson loop expectation values.

  \appendix{C}{ Remarks on the dispersion relation }

    In this appendix we would like to make a couple of remarks on the dispersion relation for excitations
    around the flux vacuum described in the above appendix.
    Let us discuss it first at weak coupling. In the free theory, the particles do not feel the flux and
    the spectrum is the same as in the absence of flux. It turns out that the energies are quantized and
    independent of the momentum, due to the $SL(2,R)\times SL(2,R)$ representation theory
    of the problem \AMtwo . At first order in the coupling we get a correction.
    The states in question can be viewed as excitations around an infinite ``sea of derivatives''. In other
    words, high spin operators have the rough form $Tr[ \Phi \partial^S_+ \Phi]$.  In the large $S$ limit, the two field
    insertions $\Phi$ become displaced by the derivatives giving rise to the Wilson lines. Thus, we are left with the sea
    of derivatives. We can then insert other fields among this sea of derivatives. In the free theory
    the total twist, which we are calling ``energy'', is just the twist of the extra fields, which could
    be 1,2,3, etc. At first order in the coupling we see that the degeneracy is broken and we get a
    non-trivial dispersion relation. A similar phenomenon occurs around other large charge ``vacuua'' such
    as the BMN vacuum \BMN .  At one loop the twist one fields  give rise to well defined excitations
    with a dispersion relation $\epsilon(p) = 1 + \lambda \gamma(p)$. For ${\cal N}=4$ SYM these
    excitations are two gauge field insertions $F_{+i}$ and six scalars $\Phi^I$, and eight fermionic
    excitations $\psi_{+{1 \over 2} , \alpha }$. Since these fields are well defined particles at weak
    coupling, we expect that they will survive at all values of the coupling.
    At twist two we have several possibilities, one of them is $F_{+-}$, for example. We do not know if
   it  survives as a well defined particle for all values of the coupling.
    At strong coupling we have an accidental relativistic symmetry, as we have around the BMN vacuum \BMN .
    This is only present at strong coupling and is broken as we go away from the strong coupling limit.
     We have \FrolovAV\  particles with mass $m = \sqrt{2}$ that corresponds to the $F_{+i}$ particles. We also have
    eight massive fermions of mass $m =1$. We have five massless scalars from the five sphere.  Due to strong
    IR effects these give rise to six massive particles of very tiny mass $m \propto  e^{- { 1\over 4 }  \sqrt{\lambda} } $ \AMtwo.
    Finally, there is a particle of mass $m=2$.

    It turns out that some of these particles can be interpreted as ``massive goldstone'' particles.

   Notice that after the analytic continuation in \newpar\ the coordinates \explsym\ become
   \eqn\newcoord{
   ds^2 = d \tau^2 + d\sigma^2 + 2 \sin \tilde 2 \beta d\tau d\sigma - d\tilde \beta^2 + d\phi^2
   }
   We see that we have a symmetry under $\tau \leftrightarrow \sigma$ and $\tilde \beta \to - \tilde
   \beta$.
   This leaves the flux essentially invariant.
   This implies that the dispersion relation is expected to have the following symmetry.
   If we write the dispersion relation as $f(\epsilon, p, \lambda) =0$, then if $\epsilon$ and
   $p$ are a solution, then so is $ \epsilon' = ip ~,p' = i \epsilon$. Thus we have a Wick rotation symmetry.

    \subsec{Massive Goldstone particles }

    The flux vacuum that we are considering in this paper is
     breaking
    some symmetries spontaneously.
    When we have spontaneously broken symmetries  we can act with
    the broken symmetry generators and generate some particular excitations. These excitations have
    particular energies and momenta which are fixed by the commutation relations between the broken symmetry
    generator and the energy and momentum generators.
    Let us see what this implies for our case \AMtwo .
     For example, the flux breaks supersymmetry.
    Under the $SL(2)_L \times SL(2)_R \subset SO(2,4)$ generators the
     supercharges transform as $({1 \over 2 },0)+(0,{1 \over 2 })$.
    Thus they have either $\epsilon = \pm 1$ and $p=0$ or $\epsilon =0, ~p=\pm i $ (recall that the energy
    is a generator inside $SL(2)_L$ and the momentum is a generator inside $SL(2)_R$).
    Thus the dispersion relation of the fermions should obey
    \eqn\disperl{
    \epsilon(\lambda , p=0) =1 ~~~~~~,~~~~~~~~\epsilon(\lambda , p=\pm i ) = 0 ~,~~~~~~~~{\rm fermions}
    }
    This should be true for all values of the coupling. E.g., this argument determines the mass of the
    fermions at strong coupling.

    For the bosons associated to $F_{+i}$ there is a similar prediction. In this case the corresponding
    generators have spin $({1\over 2 } ,{ 1 \over 2 } )$ under the $SL(2) \times SL(2)$ of $AdS_3$. Thus, in this case the
    Goldstone bosons have $\epsilon = \pm 1 $, $p = \pm i $. Thus, in this case we have the constraint
     \eqn\disperl{
    \epsilon(\lambda , p=\pm i ) = 1 ~,~~~~~~~~~~~~F_{+i}~~~~ {\rm bosons }
    }
    One is tempted to make the same argument for the broken generator involving the broken $SL(2,R)$ generators.
    In this case one would be lead to a particle with energy $\epsilon =2,~ p=0$. This is indeed present
    at strong coupling, it is the $m=2$ particle. On the other hand, it might be that this particle
    decays into pairs of fermions, which would mean that the symmetry generator would create pairs of
    fermions rather than a single boson. Notice that the prediction \disperl\ is quite general. It should hold in any CFT in dimensions $d\geq 3$. Superficially this relation seems to be contradicted by \gammaexp; we believe
     that the limit $p\to i$ does not commute with perturbation theory. A closer analysis of the finite coupling dispersion of \Basso\ should hopefully confirm this.

    This non-relativistic version of the  Goldstone theorem constrains the dispersion relation only at
    particular values of the momentum. In these cases the broken symmetries do not commute with the Hamiltonian
    and thus they give rise to excitations with finite energies.

    Notice that the excitations around the flux vacuum have a gap, both at weak and strong coupling, though
    the gap becomes non-perturbatively small at strong coupling. In particular,   the anomalous
    dimensions of the $\Phi^I$ insertions at zero momentum is negative, as naively expected from the fact that they become
    very small at strong coupling \refs{\KorchemskySpin,\Basso}.

\appendix{D}{On the expansion for Wilson loops in the $U(1)$ theory}

In this appendix we aim to give some more details about the
computation of $r_{U(1)}$ for a general polygon.
It is convenient to rewrite the contour integrals in terms of
projective coordinates $Z$
as
\eqn\rbdsZpath{
\log r_{U(1)} =  \oint\limits_{{}^{W_{square}-}_{\,\,\,\, W_{top}}} \!\!\!\! ds
\oint\limits_{{}^{W_{square}-}_{\,\, W_{bottom}}}\!\!\!\! ds' \,\,\frac{\dot Z(s) \cdot \dot
Z'(s')}{Z(s) \cdot Z'(s')}}
Notice that a position-dependent real rescaling $Z(s) \to Z(s)
\lambda(s)$ changes the integrand up by a total derivative only, which
integrates to zero. Be $A$,$D$ the top vertices of the square, $B$,$C$
the bottom vertices. $W_{square}-W_{top}$ consists of a path
$\gamma$ going from $A$ to $D$ minus the straight segment $AD$, see \Familytopbot (b).
$W_{square}-W_{bottom}$ consists of a path
$\gamma'$ going from $B$ to $C$ minus the straight segment $BC$. We can write
$\gamma(s) = a(s) A + b(s) B + c(s) C + d(s) D + n(s)$, where
$a,b,c,d$ are scalar functions and $ n$ is a vector in the plane
orthogonal to the
square. We can also write $\gamma'(s') = a'(s') A + b'(s') B + c'(s')
C + d'(s') D + n'(s')$.
It is useful to parameterize the segment $AD$ as $Z_0(s) = a(s) A +
d(s) D$ and write $\gamma(s) = Z_0(s) + \delta(s)$.
Without loss of generality, we can pick $a(s) = e^{- s}$ and $d(s) =
e^s$. We will also parameterize the segment
$BC$ as $Z'_0(s') = b'(s') B + c'(s') C$ and write $\gamma'(s') =
Z'_0(s') + \delta'(s')$.
Without loss of generality, we can pick $b'(s') = e^{- s'}$ and
$c'(s') = e^{s'}$. We can also normalize $A\cdot C=B \cdot D=1/2$.

Notice that $\delta \cdot Z' = \delta' \cdot Z=0$ and their
derivatives are also orthogonal. After some simple algebra
\eqn\rbdsZpath{
\log r_{U(1)} =  \int_{-\infty}^\infty ds \int_{-\infty}^\infty ds'
\frac{\delta(s) \cdot \delta'(s')+ \dot \delta(s) \cdot \dot
\delta'(s')}{\cosh (s-s') + \delta(s) \cdot \delta'(s')}}

The action of $M$ on $\gamma'(s')$ can be combined with a
trivial rescaling and shift of the integration variable as
$e^{-\tau} M\gamma'(s'-\sigma) = e^{-2\tau-\sigma}
a'(s'-\sigma) A + e^{- s'} B + e^{s'}C +  e^{-2\tau+\sigma}
d'(s'-\sigma) D + e^{-\tau}  n'_\phi(s'-\sigma)$. We denoted
the action of the rotation in the transverse plane on $ n'$ as
$n'_\phi$. Hence in order to write the OPE we want to redefine
$\delta'(s'-\sigma,\tau,\sigma,\phi) = e^{-2\tau-\sigma}
a'(s'-\sigma) A +  e^{-2\tau+\sigma} d'(s'-\sigma) D +
e^{-\tau} n'_\phi(s'-\sigma)$, and \eqn\rbdsZpath{ \log
r_{U(1)}=  \int_{-\infty}^\infty ds \int_{-\infty}^\infty ds'
\frac{\delta(s) \cdot \delta'(s'-\sigma)+ \dot \delta(s) \cdot
\dot \delta'(s'-\sigma)}{\cosh (s-s') + \delta(s) \cdot
\delta'(s'-\sigma)}} Now we can expand the denominator and
Fourier transform term by term. The leading piece is
\eqn\rbdsZpath{\eqalign{ \log r_{U(1)} \sim & e^{- \tau}
\int_{-\infty}^\infty ds \int_{-\infty}^\infty ds'  \frac{n(s)
\cdot n'_\phi(s'-\sigma)+ \dot n(s) \cdot \dot
n'_\phi(s'-\sigma)}{\cosh (s-s')} \sim \cr \sim & \int dp e^{i
p \sigma-\tau} \tilde n(p) \cdot \tilde
n'_\phi(p)\frac{p^2+1}{\cosh { \pi p \over 2 } } }} For the
hexagon we find that $n(s) \sim e^{-|s| } n_0$ where $n_0$ is a
constant vector. We have a similar expression for $n'$. Its
Fourier transform is then proportional to $ (1 + p^2)^{-1}$. We
get a similar factor from $n'$. Together these to factors give
the function in \fourtrmain .

   \appendix{E}{The modified TBA and Yang-Yang functional for $AdS_3$ null polygons}

In this appendix we present  a new version for the integral equations that determine the result at strong
coupling in the case that the polygon can be embedded in $R^{1,1}$ and the surface in $AdS_3$.
This modified integral equations, or TBA equations,  involve only the physical spacetime cross ratios, in
contrast with the ones in \AMSV\ which involved some other auxiliary parameters.
In addition, we will find that the expression for the area can be written
as the critical value of the associated  Yang-Yang functional \refs{\AlexandrovPP,\NekrasovRC}.
The computation of the regularized area (or better, the remainder function) for the minimal surface
 in $AdS_3$ ending on
a given null polygon on the boundary is done in \refs{\octagon,\AMSV}
through three basic steps. The first step is to promote
the cross-ratios of the null polygon to holomorphic functions of a spectral parameter $\zeta$, which capture
the higher conserved charges of the classical integrable system. The second step is to
derive a set of functional equations, which are then converted into convenient integral equations.
 The third step is to
compute the regularized area from the higher conserved charges or $\zeta$ dependent cross ratios.

The specific problem of minimal surfaces ending
on null polygons in $AdS_3$ is a special case of a general theory of Hitchin systems on a Riemann surface.
The results of this appendix apply to any such system, hence we will try to keep our integral equations
as general as possible. Fortunately, the integral equations given for general Hitchin systems
\refs{\GMNone,\GMNtwo} and the integral equations which are optimized for the case of polygons in $AdS_3$
\AMSV\
actually coincide in the simplest kinematic region.
There is a large amount of freedom in setting up the Riemann Hilbert problem,
which leads to a variety of different forms for the integral equations, each with its own
advantages and disadvantages.
The purpose of this appendix is to present the integral equations in a way which
is well suited for the analysis of soft limits, and make manifest the
cancellation of spurious terms between the various contributions to the regularized area.

The TBA-like integral equations for general Hitchin systems can be written as
\eqn\TBA{\ln X_\gamma(\zeta) = \frac{Z_\gamma}{\zeta} + i \theta_\gamma + \bar Z_\gamma \zeta
- \frac{1}{4\pi i} \sum_{\gamma' \in \Gamma} \Omega(\gamma') \langle \gamma,\gamma' \rangle \int_{\ell_{\gamma'}} \frac{d\zeta'}{\zeta'} \frac{\zeta' + \zeta}{\zeta' - \zeta} \log \left( 1+X_{\gamma'}(\zeta') \right)
}

The equations depend on some discrete data: the set $\Gamma$ of possible labels  $\gamma,\gamma'$,
certain integer numbers $\Omega(\gamma)$, an antisymmetric pairing $\langle \gamma,\gamma' \rangle$.
This data is the final product of a careful WKB analysis of some differential equations on the surface,
and for the purpose of this appendix we only need to use some basic facts about it.
The $Z_\gamma$ are auxiliary complex numbers which we wish to eliminate from the equations.
The $\theta_\gamma$ are angles which will presently be set to zero (or $\pi$, in which case they can be reabsorbed by
changing the sign of some $X_\gamma$ in the equations): in the language of Hitchin systems, we are restricting to a real section.
The lines of integration $\ell_\gamma$ are straight rays from $\zeta'=0$ to $\zeta'=\infty$, which for convergence reasons should
lie in the half plane where $Z_\gamma/\zeta'$ has a negative real part. A canonical choice is to set them to $Z_\gamma/\zeta'$ real and negative.
The relative ordering of the lines is important, in the sense that
if two lines are moved across each other, the integration contours will cross poles of the integration kernels. The result of such a ``wallcrossing''
is captured by a certain specific change in the discrete data.

Once the angles $\theta_\gamma$ are set to zero, a simplification occurs: $X_{-\gamma}(-\zeta) =X_\gamma(\zeta)$. This is due to a certain symmetry of the
equations: labels in $\Gamma$ come in pairs, which we can denote as $\gamma$ and $-\gamma$,
such that  $\Omega(\gamma) = \Omega(-\gamma)$, $\langle-\gamma, \gamma'\rangle = \langle\gamma, -\gamma'\rangle = -\langle\gamma, \gamma'\rangle$,
$Z_{-\gamma} = -Z_\gamma$.

We can combine the contributions from $\gamma'$ and $-\gamma'$, and have the sum
 run only half of the set of $\gamma$'s, say $\Gamma^+$.
From now all sums will be over this subset.
We will also write $\zeta = e^\theta$.
\eqn\TBAreal{\ln X_\gamma(\theta) = Z_\gamma e^{- \theta} + \bar Z_\gamma e^\theta
+ \frac{1}{2\pi i} \sum_{\gamma' \in \Gamma^+}  \Omega(\gamma') \langle \gamma,\gamma' \rangle \int_{\ell_{\gamma'}} \frac{d\theta'}{\sinh (\theta' - \theta)} \log \left( 1+X_{\gamma'}(\theta') \right)
}

In order to make contact with the equations of \AMSV\
for $2N$ gluons we can take the labels $\gamma=s$ to run over integers $1\cdots N-3$,
with $\Omega(s)=1$ and $\langle s, s+1 \rangle =-1$. We also need to set $Z_s = -\frac{i^s}{2} m_s$, and use the canonical choice of lines.
The functions $X_s$ and $Y_s$ agree on the integration lines, up to an appropriate shift of $\theta$ by $\arg Z_s$.\foot{The reader should be cautious
in extending this relation away from integration lines: the $Y_s$ functions are usually defined as analytic continuations
from the integration lines, while the $X_s$ are defined by the integral equations for all $\theta$ and therefore have discontinuities.}

We want to rewrite the equations in terms of $x_\gamma^+$ and $x_\gamma^-$, the values of
 the cross-ratios $X_\gamma$ at the physical values $\zeta=1$
and $\zeta=i$. This is easy: we set $\zeta =1$ and $\zeta =i$ in \TBAreal , solve for $Z_\gamma$ and $\bar Z_\gamma$
and then insert the result back into \TBAreal . We find
\eqn\TBAadsthree{\eqalign{
\ln X_\gamma(\theta) = &  \cosh \theta \ln x^+_\gamma - i \sinh \theta \ln x^-_\gamma + \cr
&+
 \frac{1}{2\pi i} \sum_{\gamma' \in \Gamma^+}  \Omega(\gamma') \langle \gamma,\gamma' \rangle \int_{\ell_{\gamma'}} \frac{d\theta' \sinh 2 \theta}{\sinh (\theta' - \theta)\sinh 2 \theta'} \log \left( 1+X_{\gamma'}(\theta') \right)
}}
Notice that the auxiliary parameters $Z_\gamma$ have
disappeared from the equations and we only have the
spacetime cross ratios $x_\gamma^\pm$.
Notice that the kernel of these modified TBA equations is not symmetric in $\theta$ and $\theta'$. It becomes symmetric if we
pick a different choice of rapidity variable, $u =  \frac{\cosh 2 \theta}{\sinh 2 \theta}$, so that $-2 d\theta' = du' \sinh^2 2 \theta'$.
The usefulness of this alternative rapidity variable will become evident momentarily. General TBA equations with a symmetric kernel
\eqn\genTBA{\ln X_a(u)= L_a(u)
 + \frac{1}{2\pi} \sum_{b}  \int_{\ell_b} du' K_{ab}(u,u') \log \left( 1+X_b(u')\right)
}
can be recast as the conditions for a Yang-Yang functional to be extremized
\eqn\YY{\eqalign{
YY= & \frac{1}{2 \pi} \sum_a  \int_{\ell_a} du
\left( \rho_a(u) \phi_a(u)-Li_2(-e^{L_a(u)- \phi_a(u)})\right)+
\cr
&+  \frac{1}{8 \pi^2} \sum_{a,b} \int_{\ell_a} du \int_{\ell_b} du' K_{ab}(u,u')\rho_a(u) \rho_b(u')
}}
Indeed the variation with respect to $\rho$ sets
\eqn\YYrho{ \phi_a(u)+ \frac{1}{2 \pi} \sum_{b} \int_{\ell_b} du' K_{ab}(u,u') \rho_b(u')=0 }
while the variation with respect to $\phi_a$ sets
\eqn\YYphi{\rho_a(u) =\log \left( 1+e^{L_a(u)- \phi_a(u)}\right) }

If we set $\log X_a = L_a - \phi_a$ we recover the TBA equations \genTBA .
 We would like to show now that the interesting
part of the regularized area (or better, reminder function) of the minimal surface in $AdS_3$ coincides with the extremum
of the Yang-Yang functional for the modified TBA equations. In a sense, this result makes manifest an important property of the area:
it should be the extremum of an action functional with fixed boundary conditions given by the choice of physical cross-ratios.

For convenience, let's take $N$ to be odd, where the polygon has $2N$ sides.
 For even $N$ there are some slight complications, which
pop out in various places in the calculation, only to cancel out at the very end in the remainder function.
For that reason, it is simpler to just do formal computations for odd $N$, and
then recover the even $N$ results by a soft collinear limit. If $N$ is odd, there is an antisymmetric matrix $w_{\gamma, \gamma'}$
which roughly speaking inverts $\langle \gamma,\gamma' \rangle$.\foot{In some cases the set of $\gamma$'s is an
over complete basis, that is why we cannot find a proper inverse. In the kinematic region considered in
\AMSV\ one can find a proper inverse.  } More precisely,
it satisfies $\sum_{\gamma, \gamma'} Z_\gamma w_{\gamma, \gamma'} \langle \gamma',\gamma''
\rangle = Z_{\gamma''}$ and
$\sum_{\gamma', \gamma''} \langle \gamma,\gamma''\rangle w_{\gamma', \gamma''}
 \langle \gamma'',\gamma''' \rangle = \langle \gamma,\gamma''' \rangle$.

The higher conserved charges of the classical integrable system are hidden in the large positive $\theta$ asymptotic expansion of
$\log X_\gamma \sim \sum_{n=-1}^\infty c_{n,\gamma} e^{-n \theta}$. An alternative set of charges appear at large negative $\theta$:
$\log X_\gamma \sim \sum_{n=-1}^\infty  \tilde c_{n,\gamma} e^{n \theta}$.
The contribution to the regularized area denoted as $A_{periods} + A_{free}$ in \AMSV \foot{ Recall that
$A$ denotes different pieces of the area. The Wilson loop expectation values are then obtained by
$ \langle W \rangle \sim  e^{ - { R^2_{AdS} \over 2 \pi \alpha' }({\rm Area}) } \sim  e^{ - { \sqrt{\lambda}
 \over 2 \pi }({\rm Area}) }$.}
can be computed from the conserved charges as
$i \sum_{\gamma, \gamma'} w_{\gamma, \gamma'} c_{-1,\gamma} c_{1,\gamma'}$. An alternative expression
is  $-i \sum_{\gamma, \gamma'} w_{\gamma, \gamma'} \tilde c_{-1,\gamma} \tilde c_{1,\gamma'}$.
The cleanest formulae
usually arise by averaging the two expressions.

From the TBA equations \TBAadsthree\ we can compute
\eqn\conetsws{c_{-1,\gamma}=  \half (\ln x^+_\gamma - i  \ln x^-_\gamma)
 - \frac{1}{2\pi i} \sum_{\gamma' \in \Gamma^+}  \Omega(\gamma') \langle \gamma,\gamma' \rangle \int_{\ell_{\gamma'}} \frac{d\theta' e^{\theta'}}{\sinh 2 \theta'} \log \left( 1+X_{\gamma'}(\theta') \right)
}
\eqn\conetsws{c_{1,\gamma}=  \half (\ln x^+_\gamma + i  \ln x^-_\gamma)
 - \frac{1}{2\pi i} \sum_{\gamma' \in \Gamma^+}  \Omega(\gamma') \langle \gamma,\gamma' \rangle \int_{\ell_{\gamma'}} \frac{d\theta' e^{3\theta'}}{\sinh 2 \theta'} \log \left( 1+X_{\gamma'}(\theta') \right)
}
\eqn\conetsws{\tilde c_{-1,\gamma}=  \half (\ln x^+_\gamma + i  \ln x^-_\gamma)
 - \frac{1}{2\pi i} \sum_{\gamma' \in \Gamma^+}  \Omega(\gamma') \langle \gamma,\gamma' \rangle \int_{\ell_{\gamma'}} \frac{d\theta' e^{-\theta'}}{\sinh 2 \theta'} \log \left( 1+X_{\gamma'}(\theta') \right)
}
\eqn\conetsws{\tilde c_{1,\gamma}=  \half (\ln x^+_\gamma - i  \ln x^-_\gamma)
 - \frac{1}{2\pi i} \sum_{\gamma' \in \Gamma^+}  \Omega(\gamma') \langle \gamma,\gamma' \rangle \int_{\ell_{\gamma'}} \frac{d\theta' e^{-3\theta'}}{\sinh 2 \theta'} \log \left( 1+X_{\gamma'}(\theta') \right)
}

The piece of $A_{periods} + A_{free}$ with no integrals is
\eqn\azero{
A_0 = - \half \sum_{\gamma, \gamma'} w_{\gamma, \gamma'} \ln x^+_\gamma \ln x^-_{\gamma'}
}
The piece with one integral, averaged, is
\eqn\atemp{
A_{temp}= - \frac{1}{2\pi } \sum_{\gamma \in \Gamma^+}  \Omega(\gamma) \int_{\ell_\gamma} \frac{d\theta \cosh 2 \theta}{\sinh 2 \theta} \left( \sinh \theta \ln x^+_\gamma - i \cosh \theta \ln x^-_\gamma \right)\log \left( 1+X_{\gamma}(\theta) \right)
}
Notice that the term in parenthesis is
$\partial_\theta \left( \cosh \theta \ln x^+_\gamma - i \sinh \theta \ln x^-_\gamma \right)$.
We can use the modified TBA \TBAadsthree\ to trade it for
$\partial_\theta \ln X_\gamma(\theta)$ up to terms which we will combine with the other two-integral pieces of $A_{periods} + A_{free}$.
Then we can write $\partial_\theta \ln X_\gamma(\theta)\log \left( 1+X_{\gamma'}(\theta') \right) = \partial_\theta \left( - Li_2(-X_\gamma) \right)$ and integrate by parts to
\eqn\aone{
A_1 = \frac{1}{\pi } \sum_{\gamma \in \Gamma^+}  \Omega(\gamma) \int_{\ell_\gamma} \frac{d\theta}{\sinh^2 2 \theta} Li_2(-X_\gamma)
}
The remaining term is
\eqn\aremtadf{\eqalign{
A_{temp}-A_1= &\frac{1}{4\pi^2 i }  \sum_{\gamma,\gamma' \in \Gamma^+}  \Omega(\gamma)  \Omega(\gamma')
\langle \gamma,\gamma' \rangle
%\times \cr &
\int_{\ell_{\gamma'}}\frac{d\theta'}{\sinh 2 \theta'}  \int_{\ell_\gamma}
\frac{d\theta }{\sinh 2 \theta} \times
\cr & \times \cosh 2 \theta \partial_\theta
\left( \frac{ \sinh 2 \theta}{\sinh (\theta' - \theta)}\right)
%\cr &
  \log \left( 1+X_{\gamma'}(\theta') \right) \log \left( 1+X_{\gamma}(\theta) \right)
}}

We can symmetrize the kernel to
 $\frac{1}{\sinh (\theta'-\theta)} + \sinh (\theta'-\theta) \cosh 2 (\theta + \theta')$.
  The second piece cancels out against  the remaining two integral terms
and we are left with $A_{periods} + A_{free}=A_0 + A_1 + A_2$, where
\eqn\atwo{\eqalign{
A_2= &\frac{1}{4\pi^2 i }  \sum_{\gamma,\gamma' \in \Gamma^+}  \Omega(\gamma)  \Omega(\gamma') \langle \gamma,\gamma' \rangle \int_{\ell_{\gamma'}}\frac{d\theta'}{\sinh 2 \theta'}  \int_{\ell_\gamma} \frac{d\theta }{\sinh 2 \theta} \frac{1}{\sinh (\theta'-\theta)}  \cr & \log \left( 1+X_{\gamma'}(\theta') \right) \log \left( 1+X_{\gamma}(\theta) \right)
}}
We see that $A_{periods} + A_{free}=A_0 + Y Y_{cr}$, where $YY_{cr}$ is the critical value of the $YY$ functional.
In summary, the full area of the surface can be written as $A = A_{div} + A_{BDS-like} + A_0 + YY_{cr}$,
where $YY_{cr} = A_1 + A_2$ in \aone\ and \atwo . $A_{div}$ are the divergent terms and $A_{BDS-like}$ can
be found in formula (5.10) in \octagon .
We see that $YY_{cr}$ is manifestly small in the region where the $X_\gamma$ are small, which is when
the $x^\pm_\gamma$ are small. We also expect that $A_{BDS-like} - A_{BDS} + A_0 $
 becomes small in this region. This is an expression written purely in terms of the spacetime cross ratios.

\subsec{Evaluating the OPE for general configurations in $AdS_3$}

In this subsection we expand the OPE with $AdS_3$ kinematics in the case that
we split the polygon into two general subpolygons.

If we are given any functional $F[x]$ and a small deformation $F[x] + f[x,y]$,
we can ask what is the relation between the critical values $F_{cr}$ and $(F+f)_{cr}$.
It is elementary to show that the difference of the two critical values can be computed as the critical value of $f[x_{cr},y]$,
where $x_{cr}$ extremizes $F(x)$.  We can immediately apply this to the case of the Yang-Yang functionals. We take
 $F=YY[c]$, the Yang-Yang functional with a certain label $c\in\Gamma^+$ erased, and $f$ is the remaining part (notice that kernel $K_{cc}$ is always zero for us)
\eqn\f{\eqalign{
f=& \frac{1}{2 \pi}  \int_{\ell_c} du \left( \rho_c(u) \phi_c(u)-Li_2(-e^{L_c(u)- \phi_c(u)})\right)+
\cr & + \frac{1}{4 \pi^2} \sum_{b} \int_{\ell_c} du \int_{\ell_b} du' K_{cb}(u,u')\rho_c(u) \rho_b(u') }}
The extremum of $f$ with respect to $\rho_c$ and $\phi_c$ is simply
\eqn\f{f_{cr}=YY_{cr} - YY[c]_{cr} \sim -\frac{1}{2 \pi}  \int_{\ell_c} du Li_2(-X_c(u))
 \sim \frac{1}{2 \pi}  \int_{\ell_c} du X_c(u)}
where
\eqn\genTBA{\ln X_c(u)= L_c(u)
 + \frac{1}{2\pi} \sum_{b}  \int_{\ell_b} du' K_{cb}(u,u') \log \left( 1+X_b(u')\right)
}

The soft limit OPE we consider in this paper leads exactly to this sort of decoupling limits. For simplicity,
we can consider the simple kinematic region of \AMSV. In the $AdS_3$ case,
if we decompose a $2N$-gon into a $2n+2$-gon and $2N-2n+2$-gon by applying the $\tau$ rescaling to the $x^+$ coordinates,
it is easy to see that $Z_{n-1}$ acquires a large real part, so that $x^+_{n-1}$ becomes very small, scaling as $e^{-\tau}$, while
all the other $x^+_\gamma$ and $x^-_\gamma$ are held fixed. Indeed, $YY[n-1]$ is the sum of the Yang-Yang functionals for the
$2n+2$-gon and $2N-2n+2$-gon. Then
the leading exponential behavior is controlled by
\eqn\f{f_{cr}=YY_{cr} - YY[n-1]_{cr} \sim \frac{1}{2 \pi}  \int_{\ell_{n-1}} du e^{\ln x^+_{n-1} \cosh \theta -i \ln x^-_{n-1} \sinh \theta  }  C_1(\theta) C_2(\theta) }
where \eqn\Cone{\ln C_1(\theta) = \frac{1}{2\pi} \sum_{s<n-1}  \int_{\ell_b} du' K_{cb}(u,u') \log \left( 1+X_b(u')\right)} and
\eqn\Ctwo{\ln C_2(\theta) = \frac{1}{2\pi} \sum_{s>n-1}  \int_{\ell_b} du' K_{cb}(u,u') \log \left( 1+X_b(u')\right)} are interpreted naturally as the
density of one-particle excitations on the GKP string \GKP\
 created by the   $2n+2$-gon and $2N-2n+2$-gon respectively.

\appendix{F}{The modified TBA and Yang-Yang functional for $AdS_5 $ null polygons}

In this section we generalize the results of the previous section to the most general $AdS_5$ kinematics. For convenience, we take the number of gluons $n$ to be odd. The case were $n$ is even can be recover by a soft collinear limit.
In the notations of \AMSV\  the TBA equations take the form
\eqn\tbantwo{\eqalign{
\log Y_{2,s} (\theta)&= - |m_s| \sqrt{2} \cosh (\theta- i \phi_s)- K_2 \star \alpha_s -  K_1 \star \beta_s \cr
\log Y_{1,s}  (\theta)&= - |m_s| \cosh (\theta- i \phi_s) -C_s- \half K_2 \star \beta_s - K_1 \star \alpha_s- \half K_3 \star \gamma_s \cr
\log Y_{3,s}  (\theta)&= - |m_s| \cosh (\theta- i \phi_s)+ C_s- \half K_2 \star \beta_s - K_1 \star \alpha_s+ \half K_3 \star \gamma_s}}
where $\star $ denotes convolution. The kernels are
\eqn\kernels{
K_1(\theta) =\frac{1}{2 \pi \cosh \theta} \,\,\, , \,\,\, K_2(\theta) = \frac{\sqrt{2} \cosh \theta}{\pi \cosh 2\theta} \,\,\, , \,\,\, K_3(\theta) = \frac{i}{\pi} \tanh 2\theta}
and
\eqn\abc{\eqalign{
&\alpha_{s} \equiv \log \frac{\left(1+  Y_{1,s}\right)  \left(1+ Y_{3,s}\right) }{\left(1+Y_{2,s-1}\right)\left(1+Y_{2,s+1}\right)} \,\, , \,\,
\gamma_{s} \equiv \log \frac{ \left(1+ Y_{1,s-1}\right)\left(1+Y_{3,s+1}\right)}{ \left(1+  Y_{1,s+1}\right)\left(1+ Y_{3,s-1}\right) } \,,   \cr
&\beta_{s} \equiv \log \frac{\left(1+Y_{2,s}\right)^2 }{\left(1+ Y_{1,s-1}\right)\left(1+ Y_{1,s+1}\right)\left(1+ Y_{3,s-1}\right)\left(1+ Y_{3,s+1}\right)} \,.
}}
We now have three parameters per column $s$: the magnitude of the {\it mass} $|m_s|$, its phase $\phi_s$ and the {\it chemical potential} $C_s$. As explained in the $AdS_3$ case we can eliminate them in favor of physical cross-ratios $y_{a,s}\equiv \widehat Y_{a,s}(0)$ where $a=1,2,3$. The hatted $Y$-functions are defined in the appendix D of \AMSV\ and are given by $\widehat Y_{a,s}(\theta)=Y_{a,s}(\theta)$ if $a+s$ is even and $\widehat Y_{a,s}(\theta)=Y_{a,s}(\theta-i\pi/4)$ for $a+s$ odd.
The modified TBA equations then read
\eqn\modeqs{\eqalign{
&\log\widehat Y_{2,s} -E_s= - \widetilde K_2 \diamond \hat\alpha_s -  \widetilde K_1 \diamond \hat\beta_s\,, \cr
&\log\widehat Y_{1,s} \widehat Y_{3,s}-\sqrt{2} E_s^{[(-)^{s+1}]}= - \widetilde K_2^{[2(-)^{s+1}]}\diamond \hat\beta_s -2 \widetilde K_1 \diamond \hat\alpha_s\,, \cr
&\log \widehat Y_{1,s}/\widehat Y_{3,s} -\log y_{1,s}/y_{3,s}=   -  \widetilde K_3 \diamond \hat\gamma_s \,,
}}
where $f^{[m]}(\theta,\theta')=f(\theta+im\pi/4,\theta')$, $\{\hat\alpha_s,\hat\beta_s,\hat\gamma_s\}$ are $\{\alpha_s,\beta_s,\gamma_s\}$ evaluated on the hated $Y$'s and $E_s=-i(-1)^s\left( \sqrt{2} \sinh[\theta+(-)^si \pi/4]\log y_{2,s}- \sinh [\theta]\log y_{1,s}y_{3,s}\right)$ is the new source which depends only on the physical cross ratios. The modified kernels read
\eqn\abc{\widetilde K_1=-\frac{1}{2\pi} \frac{\sinh(2\theta) }{\sinh(2\theta') \cosh (\theta-\theta')}\,\qquad\widetilde K_3 = \frac{i}{\pi} \frac{\sinh(2\theta)}{\sinh (2\theta-2\theta') \sinh (2 \theta')} }
and
\eqn\Ktwo{K_2 =i\sqrt{2} \sinh[\theta-\theta'+(-)^si \pi/4]\widetilde K_3\,.}
%\eqn\abc{\eqalign{
%&\widetilde K_3 = \frac{i}{\pi} \frac{\sinh(2\theta)}{\sinh (2\theta-2\theta') \sinh (2 \theta')} \,, \qquad\widetilde K_2 =i\sqrt{2} \sinh[\theta-\theta'+(-)^si \pi/4]\widetilde K_3 \cr&\widetilde K_1=-\frac{1}{2\pi} \frac{\sinh(2\theta) }{\sinh(2\theta') \cosh (\theta-\theta')}\,.}}
Notice that in \modeqs, $\diamond$ no longer denotes a convolution but instead an application of an integral kernel $\widetilde K(\theta, \theta')$.
Let us denote the right hand side of the three equations in \modeqs\ by $\widehat{ \cal A}_{2,s}$, $\widehat{ \cal A}_{1,s}+ \widehat{\cal A}_{3,s}$ and $\widehat{ \cal A}_{1,s}-\widehat{ \cal A}_{3,s}$ respectively.
The contribution $A_{periods}+A_{extra}$ can then be computed as explained in the previous sections.
The result turns out to be simply given by
\eqn\YangAdSFive{
A_{periods}+A_{extra}=A_0+YY_c~,}
where
\eqn\YYc{YY_c=
 \sum_{a,s} \int \frac{d\theta}{\pi \sinh^2(2\theta)}\left[ Li_2(-\widehat Y_{a,s})
 -\frac{1}{2}\log(1+\widehat Y_{a,s})\widehat {\cal A}_{a,s}\right]}
is the value of the Yang-Yang functional at the extremum. Note that the natural particle rapidities are again given by $u$.

The pice $A_0$ is very much like $A_{periods}$ and is given by
$A_0=-\frac{i}{2} {\bar v}_{a,s} \omega_{a,s;a's'} v_{a's'} $
where
\eqn\vectorv{v_{a,s} = \frac{i}{4}\left[2 \log y_{2,s}-(1+i(-1)^{s+1}) \log y_{1,s}y_{3,s}\right](-1)^{s+1}(1-(-1)^{s+1}i)^{\delta_{a,2}}\,,
}
$\bar v_{a,s}$ is the complex conjugate vector, and $\omega_{a,s;a's'}$ is the inverse of the intersection form of cycles and is given in \AMSV\ .

\subsec{Explicit details for the hexagon}

In order to illustrate how the kernels of the modified TBA and the Yang-Yang functional emerge in a concrete, relatively simple example,
we can consider the case of the hexagon in full, tedious detail.  We are interested in a limit where one of the three
cross-ratios, conventionally $u_2$, goes to zero,
while the other two remain finite, and their sum $u_1 + u_3$ approaches $1$.
At strong coupling, this conditions constrain the behavior of the conserved spin four current $P(z)$: the two zeroes of the
polynomial $P(z)$ are far from each other in a specific direction.
In the language of \AGM\ this limit corresponds to a large absolute value of the
``period'' $Z$ with phase $\varphi \sim -\pi/4$,
$\hat \varphi=\varphi+\pi/4\sim 0$ (more precisely, we will take it to be slightly positive).

Compared to the previous section, we now have only one value of $s$ and only three $Y$ functions
$Y_{a} = Y_{1,a}$.  One of the three TBA equations is trivially solved,
since the ratio $Y_1/Y_3$ is constant. The other two equations for $\log Y_2$ and $\log Y_1 Y_3$
reduce to the equations in \AGM\  (see (3.6) and (3.7) in \AGM )
   \eqn\tbaone{
  \epsilon(\theta-i \hat \varphi) =  Z e^{i \pi/4- \theta} + \bar Z e^{\theta-i \pi/4 } + \int d\theta'  K_2(\theta-\theta') \tilde L(\theta')
   +\int d\theta' K_1(\theta-\theta') L(\theta')
   }
      \eqn\tbatwo{
 \tilde \epsilon(\theta-i \hat \varphi)  = \sqrt{2} (Z e^{i \pi/4- \theta} + \bar Z e^{\theta-i\pi/4}) + \int d\theta' 2 K_1(\theta-\theta') \tilde L(\theta')
   + \int d\theta' K_2(\theta-\theta')L(\theta')
   }
We denote as $\tilde L(\theta) =\log(1+e^{- \tilde \epsilon(\theta-i \hat \varphi)})$ and
$L(\theta) =   \log (1+\mu e^{- \epsilon(\theta-i \hat \varphi)})(1+\mu^{-1} e^{- \epsilon(\theta-i \hat \varphi)})$.
We shifted the integration variables for convenience. Remember the correct integration paths: the real part of $\theta'$
runs from $-\infty$ to $\infty$, the imaginary part is fixed at $\hat \varphi$.

We can evaluate the equations at $\theta=0$ and $i \pi/4$ in order to trade $Z$ for cross-ratios.
     \eqn\bone{
  Z+ \bar Z  = \log(b_1) -\int d\theta' K_2(\theta' - i \pi/4)  \tilde L(\theta')  - \int d\theta'K_1(\theta'- i \pi/4) L(\theta')
   }
   and
    \eqn\utwo{
   \sqrt{2} (Z e^{i \pi/4} + \bar Z e^{-i\pi/4}) =  \log(1/u_2-1)  - \int d\theta' 2 K_1(\theta') \tilde L(\theta')
   - \int d\theta'  K_2(\theta') L(\theta')
   }

 We need to decompose
    \eqn\ZbZ{ Z e^{i \pi/4- \theta} + \bar Z e^{\theta-i\pi/4} = - i (Z+\bar Z)\sqrt{2} \sinh(\theta) + i  (Z e^{i \pi/4} + \bar Z e^{-i\pi/4})\sqrt{2} \sinh(\theta-i\pi/4)}
so that we are ready to substitute
   \eqn\ubis{
    Z e^{i \pi/4- \theta} + \bar Z e^{\theta-i\pi/4}  = E(\theta)
- {\sqrt{2} \over \pi}\int d\theta' \frac{\sinh(\theta+ \theta' )}{\sinh 2\theta'} \tilde L(\theta')
   - \frac{1}{\pi}\int d\theta' \frac{\cosh (\theta'+\theta)}{\cosh 2 \theta'} L(\theta')
   }
  in the TBA equations. Here we define
  \eqn\E{E(\theta) = - i \sqrt{2} \sinh \theta \log(b_1)+ i\sinh (\theta-i\pi/4)  \log(1/u_2-1). }

  We get the new kernels
      \eqn\tbathree{\eqalign{
  \epsilon(\theta-i \hat \varphi) = E(\theta) -& {\sqrt{2} \over \pi}\int d\theta' \frac{\cosh 2 \theta \sinh(\theta-\theta')}{\cosh 2(\theta-\theta')\sinh 2 \theta'} \tilde L(\theta')
  \cr -&\frac{1}{2\pi}\int d\theta' \frac{\cosh 2 \theta}{\cosh (\theta - \theta') \cosh 2 \theta'} L(\theta')
   }} and
      \eqn\tbafour{\eqalign{
 \tilde \epsilon(\theta-i \hat \varphi)  = \sqrt{2}E(\theta)  -& {1 \over \pi}\int d\theta' \frac{\sinh 2 \theta}{\cosh(\theta- \theta') \sinh 2 \theta'} \tilde L(\theta')
   \cr-&\frac{\sqrt{2}}{\pi}\int d\theta' \frac{\sinh 2 \theta \sinh (\theta-\theta')}{\cosh 2 (\theta-\theta') \cosh 2 \theta'} L(\theta')
   }}

   In order to compute $A_{periods}+A_{free}$ it is useful to go back to the original expression for the regularized area, involving the conserved quantities
   hidden in the large $\theta$ asymptotic expansion of the
   $\epsilon$ and $\tilde \epsilon $ functions.
   If we denote the coefficients of the expansion as
   \eqn\asy{\epsilon(\theta) \sim \sum_{n=-1}^\infty \epsilon_n e^{- n \theta}  \qquad  \tilde \epsilon(\theta) \sim \sum_{n=-1}^\infty \tilde \epsilon_n e^{- n \theta},} then
\eqn\Aasy{A_{periods}+A_{free} = - \frac{i}{2}\left( e^{i \pi/4} \epsilon_{-1} \tilde \epsilon_1- e^{-i \pi/4}\tilde \epsilon_{-1} \epsilon_1\right).}

We can expand in powers of $e^{-\theta }$
         \eqn\tbafive{ \eqalign{
  \epsilon(\theta-i \hat \varphi) \sim    E(\theta)
  - &{\sqrt{2} \over \pi}\int d\theta' \left( \frac{1}{4}(\frac{1}{\sinh \theta'} +\frac{1}{\cosh\theta'}) e^\theta - \frac{e^{3 \theta'}}{2\sinh 2\theta'}e^{-\theta} \right)\tilde L(\theta') \cr
   -&\frac{1}{2\pi}\int d\theta' \left( \frac{e^{\theta+\theta'}}{\cosh 2 \theta'} - \frac{e^{3 \theta'}}{\cosh 2 \theta'} e^{- \theta} \right)L(\theta')
   }}
      \eqn\tbasix{ \eqalign{
 \tilde \epsilon(\theta-i \hat \varphi)  \sim  \sqrt{2}E(\theta)
  - &{1 \over \pi}\int d\theta' \left( \frac{1}{2}(\frac{1}{\sinh \theta'} +\frac{1}{\cosh\theta'}) e^\theta - \frac{e^{3 \theta'}}{\sinh 2\theta'}e^{-\theta} \right) \tilde L(\theta') \cr
   -&\frac{\sqrt{2}}{\pi}\int d\theta' \frac{1}{2} \left( \frac{e^{\theta+\theta'}}{\cosh 2 \theta'} - \frac{e^{3 \theta'}}{\cosh 2 \theta'} e^{- \theta} \right) L(\theta')
   }}
   Notice that at the order which matters, the expansions of $\epsilon$ and $\tilde \epsilon$ are simply proportional: $\tilde \epsilon_{-1} \sim \sqrt{2} \epsilon_{-1}$ and $\tilde \epsilon_1 \sim \sqrt{2} \epsilon_1$. The proportionality implies that  $A_{periods}+A_{free} = \frac{i}{2} \epsilon_{-1}  \epsilon_1 \sqrt{2} (e^{i \pi/4}- e^{-i \pi/4}) = - \epsilon_{-1}  \epsilon_1$
   We can read off the relevant pieces of the expansion
          \eqn\tbamone{
  \epsilon_{-1}= E_{-1}
  - {1 \over \sqrt{2}\pi}\int d\theta' \frac{e^{\theta'}}{\sinh 2\theta'} \tilde L(\theta')
   -\frac{1}{2\pi}\int d\theta'  \frac{e^{\theta'}}{\cosh 2 \theta'} L(\theta')
   }
             \eqn\tbapone{
  \epsilon_{1} = E_1
  + {1 \over \sqrt{2} \pi}\int d\theta' \frac{e^{3 \theta'}}{\sinh 2\theta'}\tilde L(\theta')
   +\frac{1}{2\pi}\int d\theta' \frac{e^{3 \theta'}}{\cosh 2 \theta'} L(\theta')
   }
where \eqn\Emone{E_{-1} =
 - { i  \over \sqrt{2}}  \log(b_1)+ { i \over 2 }
  e^{-i\pi/4} \log(1/u_2-1)} and \eqn\Eone{E_1 =  {i  \over \sqrt{2}} \log(b_1)-  { i \over 2}
   e^{i\pi/4} \log(1/u_2-1)}
Actually, the most useful expression for $A_{periods}+A_{free} $ is the average of the expression
derived from the large $\theta$ asymptotics and the expression derived from the small $\theta$ asymptotics.
Setting $\epsilon \sim \epsilon'_{1} e^{-\theta} + \epsilon'_{-1} e^{\theta} + \cdots$ and a similar expression for $\tilde \epsilon$, which again equals $\sqrt{2} \epsilon$ at this expansion order,
        \eqn\tbamtwo{
  \epsilon'_{-1}\sim  E_{-1}
  - {1 \over \sqrt{2}\pi}\int d\theta' \frac{e^{- 3 \theta'}}{\sinh 2\theta' } \tilde L(\theta')
   +\frac{1}{2\pi}\int d\theta'  \frac{e^{-3\theta'}}{\cosh 2 \theta'} L(\theta')
   }
             \eqn\tbaptwo{
  \epsilon'_{1} \sim E_1
  + {1 \over \sqrt{2} \pi}\int d\theta'  \frac{e^{- \theta'}}{\sinh 2\theta'}\tilde L(\theta')
   -\frac{1}{2\pi}\int d\theta' \frac{e^{- \theta'}}{\cosh 2 \theta'} L(\theta')
   }

%%%%%%%%%%%%%%added%%%%%%%%%%%%%%%%%%%
Putting all together, we can organize
the final result in terms with two, one or zero contour integrals:
$\epsilon_1 \epsilon_{-1} = A_2 + A_1 + A_0$.
\eqn\tbaAzero{
A_0 = ( - i /\sqrt{2}  \log(b_1)+ i/2 e^{-i\pi/4} \log(1/u_2-1) )(i /\sqrt{2} \log(b_1)- i/2 e^{i\pi/4} \log(1/u_2-1) )}
simplifies to \eqn\Azerobis{
A_0 =  \frac{1}{4}  \log^2 b_1 + \frac{1}{4} \log^2 (b_3-1/b_1)
 }
In the next term, we can average the two available expressions and split again $A_1 = A_{1,1} + A_{1,2}$
% \eqn\tbaAoneone{A_{1,1} =
%{1 \over \sqrt{2} \pi}\int d\theta'  \left( ( - i /\sqrt{2}  \log(b_1)+ i/2 e^{-i\pi/4} \log(1/u_2-1) )e^{ \theta'}- ( i /\sqrt{2} \log(b_1)- i/2 e^{i\pi/4} \log(1/u_2-1) )e^{- \theta'}  \right) \frac{\cosh 2 \theta'}{\sinh 2\theta'}\log(1+e^{- %\tilde \epsilon(\theta'-i \hat \varphi)})}
\eqn\tbaAoneone{A_{1,1} =
{1 \over \sqrt{2} \pi}\int d\theta'  \left( - i \sqrt{2}  \log(b_1) \cosh \theta' + i \log(1/u_2-1) )\cosh (\theta' - i \pi/4)  \right) \frac{\cosh 2 \theta'}{\sinh 2\theta'}\tilde L(\theta')}
%   \eqn\tbaAonetwo{A_{1,2} =  ( - i /\sqrt{2}  \log(b_1)+ i/2 e^{-i\pi/4} \log(1/u_2-1) )
 %\left(
 %  +\frac{1}{2\pi}\int d\theta' \frac{e^{ \theta'}\sinh 2 \theta'}{\cosh 2 \theta'} L(\theta') \right)
 %+( i /\sqrt{2} \log(b_1)- i/2 e^{i\pi/4} \log(1/u_2-1) )\left(
  % -\frac{1}{2\pi}\int d\theta'  \frac{e^{-\theta'}\sinh 2 \theta'}{\cosh 2 \theta'} L(\theta') \right)
   %}
 \eqn\tbaAonetwo{A_{1,2} =  \frac{1}{2\pi}\int d\theta'  \left( - i \sqrt{2}  \log(b_1) \cosh \theta' + i \log(1/u_2-1) )\cosh (\theta' - i \pi/4)  \right)  \frac{\sinh 2 \theta'}{\cosh 2 \theta'} L(\theta') }

 Next we would like to replace the term in parenthesis with the $\theta$ derivatives of $\tilde \epsilon$ and $\epsilon$ respectively, to allow for a useful integration by parts.
 The difference between the terms in parenthesis and the $\theta$ derivatives will generate new terms to be added to $A_2$: $A_{1,1} = A_{1,1}' + A_{2,1}+ A_{2,2}$ and $A_{1,2} = A_{1,2}' + A_{2,3}+ A_{2,4}$.
 \eqn\tbaAoneone{A_{1,1}' =
{1 \over 2 \pi}\int d\theta' \partial_{\theta'} \tilde \epsilon(\theta' -i \hat \varphi)  \frac{\cosh 2 \theta'}{\sinh 2\theta'}\tilde L(\theta')}
  \eqn\tbaAonetwo{A_{1,2}' =  \frac{1}{2\pi}\int d\theta' \partial_{\theta'} \epsilon(\theta' -i \hat \varphi) \frac{\sinh 2 \theta'}{\cosh 2 \theta'} L(\theta') }

   \eqn\tbaAzeroone{A_{2,1} =
{1 \over \sqrt{2} \pi}\int d\theta  {1 \over \sqrt{2} \pi}\int d\theta' \partial_\theta \left( \frac{\sinh 2 \theta}{\cosh(\theta- \theta') \sinh 2 \theta'}\right) \tilde L(\theta')\frac{\cosh 2 \theta}{\sinh 2\theta}\tilde L(\theta)}
     \eqn\tbaAzerotwo{A_{2,2} =
{1 \over \sqrt{2} \pi}\int d\theta \frac{1}{\pi}\int d\theta' \partial_\theta \left( \frac{\sinh 2 \theta \sinh (\theta-\theta')}{\cosh 2 (\theta-\theta') \cosh 2 \theta'} \right)L(\theta') \frac{\cosh 2 \theta}{\sinh 2\theta}\tilde L(\theta)}
  \eqn\tbaAzerothree{A_{2,3} =  \frac{1}{2\pi}\int d\theta {\sqrt{2} \over \pi}\int d\theta' \partial_\theta \left( \frac{\cosh 2 \theta \sinh(\theta-\theta')}{\cosh 2(\theta-\theta')\sinh 2 \theta'}\right) \tilde L(\theta')  \frac{\sinh 2 \theta}{\cosh 2 \theta} L(\theta) }
\eqn\tbaAzerofour{A_{2,4} =  \frac{1}{2\pi}\int d\theta \frac{1}{2\pi}\int d\theta' \partial_\theta \left( \frac{\cosh 2 \theta}{\cosh (\theta - \theta') \cosh 2 \theta'}\right) L(\theta') \frac{\sinh 2 \theta}{\cosh 2 \theta} L(\theta) }

On the other hand, we can decompose $A_2 = A_{2,5} + A_{2,6} + A_{2,7}$, averaging the two expressions and symmetrizing
     \eqn\tbaAzerofive{A_{2,5} =   - {1 \over \sqrt{2}\pi}\int d\theta {1 \over \sqrt{2} \pi}\int d\theta' \frac{\cosh2(\theta+\theta') \cosh(\theta - \theta')}{\sinh 2\theta \sinh 2\theta'}
     \tilde L(\theta')\tilde L(\theta)}

        \eqn\tbaAzerosix{A_{2,6} = - {\sqrt{2} \over \pi}\int d\theta \frac{1}{2\pi}\int d\theta' \frac{\sinh 2(\theta + \theta') \cosh(\theta-\theta')}{\cosh 2 \theta' \sinh 2\theta}
        L(\theta') \tilde L(\theta)}
   \eqn\tbaAzeroseven{A_{2,7} = -\frac{1}{2\pi}\int d\theta  \frac{1}{2\pi}\int d\theta'   \frac{\cosh2(\theta+\theta') \cosh(\theta - \theta')}{\cosh 2\theta \cosh 2\theta'}   L(\theta') L(\theta)}

 Combining pieces together vast simplifications occur
      \eqn\tbaAzerofveone{A_{2,1} + A_{2,5} =   {1 \over \sqrt{2}\pi}\int d\theta {1 \over \sqrt{2} \pi}\int d\theta' \frac{1}{\sinh 2\theta \sinh 2\theta' \cosh (\theta - \theta') }
     \tilde L(\theta')\tilde L(\theta)}

        \eqn\tbaAzerotwothreesix{A_{2,2} + A_{2,3}+A_{2,6} =  {\sqrt{2} \over \pi}\int d\theta \frac{1}{\pi}\int d\theta' \frac{\sinh (\theta - \theta')}{\cosh 2 \theta' \sinh 2\theta \cosh(2 \theta - 2 \theta')}
        L(\theta') \tilde L(\theta)}
   \eqn\tbaAzerofourseven{A_{2,4} + A_{2,7} = -\frac{1}{2\pi}\int d\theta  \frac{1}{2\pi}\int d\theta'   \frac{1}{\cosh 2\theta \cosh 2\theta' \cosh (\theta - \theta')}   L(\theta') L(\theta)}

On the other hand, the pieces to be integrated by parts give
 \eqn\tbaAoneoneli{A_{1,1}' =
{1 \over \pi}\int d\theta'  \frac{1}{\sinh^2 2\theta'}Li_2(-e^{- \tilde \epsilon(\theta'-i \hat \varphi)})}
  \eqn\tbaAonetwoli{A_{1,2}' =  - \frac{1}{\pi}\int d\theta' \frac{1}{\cosh^2 2 \theta'} \left( Li_2(-\mu e^{- \epsilon(\theta'-i \hat \varphi)}) + Li_2(-\mu^{-1} e^{- \epsilon(\theta'-i \hat \varphi)}) \right)}
  Here $Li_2$ indicates Mathematica $PolyLog[2,x]$.

 % The final expression for $-\epsilon_1 \epsilon_{-1} $ has some interesting features. It has a structure which resembles
 % the critical value of the Yang-Yang functional (as in $1003.3964$), especially if we rewrite the integrals in terms of a new rapidity variable.
 %$v = \frac{\cosh 2 \theta}{\sinh 2 \theta}$. We should shift back the integration variables in the integrals involving $\epsilon(\theta - i \hat \phi)$ as $\theta \to \theta+ i \pi/4$
 %so that $\zeta = e^\theta$ uniformly. Then the polylog terms have uniform integration measure  $d v = -\frac{2 d\theta}{\sinh^2 2 \theta}$.
 %The change in the measure also makes the integration kernels symmetric: $\half dv' \frac{\sinh 2 \theta  \sinh 2 \theta'}{\cosh (\theta - \theta')}$,
 % $ \sqrt{2} dv' \frac{\sinh 2 \theta \sinh 2 \theta' \sinh(\theta-\theta'+ i \pi/4)}{\sinh 2(\theta-\theta')}$,    $ \sqrt{2} dv' \frac{\sinh 2 \theta' \sinh 2 \theta \sinh(\theta-\theta'- i \pi/4)}{\sinh 2(\theta-\theta')}$,
 % $ d v' \frac{\sinh 2 \theta  \sinh 2 \theta'}{\cosh (\theta - \theta')}$.
% Also, $\epsilon_1 \epsilon_{-1} $ is exactly the critical value of the Yang functional for the modified TBA.
In summary, the final result for $A_{periods} + A_{free}$ is given by \Azerobis\ plus
the critical value of the Yang-Yang functional which is the sum of\tbaAzerofveone -\tbaAonetwoli .
This is the Yang-Yang functional for the integral equations in \tbathree \tbafour , written in terms of
the new variable $du = - { 2 d\theta \over \sinh^2 2 \theta }$ which makes the kernels symmetric.
A shift in the integration contour in \tbaAonetwoli\ takes it to the form in \YYc .

     \subsec{Easy pieces of the hexagon}

   Here we focus on the   terms of the strong coupling answer for the hexagon which contain no integrals.
In particular we focus on such terms for the reminder function. These terms come from \Azerobis\ plus
  $A_{BDS-like}-A_{BDS}$, given in \AGM .
  This gives\foot{Recall the relative sign between the remainder function and
  the area $R \sim \log W \propto - {\rm Area} $. We are ignoring a factor of $\sqrt{\lambda}/(2 \pi)$. }
   \eqn\easyfirst{
R_{easy} = -{1 \over 4} \left( \log^2 b_1 +\log^2(b_3-1/b_1)\right)+{1 \over 8} \sum_i \left( \log^2 u_i +2 Li_2(1-u_i)\right)
 }
 we have $u_1={1 \over b_2 b_3},~~u_2={1 \over b_1 b_3},~~u_3={1 \over b_1 b_2}$. We can write everything in terms of $b_1,b_3$ and $\mu=-2\cos \phi$ by using $b_2={b_1+b_3-2 \cos \phi \over b_1 b_3-1}$. The limit we are interested in is then $b_1$ and $b_3$ of the same order and very large with  $\phi$ fixed. One can explicitly check that the above combination is finite in such limit. One question we would like to answer is whether we could
 get problematic  terms of the form $u_2^p \log^q u_2$ in the small $u_2$ expansion. Such terms could only come from the following piece in the above answer (remember that $u_1+u_3=1$ in the limit.):
 \eqn\easyparts{
R_{easy-part}= -{1 \over 4} \left( \log^2 b_1 +\log^2(b_3-1/b_1)\right)+{1 \over 8}\left( \log^2 u_2 +2 Li_2(1-u_2)\right)
 }
Note that this can be entirely written in terms of $b_1$ and $b_3$. In order to show that this does not contain the kind of terms mentioned above we use: $Li_2(1-u_2)=-Li_2(u_2) -\log u_2 \log(1-u_2)+{\pi^2 \over 6}$. Furthermore, we call $z^2={b_1 \over b_3}$, then we obtain
 \eqn\easyparts{
R_{easy-part}={\pi^2 \over 24}-{1 \over 4} \left( \log^2(1-u_2) -2 \log(1-u_2) \log z+2 \log^2 z-Li_2 u_2 \right)
 }
 From this is clear that in the small $u_2$ expansion we will not get problematic terms, just simple power series. Furthermore, in this form it is very easy to expand.

\subsec{Expanding the hexagon at strong coupling}

We choose the following six points, $P=( Z_{-1},Z_0,Z_1,Z_2,Z_3,Z_4) $, as
\eqn\sixpoints{
\eqalign{ &  P_1= ( 1,   e^{- 2 \tau } ,   e^{- 2 \tau}, 1,0,0) ~,~~~~~~~ P_3 =
 ( e^{-2 \tau} ,1,-1,  - e^{- 2 \tau }     ,0,0)
  ~,~~~~~~~~
  \cr
  & P_2 = ( \sinh (\tau+\sigma) , \sinh (\tau-\sigma), -\cosh (\tau-\sigma) , \cosh (\tau+\sigma)   , - i \sqrt{2} \cos \phi ,
   i \sqrt{2} \sin \phi)
 \cr
 & P_6= ( 1,   1 ,  1, 1,0,0) ~,~~~~~~~P_4 =
 ( 1 ,1,-1,  - 1 ,    ,0,0)
  ~,~~~~~~~~
    P_5 = ( 0 ,0, 1 , -1  , i \sqrt{2},0)
  }}
 We then define $u_{i} = { d^2_{i+1,i+5} d^2_{i+2,i+4} \over d^2_{i+1,i+4} d^2_{i+2,i+5}} $. We
 obtain \writecr .
 We also define $b_i = \sqrt{ u_i \over u_{i+1} u_{i+2} } $. These are
\eqn\bdefi{ b_2 =  2 (  \cosh \tau \cosh \sigma -\cos \phi )/\sinh^2 \tau ~,~~~~~b_1 = e^\sigma \cosh \tau ~,~~~~
b_3 = e^{-\sigma} \cosh \tau
 }
This implies that $\mu$ defined via $\mu + \mu^{-1} = b_1 b_2 b_3 - b_1 -b_2- b_3 = -2 \cos \phi$, is
$\mu = - e^{i \phi}$.

%We now want to expand the Yang Yang function to order $e^{-2 \tau}$. This implies that we will
%keep only terms that have a single power of $e^{-\epsilon}$ and $e^{-\tilde \epsilon}$. Thus, only
%only the term involving the dilogarithm in the Yang Yang function REF contributes. Expanding the di-logarithm
%to first order and summing over the two possibilities we find the pieces involving integrals over $\theta$
%in
%$R_{\sqrt{2}}$ and $R_2$ in \resex .
%To get the terms in $R_1$ and the terms with no integrals in $R_2$ we simply need to expand
The remainder function is
$R = - { \sqrt{ \lambda} \over 2 \pi }  ({\rm Area } )$.
 We will suppress the factor of ${\sqrt{\lambda } \over
2 \pi }$ but we will take into account the minus sign.
 Let us now expand the various terms to the desired order.
 Inserting \bdefi\ and \writecr\ into
 \easyfirst\ and expanding we get
\eqn\expansrz{ \eqalign{
R_{\rm easy } \sim &  - \cos \phi e^{-\tilde \tau} [ \cosh \sigma \log[ 2 \cosh \sigma] - \sigma \sinh \sigma )
+
\cr
  & + e^{ - 2 \tilde \tau} \left[ { \log[ 2 \cosh \sigma] - \sigma \over 2 }
 +  \cos 2 \phi  \, g(\sigma) \right]
\cr
& g(\sigma) = - { 1 \over 4 } \cosh 2 \sigma \log[ 2 \cosh \sigma] + { 1 \over 4 } \sigma \sinh 2 \sigma +
{1 \over 8 }
}}
where $e^{ - \tilde \tau} = e^{-\tau}/2$. There is a constant term that we neglected. Note
that individual terms in \easyfirst\ diverge, but the combination is finite, which is what we
expected for the remainder function.

 The leading order expansion of the Yang-Yang functional comes only from the di-logarithm piece. Expanding
 the di-logarithm in \tbaAoneoneli\ \tbaAonetwoli\ to first order we get
 \eqn\remayang{ \eqalign{
 R_{YY} = & - YY_{cr} \sim { 1 \over \pi } \int d\theta \left[ { 1 \over \sinh^2 2 \theta } e^{-\sqrt{2} E(\theta) }
 -{ ( \mu + \mu^{-1} ) \over \cosh^2 2 \theta} e^{ - E(\theta) } \right]
\cr
  &  E(\theta) = \sqrt{2} [ \tilde \tau \cosh \theta - i \sigma \sinh \theta ]
  }}
  We first used the leading order expressions for $\epsilon(\theta - i \hat \varphi)$ in
  \tbathree ,\tbafour , to get the first line. The second line is the approximate
  expression for \E\  using  the leading order expressions for the cross ratios
\eqn\leador{ \eqalign{
 - \log y_2 =&  \log [ {1 \over u_2 } -1 ] \sim - \log u_2 \sim   - 2 \tilde \tau
  \cr
  \log b_1 = -  & { 1 \over 2 } \log y_1 y_3 \sim  \tilde \tau + \sigma
  }}
Of course, in deriving \expansrz\ it was important to go beyond this leading order expression.
\expansrz\ and \remayang\ are the expansions quoted in the main text \resex , up to the trivial relabeling
$\tilde \tau \to \tau$.

\listrefs

 \bye